\documentclass[structabstract]{aa}
\usepackage{graphicx}
\usepackage{txfonts}
\def\nbase{n_\mathrm{B}}
\def\nlambda{n_\lambda}

\def\Vobs{V^\mathrm{obs}}

\def\Fobs{F^\mathrm{obs}}
\def\sigmaV{\sigma_{V,k}}
\def\sigmaF{\sigma_{F,j}}

\def\chir{\chi_\mathrm{r}}
\def\chirmin{\chi_\mathrm{min,r}}

\def\Rin{R_\mathrm{in}}
\def\Tin{T_\mathrm{in}}
\def\Rout{R_\mathrm{out}}

\def\rhoin{n_\mathrm{in}}
\def\thetadisc{\Delta\theta_\mathrm{d}}

\def\micron{\,\mu\mathrm{m}}
\def\Rs{R_\mathrm{s}}

\def\Isl{I^{\mathrm{s}}_\lambda}
\def\Islref{I^\mathrm{s}_{\lambda_0}}
\def\fs{F^\mathrm{s}_{\lambda_0}}
\def\fd{F_\mathrm{d}}
\def\ftot{F_\mathrm{tot}}
\def\PAd{\mathrm{PA}_\mathrm{d}}
\def\W{\mathrm{W}}
\def\m{\mathrm{m}}
\def\Rsun{R_{\sun}}
\def\Lsun{L_{\sun}}
\def\m{\mathrm{m}}

\def\W{\mathrm{W}}
\def\str{\mathrm{str}}

\begin{document}
   \title{Fast ray-tracing algorithm for circumstellar structures (FRACS)}

   \subtitle{II. Disc parameters of the B[e] supergiant CPD-57\degr\,2874 from VLTI/MIDI data \thanks{Based on VLTI/MIDI
observations collected at the European Southern Observatory (ESO), Paranal,
Chile under ESO Programmes 074.D-0101 and 078.D-0213. Also based on observations at the ESO 2.2-m telescope, La Silla, Chile, under agreement with the Observat\'orio Nacional-MCT (Brazil).
}}

\titlerunning{FRACS modelling - disc parameters of CPD-57\degr\,2874 from VLTI/MIDI}
\authorrunning{A. Domiciano de Souza et al.}

\author{
A.~Domiciano~de~Souza\inst{1}
\and
P.~Bendjoya\inst{1}
\and
G.~Niccolini\inst{1}
\and
O.~Chesneau\inst{1}
\and
M.~Borges Fernandes\inst{1,2}
\and \\
A.C.~Carciofi\inst{3}
\and
A.~Spang\inst{1}
\and
P.~Stee\inst{1}
\and
T.~Driebe\inst{4,5}
}
\offprints{A. Domiciano de Souza}
\mail{Armando.Domiciano@obs-azur.fr}

\institute{Laboratoire H. Fizeau - UMR CNRS 6525 - Universit\'{e} de Nice-Sophia
Antipolis (UNS), Observatoire de la C\^{o}te d'Azur (OCA), Campus Valrose,
F-06108 Nice cedex 2, France
\and Observat\'orio Nacional, Rua General Jos\'{e} Cristino, 77, 20921-400, S\~{a}o Cristov\~{a}o, Rio de Janeiro, Brazil
\and Instituto de Astronomia, Geof\'{\i}sica e Ci\^{e}ncias
Atmosf\'{e}ricas, Universidade de S\~{a}o Paulo (USP), Rua do Mat\~{a}o 1226,
Cidade Universit\'{a}ria, S\~{a}o Paulo, SP - 05508-900, Brazil
\and Max-Planck-Institut f\"{u}r Radioastronomie, Auf dem H\"{u}gel 69, D-53121 Bonn, Germany \and Deutsches Zentrum f\"{u}r Luft- und Raumfahrt, Raumfahrt-Agentur, K\"{o}nigswinterer Strasse 522-524, D-53227 Bonn, Germany}

\date{Received ; Accepted}


  \abstract
   {B[e] supergiants are luminous, massive post-main sequence stars exhibiting non-spherical winds, forbidden lines, and hot dust in a disc-like structure. The physical properties of their rich and complex circumstellar environment (CSE) are not well understood, partly because these CSE cannot be easily resolved at the large distances found for B[e] supergiants (typically $\ga 1$~kpc).}
   {From mid-IR spectro-interferometric observations obtained with VLTI/MIDI we seek to resolve and study the CSE of the Galactic B[e] supergiant CPD-57\degr\,2874.}
   {For a physical interpretation of the observables (visibilities and spectrum) we use our ray-tracing radiative transfer code (FRACS), which is optimised for thermal spectro-interferometric observations. }
   {Thanks to the short computing time required by FRACS ($<10$~s per monochromatic model), best-fit parameters and uncertainties for several physical quantities of CPD-57\degr\,2874 were obtained, such as inner dust radius, relative flux contribution of the central source and of the dusty CSE, dust temperature profile, and disc inclination.}
   {The analysis of VLTI/MIDI data with FRACS allowed one of the first direct determinations of physical parameters of the dusty CSE of a B[e] supergiant based on interferometric data and using a full model-fitting approach. In a larger context, the study of B[e] supergiants is important for a deeper understanding of the complex structure and evolution of hot, massive stars.}


   \maketitle
%

\section{Introduction \label{introduction}}

The B[e] stars form a heterogeneous group composed of objects at different evolutionary stages, but with many similar observational signatures characterising the so-called "B[e] phenomenon": simultaneous presence of hydrogen emission lines, low-excitation forbidden and permitted metallic lines in emission, and a significant infrared (IR) excess mainly caused by hot circumstellar dust. Another common property of B[e] stars is the presence of a non-spherical circumstellar environment (hereafter CSE; e.g. Zickgraf 2003).

Lamers et al. (1998) defined five B[e] sub-classes, one of which contains unclassified stars. Miroshnichenko (2007) proposed an additional B[e] sub-class (the FS CMa stars) to explain at least part of the unclassified B[e]-type stars as binaries at a phase of ongoing or recently ended rapid mass transfer and dust formation.

One of the B[e] sub-classes is composed of luminous ($\log (L_\star/\Lsun) \ga 4$) post-main sequence objects: the B[e] supergiant stars (hereafter sgB[e]). Previous spectroscopic and polarimetric observations of sgB[e] (e.g. Zickgraf et al. 1985; Magalh\~{a}es 1992) show that the wind of these massive and luminous stars is composed of two distinct components: (1) a wind of low density and high velocity and (2) a wind of high density and low velocity. Zickgraf et al. (1985) proposed a picture where the sgB[e] winds consist of a hot and fast radiation-driven polar wind and a slow, much cooler and denser (by a factor of $10^2$ or $10^3$) equatorial wind. This disc-like structure provides a natural explanation for the existence of dust around those objects, because the presence of dust requires regions of sufficiently high density and low kinetic temperatures. One possible explanation for this two-component CSE is that rapid rotation of the central star leads to the formation of an equatorial disc because of the combination of rotation-induced bi-stability and rotation-induced wind compression (Lamers \& Pauldrach 1991; Bjorkman 1998; Pelupessy et al. 2000). Other mechanisms, such as binarity, are evoked to explain the disc-like CSE, but in any case, rapid rotation seems to play a key role in the origin of these discs (e.g. Meynet \& Maeder 2006). Owing to their physical characteristics (fast rotation, disc-like CSE, high luminosity, evolutionary status), it has also been suggested (Vink et al. 2009) that sgB[e] might share evolutionary links with rapidly rotating O-stars and long-duration gamma-ray bursts (GRBs).

Because of the large distances of sgB[e] ($\ga 1$~kpc for the closest ones), the geometry and physical structure (e.g. density and temperature distribution) of their CSE could be only quite recently directly probed, thanks to modern high angular resolution (HAR) techniques. For example, Domiciano de Souza et al. (2008) used ESO's VLT/VISIR instrument to directly measure the typical size of the dusty CSE of the sgB[e] MWC\,300 from diffraction-limited mid-IR images. Among HAR techniques, optical/IR long baseline interferometry (OLBI) provides the highest resolving power, allowing the study of sgB[e] CSEs at angular resolutions $\sim1-10$ milliarcseconds (mas).  For example, Millour et al. (2009) combined adaptive optics (VLT/NACO) and OLBI (VLTI/AMBER and VLTI/MIDI) to detect a companion around the sgB[e] HD~87643 and to perform an extensive study of the dusty CSE of the binary system. 

In the examples above as well as in most works based on HAR data of sgB[e], two different strategies are commonly adopted to interpret the observations: (1) geometrical analytical modelling and (2) radiative transfer modelling (e.g. using the Monte Carlo approach). The geometrical analytical models have the advantage to be very fast to calculate, allowing a full model-fitting (for example a $\chi^2$ minimisation) and error estimate of the model parameters (mostly geometrical parameters of the CSE). However, these simple models do not give access to physical parameters of the targets such as temperature and density distributions, optical depths, etc. On the other hand, most radiative transfer models present a consistent description of the physical conditions of the CSE. However, because these models are quite complex, they demand a lot of computing time, which prevents one from exploring a large domain of the parameter space and also from obtaining a good estimate of the uncertainties on the fitted parameters. In this work we adopt a third approach for the data modelling, which tries to keep the advantages of the other approaches, without the drawbacks. To this aim we use our \textit{fast ray-tracing algorithm for circumstellar structures} (FRACS), which is based on a parametrised CSE combined to a simplified radiative transfer (no scattering). A complete description of this algorithm is given in Niccolini, Bendjoya \& Domiciano de Souza (2010; hereafter paper I).

In the present paper we apply FRACS to study the CSE of the Galactic sgB[e] CPD-57\degr\,2874 (also named \object{Hen 3-394}, \object{WRAY 15-535}) based on mid-IR spectro-interferometric observations performed with ESO's VLTI/MIDI beam-combiner instrument. Previous near- and mid-IR interferometric observations of CPD-57\degr\,2874 directly revealed an elongated CSE that is compatible with a disc-like structure formed by gas and dust (Domiciano de Souza et al. 2007; hereafter DS07). However, because only a limited number (four) of baselines was available and since the authors adopted simple analytical models, only geometrical parameters could be derived from this first analysis of CPD-57\degr\,2874. As shown below, the use of FRACS allowed us to confirm the previous results and, most importantly, to derive physical parameters for this Galactic sgB[e].

In Sect.~\ref{observations} we give the log of the VLTI/MIDI observations and describe the data reduction procedure. In Sect.~\ref{feros} we provide a new distance estimate of CPD-57\degr\,2874 obtained from spectroscopic observations with FEROS. A short reminder of the ray-tracing code FRACS is presented in Sect.~\ref{fracs}, followed by the results obtained from a model-fitting analysis of the VLTI/MIDI observations (Sect.~\ref{data_analysis}). A discussion of the results and the conclusions of this  work are presented in Sects.~\ref{discussion} and \ref{conclusions}, respectively.

\section{VLTI/MIDI observations \label{observations}}

\begin{figure}[th]
 \centering
 \sidecaption
  \includegraphics*[width=5.0cm,draft=false]{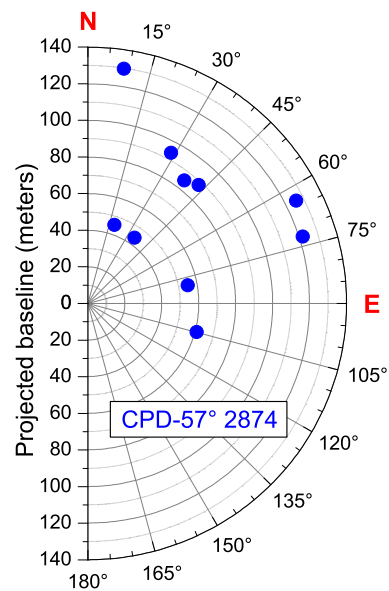}
  \caption{uv-plane coverage: projected baselines (length and position
angle) for the VLTI/MIDI observations of CPD-57\degr\,2874 (further details are given in Table~\ref{tab_midiobs}).}
  \label{fig:Bproj_PA}
%
\end{figure}

The interferometric observations of CPD-57\degr\,2874 were performed with MIDI, the mid-infrared 2-telescope beam-combiner instrument of ESO's VLTI (Leinert et al. 2004). All four 8.2~m unit telescopes (UTs) were used. The N-band spectrum as well as spectrally dispersed fringes have been recorded between $\lambda\simeq7.5\micron$ and $\lambda\simeq13.5\micron$ with a spectral resolution of $R \simeq 30$ using a prism. In total, $\nbase=10$ data sets have been obtained with projected baselines ($B_\mathrm{proj}$) ranging from $\simeq40$\,m to $\simeq130$\,m, and baseline position angles (PA) between $\simeq8\degr$ and $\simeq105\degr$ (from North to East). A summary of the VLTI/MIDI observations of CPD-57\degr\,2874 is given in Table~\ref{tab_midiobs}, and the corresponding uv-plane coverage is shown in Fig.~\ref{fig:Bproj_PA}.

%
\begin{table}[th]
\begin{minipage}[th]{\columnwidth}
\caption{Summary of VLTI/MIDI observations of CPD-57\degr\,2874: data set index, date, Coordinated Universal Time (UTC) of observation, baseline configuration, projected baseline length and position angle.}
\label{tab_midiobs}
\centering
\renewcommand{\footnoterule}{}  
\begin{tabular}{lcccrr}  
%
\hline
\#  &  date  & $t_\mathrm{obs}$ & UT config. &\multicolumn{1}{c} {$B_\mathrm{proj}$} &  \multicolumn{1}{c} {PA}    \\
    &        &     (UTC)        &            & \multicolumn{1}{c} {(m)}  & \multicolumn{1}{c} {(\degr)}  \\
\hline
1  & 2004-11-01 &  08:51:55  & UT2-UT4  &  85.1    &  37.8    \\
2\footnote{Data previously used by DS07.}  & 2004-12-29 &  05:52:12  & UT2-UT3  &  45.2    &  18.6    \\
3$^a$  & 2004-12-29 &  07:26:06  & UT2-UT3  &  43.9    &  35.1    \\
4$^a$  & 2004-12-31 &  06:04:03  & UT3-UT4  &  54.8    &  79.6    \\
5$^a$  & 2004-12-31 &  08:02:48  & UT3-UT4  &  60.9    & 104.8    \\
6  & 2006-11-09 &  07:15:36  & UT1-UT4  & 129.7    &   8.6    \\
7  & 2006-12-13 &  08:35:55  & UT1-UT3  &  93.8    &  28.6    \\
8  & 2006-12-31 &  08:13:05  & UT1-UT4  & 125.8    &  63.5    \\
9  & 2006-12-31 &  08:59:06  & UT1-UT4  & 121.8    &  72.5    \\
10 & 2007-01-05 &  08:35:09  & UT1-UT3  &  88.1    &  42.8    \\
\hline
\end{tabular}
\end{minipage}
\end{table}
%

The MIDI data were reduced with the MIA+EWS data reduction package, which includes two different sub-packages: the MIA package developed at the Max-Planck-Institut f\"{u}r Astronomie, and
the EWS package developed at the Leiden Observatory\footnote{The MIA+EWS software package is available at http://www.mpia-hd.mpg.de/MIDISOFT/ and
http://www.strw.leidenuniv.nl/\textasciitilde nevec/MIDI/index.html.}.

While MIA is based on the power spectrum analysis, which measures the total power
of observed fringes (Leinert et al. 2004), EWS coherently adds the fringes after correction for optical path differences (instrumental as well as atmospheric delays) in each scan (Jaffe et al. 2004). The data reduction results obtained with the MIA and EWS packages agree well within de uncertainties.

The instrumental transfer function at each spectral channel was obtained from the observations of calibrator stars with known uniform-disc diameters ($\diameter_\mathrm{UD}$). The calibrators used in the data reduction and the adopted angular diameters and uncertainties are listed in Table~\ref{tab:calibrators}. The calibrated visibilities were calculated from the ratio of the targets' raw visibilities and the average transfer function derived from the calibrator measurements of the corresponding night (Fig.~\ref{fig:midi_vis_cpd}). The error of the calibrated MIDI visibilities is of the order of $\simeq 5\%-10\%$ and includes the raw visibility error as well as the error of the transfer function. The uncertainties on the calibrator angular diameter are negligible compared to the standard deviation of the transfer function. Usually, $\simeq3-5$ calibrator measurements per night were available (Table~\ref{tab:calibrators}). In the few cases where only one suitable calibrator observation was available we assumed a typical transfer function error of 5\% to estimate the errors on the final calibrated visibilities.

VLTI/MIDI also provides spectral fluxes of CPD-57\degr\,2874 in the N-band. On average, all fluxes are compatible within $\simeq10\%$ with those previously presented by DS07 (MIDI and ISO-SWS spectra). We here considered the VLTI/MIDI spectrum used by DS07 with an uncertainty of 20\% (Fig.~\ref{fig:midi_flux_cpd}). This larger uncertainty ensures a complete agreement between the observed MIDI and ISO fluxes at all wavelengths. We note that the mid-IR spectrum of CPD-57\degr\,2874 does not show any clear evidence of an important silicate feature around $10\,\micron$.

%
\begin{table}[t]
\caption{Observation log and angular diameters of calibrators (values from DS07) used to derive the calibrated N-band visibilities of CPD-57\degr\,2874.}
\label{tab:calibrators}
\centering
\begin{tabular}{ccccc}
%
\hline
date & $t_\mathrm{obs}$ & UT config. & Calibrator & $\diameter_\mathrm{UD}$ \\
     &  (UTC)           &            &            &     (mas) \\
2004-11-01 & 09:28:16   & UT2-UT4 & HD94510 &  $2.16\pm0.11$   \\
2004-12-29 & 04:12:26   & UT2-UT3 & HD37160 &  $2.08\pm0.20$   \\
2004-12-29 & 05:29:32   & UT2-UT3 & HD37160 &  $2.08\pm0.20$   \\
2004-12-29 & 06:13:08   & UT2-UT3 & HD50778 &  $3.95\pm0.22$   \\
2004-12-29 & 07:47:21   & UT2-UT3 & HD94510 &  $2.16\pm0.11$   \\
2004-12-31 & 02:15:59   & UT3-UT4 & HD50778 &  $3.95\pm0.22$   \\
2004-12-31 & 03:04:33   & UT3-UT4 & HD50778 &  $3.95\pm0.22$   \\
2004-12-31 & 06:31:19   & UT3-UT4 & HD94510 &  $2.16\pm0.11$   \\
2004-12-31 & 07:19:17   & UT3-UT4 & HD107446&  $4.54\pm0.23$   \\
2004-12-31 & 07:41:22   & UT3-UT4 & HD94510 &  $2.16\pm0.11$   \\
2006-11-09 & 07:45:21   & UT1-UT4 & HD94510 &  $2.16\pm0.11$   \\
2006-12-13 & 02:04:34   & UT1-UT3 & HD23249 &  $2.33\pm0.01$   \\
2006-12-13 & 08:10:27   & UT1-UT3 & HD94510 &  $2.16\pm0.11$   \\
2006-12-13 & 08:54:41   & UT1-UT3 & HD94510 &  $2.16\pm0.11$   \\
2006-12-31 & 06:58:10   & UT1-UT4 & HD94510 &  $2.16\pm0.11$   \\
2006-12-31 & 07:44:55   & UT1-UT4 & HD94510 &  $2.16\pm0.11$   \\
2006-12-31 & 08:37:22   & UT1-UT4 & HD94510 &  $2.16\pm0.11$   \\
2007-01-05 & 08:08:04   & UT1-UT3 & HD94510 &  $2.16\pm0.11$   \\
\hline
\end{tabular}
\end{table}
%

\section{FEROS observations and distance estimate \label{feros}}

In addition to our mid-IR interferometric observations, we obtained high-resolution optical spectra of CPD-57\degr\,2874. The spectra were recorded with the high-resolution Fiber-fed Extended Range Optical Spectrograph (FEROS), attached to the 2.2-m telescope at ESO in La Silla (Chile). FEROS is a bench-mounted Echelle spectrograph with fibers, covering a sky area of $\simeq2\arcsec\times2\arcsec$ and a wavelength range from 3600\,\AA \ to 9200\,\AA. Its spectral resolution is $R \simeq 55\,000$ (around 6000 \AA). We have adopted its complete automatic online reduction, which includes the heliocentric correction. The FEROS spectra were obtained on 2008 December 21. We recorded two exposures of 1000 seconds with S/N of $\sim60$ in the 5500\,\AA \ region.

These observations were used to estimate the distance of CPD-57\degr\,2874. A previous distance estimate of $d=2.5$~kpc has been proposed by McGregor et al. (1988), assuming that this star belongs to the Carina OB association.

From our FEROS high-resolution spectra it is possible to estimate the distance of CPD-57\degr\,2874, based on the statistical relation cited by Allen (1973). This relation uses the equivalent widths of interstellar \ion{Na}{I} lines. In our data, each \ion{Na}{I} line is composed of two absorption components. However, owing to the lack of data from different epochs, it is impossible to see any temporal changes, which would allow us to derive a possible circumstellar contamination. We have therefore assumed that both \ion{Na}{I} components are of interstellar origin. The measured equivalent widths are given in Table \ref{tab:lines}. Our estimated distance for CPD-57\degr\,2874 is $d=1.7$~kpc with an uncertainty of 0.7~kpc. This large error is firstly due to a possible contamination from the circumstellar emission component and the saturation of the absorption one, and secondly to a systematic error caused by the statistical relation used. Within the error bars, our distance estimate is roughly compatible with the result of McGregor et al. (1988). We have considered both distances in our analysis: 1.7 and 2.5~kpc.


\begin{table}[t!]
\caption{Equivalent widths (EW) of the \ion{Na}{I} absorption lines ($\lambda$$\lambda$ 5890\,\AA, 5896\,\AA), obtained from our FEROS data. The relative uncertainty of these measurements is about 20\%.}
\label{tab:lines}
\begin{center}
\begin{tabular}{ccccc}
    \hline
 Line & \multicolumn{2}{c}{\textrm \protect{5890$\AA$}} &  \multicolumn{2}{c}{\textrm \protect{5896$\AA$}} \\
      &  abs. comp. 1 & abs. comp. 2 & abs. comp. 1 & abs. comp. 2 \\
    \hline
 W($\AA$) & 0.11 & 0.80 & 0.05 & 0.70 \\
    \hline
\end{tabular}
\end{center}
\end{table}

\section{Description of the model FRACS \label{fracs}}

Here we present a short description of our numerical model FRACS  and the parametrisation adopted to describe a sgB[e]. A full description of FRACS is given in paper I. 

FRACS is based on the ray-tracing technique using quadtree meshes for the grid and the full symmetries of the problem to hand to significantly decrease the computing time necessary to obtain monochromatic images (calculated with $300\times300$ pixels) within seconds. Complex visibilities and fluxes can be directly derived from these monochromatic images. FRACS neglects scattering, an approximation well suited to interpret spectro-interferometric observations in the thermal IR. Indeed, compared to absorption of light from dust, scattering can be neglected in the IR and beyond (paper I; Lamers \& Cassinelli 1999).

To analyse the VLTI/MIDI data of CPD-57\degr\,2874 we adopted the same parametrised description of a sgB[e] (central star and dusty CSE) as given in paper I. Below we summarise the main equations of this description in axis-symmetric spherical coordinates, and define the free parameters used in the model-fitting.

We assume the specific intensity from the central regions of the star to be a power-law with spectral index $\alpha$ and level $\Islref$ at a fiducial wavelength $\lambda_0=10\micron$:
\begin{eqnarray}
  \label{eq:csource}
  \Isl = \Islref\,\left(\frac{\lambda_0}{\lambda}\right)^\alpha \, . 
\end{eqnarray}
This emission from the central region includes a contribution from the stellar photosphere and from the continuum radiation (free-free and free-bound) of the circumstellar ionised gas. The spectral index $\alpha$ is sensitive to the nature of the central source. Panagia \& Felli (1975) and Felli \& Panagia (1981) give theoretical values for the spectral index for spherical envelopes: $\alpha\simeq4$ for a blackbody and $\alpha\simeq2.6$ for a fully ionised gas (free-free emission) with an electron density proportional to $r^{-2}$. Their estimates are valid within the Rayleigh-Jeans domain of the spectrum, which fits to our case when we consider the hot central parts of a sgB[e] in the mid-IR.

A radius $\Rs=54\Rsun$ was adopted for the central region. This value is used simply as a scaling factor and to convert $\Islref$ to the observed $10\micron$ flux from the central region:
\begin{eqnarray}
  \label{eq:fsource}
  \fs=\pi\left(\frac{\Rs}{d}\right)^2\Islref \, .   
\end{eqnarray}
We note that at distances of a few kpc the central regions of a sgB[e] are not resolved by VLTI/MIDI and can thus be considered as point sources. 

For the gas number density we adopt the bi-modal distribution used by Carciofi, Miroshnichenko, \& Bjorkman (2010) to study another B[e] star. Similar density distribution descriptions were adopted for Be stars (e.g. Stee et al. 1995). The adopted distribution is motivated by the two-wind scenario proposed by Zickgraf et al. (1985) and assumes a fast polar wind, a slow equatorial outflow, and a latitude-dependent mass loss rate. The number density of dust grains is therefore given by
\begin{equation}
  \label{eq:density}
  n(r,\theta)=\rhoin\,\left(\frac{\Rin}{r}\right)^2\,\left( \frac{1+A_2}{1+A_1}\right) \,
  \frac{1+A_1\,\left(\sin{\theta}\right)^m}{1+A_2\,\left(\sin{\theta}\right)^m}   \,  ,
\end{equation}
where $(r,\theta)$ are the radial coordinate and co-latitude, and $\rhoin$ is the dust grain number density at $\theta=90\degr$ and at $r=\Rin$, which is the inner dust radius where dust starts to survive. Dust is confined between $\Rin$ and the outer dust radius $\Rout$. The value of $\Rout$ cannot be determined from the VLTI/MIDI data and has been fixed to a high value: $750$~AU (the exact value does not affect our results).

The parameter $A_1$ controls the ratio between the equatorial and polar mass loss rates. More precisely, $(1+A_1)$ specifies the ratio between the mass loss rate per unit solid angle in the equator and the pole. In the bi-modal scenario, the poles are assumed to be much less dense than the equator, therefore $A_1$ must have a large value ($\ga10$). Our models indicate that for a wide range of values this parameter does not have a strong influence on the results because it can always be compensated by $\rhoin$ (see also discussion in paper I). Therefore, we have arbitrarily fixed $(1+A_1)$ to 50.

The parameter $A_2$ indicates how much faster the polar wind  is compared to the slow equatorial wind. $(1+A_2)$ is the equatorial to polar terminal velocity, i.e., $v_\infty(90\degr)/v_\infty(0\degr)$. This parameter is also quite uncertain and in principle can assume values ranging from $\sim1$ to $\sim100$.  We kept $A_2$ as a free parameter, although it is not well constrained from the observations as shown in the next sections.

Finally, parameter $m$ controls how fast the mass loss (and consequently the density) drops from the equator to the pole. Defining the disc opening angle $\thetadisc$ as the latitudinal range within which the mass loss rate is higher then half its equatorial value, we have
\begin{equation}
  \label{eq:thetadust}
  \thetadisc=2
  \arccos{\left(\frac{A_1-1}{2\,A_1}\right)^{\frac{1}{m}}}\simeq
2\arccos{\left(\frac{1}{2}\right)^{\frac{1}{m}}} \,  .
\end{equation}

High $m$ values correspond to thinner regions of high density around the equator. These dense, slowly flowing disc-like regions around the equatorial plane of sgB[e] stars provide favourable conditions for dust to form and survive. Different approaches exist to define regions of dust formation (e.g. Carciofi, Miroshnichenko, \& Bjorkman 2010). Here we adopt the relatively simple assumption where dust is allowed to exist only within the disc opening angle, i.e.,  at co-latitudes between $90\degr-0.5\thetadisc$ and $90\degr+0.5\thetadisc$.

The dust grain opacity was calculated in the Mie theory (Mie 1908) for silicate dust and for a dust size distribution following the commonly adopted MRN (Mathis, Rumpl \& Nordsieck 1977) power-law $\propto a^{-3.5}$, where $a$ is the dust grain radius. The Mie absorption cross sections are computed from the optical indices of astronomical silicate (Draine \& Lee 1984; see also paper I).  One possibility to reproduce the absence of a silicate feature in the N-band spectrum of CPD-57\degr\,2874 is to have relatively large grain sizes. We thus used grain radii ranging from  $a=0.5$ to 50 $\micron$, which can relatively well reproduce the observed spectrum (Fig.~\ref{fig:midi_flux_cpd}). {We checked that ignoring scattering remains a valid assumption for this dust distribution with large grains. By neglecting the dust albedo the visibilities and fluxes are affected by only a few percent ($\la3.5\%$) within the N-band (further details in paper I).}

The temperature structure of the dusty CSE is given by
\begin{equation}
  \label{eq:temperature}
  T(r)=\Tin\,\left(\frac{\Rin}{r}\right)^\gamma  \,  ,  
\end{equation}
where $\Tin$ is the dust temperature at the disc inner radius $\Rin$, i.e., the dust sublimation temperature. To be consistent with our choice of dust composition we require that $\Tin\leq1500$~K. The coefficient $\gamma$ is expected to assume values $\la1$.

Finally, because OLBI is sensitive to the projection of the object's intensity distribution onto the sky, there are two angles related to this projection:

\begin{itemize}
  \item the inclination of the disc plane towards the observer $i$ ($0\degr$ for pole-on view and $90\degr$ for equator-on view).
  \item the position angle (from North to East) of the maximum elongation of the sky-projected disc $\PAd$. This angle is defined for $i \ne 0\degr$.
\end{itemize}

Thus, the 10 free parameters ($n_\mathrm{free}$) of the model are: $\Islref$, $\alpha$, $\Tin$, $\gamma$, $\Rin$, $i$, $\PAd$, $A_2$, $\rhoin$, and $m$.

\section{Model-fitting with FRACS \label{data_analysis}}

Here we use FRACS with the parametrised sgB[e] description defined in the last section in order to interpret the VLTI/MIDI observations of CPD-57\degr\,2874 through a model-fitting procedure. 

To ensure spectrally independent observations for the model-fitting we decided to consider one data point every $\simeq0.5\micron$ between $8\micron$ and $13\micron$. This step approximately corresponds to twice the used spectral resolution width ($\Delta\lambda=\lambda/R$). Additionally, to avoid poor visibility calibration owing to the Earth signature of ozone around $9.6\micron$ we have not included observations in this spectral region in our analysis. Finally, the same spectral sampling ($\nlambda=10$  wavelengths points) was adopted for the VLTI/MIDI visibilities and spectrum. This choice also provides faster calculations because it is not necessary to compute model images at too many wavelengths. 

We have performed a $\chi^2$ minimisation simultaneously on the VLTI/MIDI visibilities and fluxes using a Levenberg-Marquardt (LM) algorithm (Markwardt 2008). In order to treat the visibilities and fluxes on the same level (similar weights) we have minimised a  $\chi^2$ like quantity defined as (see further details in paper I):
\begin{eqnarray}
  \label{eq:chi2}
  \chi^2=\sum\limits_{j=1}^{\nlambda}\!\sum\limits_{k=1}^{\nbase}\,
  \left[
    \left(\frac{\Vobs_k-V_k}{\sigmaV }\right)^2 +
  \left(\frac{\Fobs_j-F_j}{\sigmaF}\right)^2\right] ,
\end{eqnarray}
where $\Vobs_k$ and $V_k$ are the observed and modelled visibility modulus for baseline index $k$, $\Fobs_j$ and $F_j$ are the observed and modelled mid-IR fluxes for wavelength index $j$. $\sigmaV$ and $\sigmaF$ are the estimated errors on the visibilities and fluxes.

The starting parameter values for the fit were determined from physical considerations of the CSE and from the previous results from DS07. Below we consider the reduced $\chi^2$ defined by $\chir^2=\chi^2/(2\nbase\nlambda-n_\mathrm{free})$, where $n_\mathrm{free}=10$. The LM algorithm stops when the relative decrease in $\chir^2$ is less then $10^{-3}$. For the CPD-57\degr\,2874 data, the LM algorithm reaches the $\chir^2$ minimum ($\chirmin^2$) in a few hours ($\simeq2-3$~h) on a single CPU.

\begin{figure}[t!]
\centering
\includegraphics[width=0.9\hsize,draft=false]{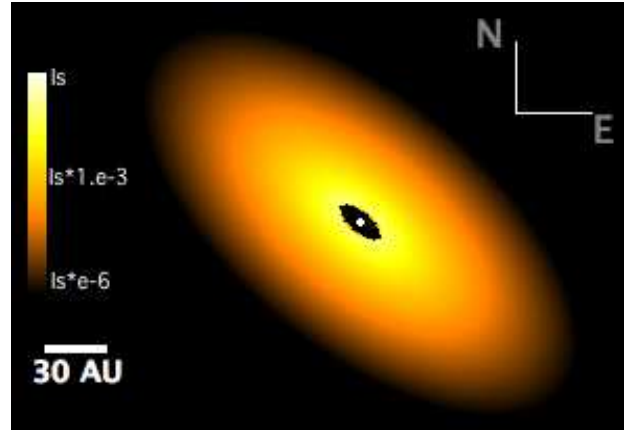}
\caption{Intensity map of CPD-57\degr\,2874  at $10\micron$ for the best-fit FRACS model obtained for a distance $d=1.7$~kpc (see Table~\ref{tab:fit_parameters}). The image scale is in log of the specific intensity $\Isl$. \label{fig:fracs_img_cpd}}
\end{figure}

\begin{figure*}[t!]
\centering
\includegraphics[width=\hsize,draft=false]{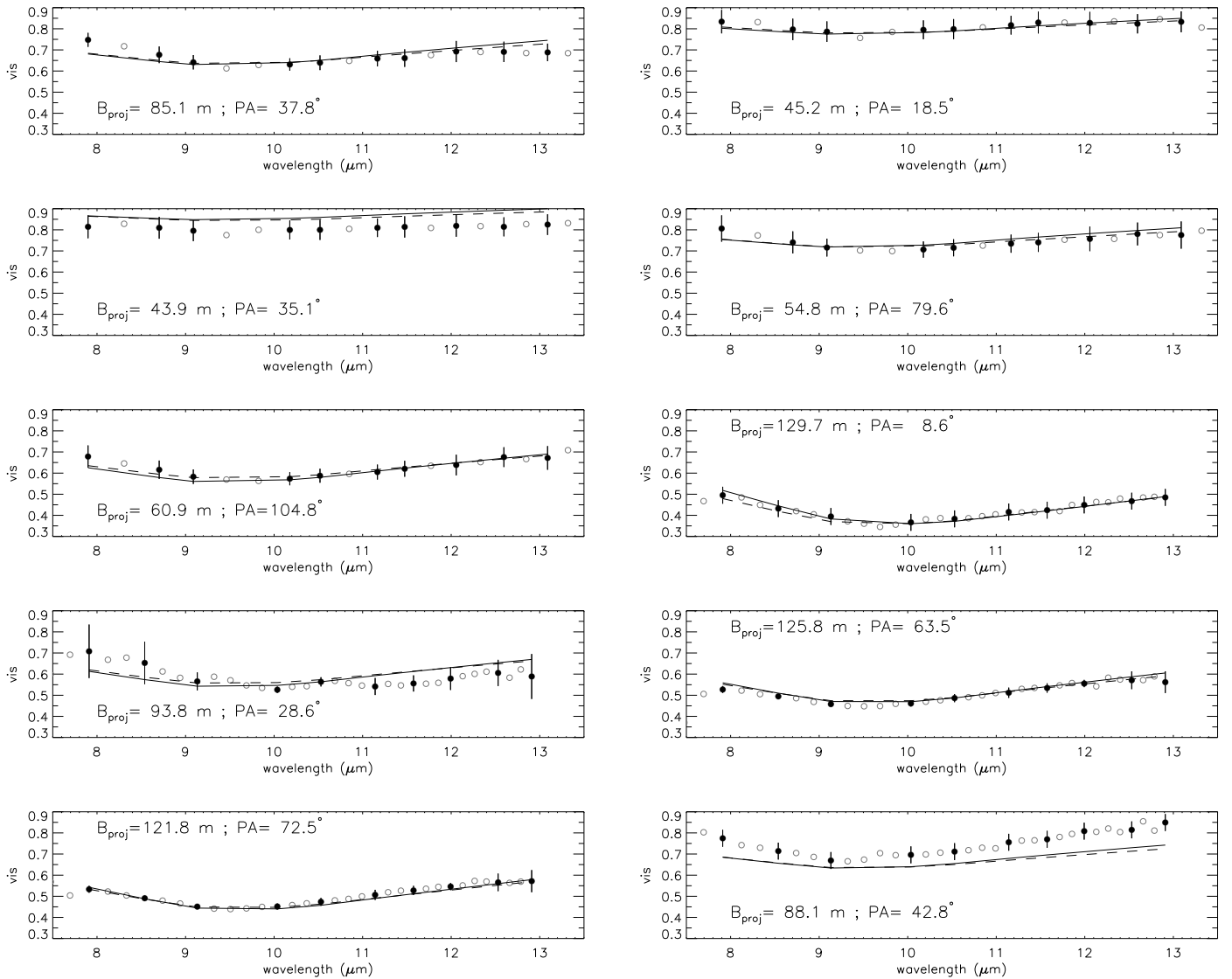}
\caption{VLTI/MIDI visibilities of CPD-57\degr\,2874 (circles) and the best-fit FRACS visibilities obtained from a $\chi^2$ minimisation for a distance $d=1.7$~kpc (solid curve) and  $d=2.5$~kpc (dashed curve). The model-fitting was performed simultaneously on the visibilities and spectral flux. The visibilities effectively used for the fit are shown as filled circles together with the corresponding visibility error bars. The good quality of this fit is reflected by a reduced $\chi^2_\mathrm{min}$ of $\simeq0.55$ for both distances (see details in Table~\ref{tab:fit_parameters}). \label{fig:midi_vis_cpd}}
\end{figure*}

\begin{figure}[t!]
\centering
\includegraphics[width=0.9\hsize,draft=false]{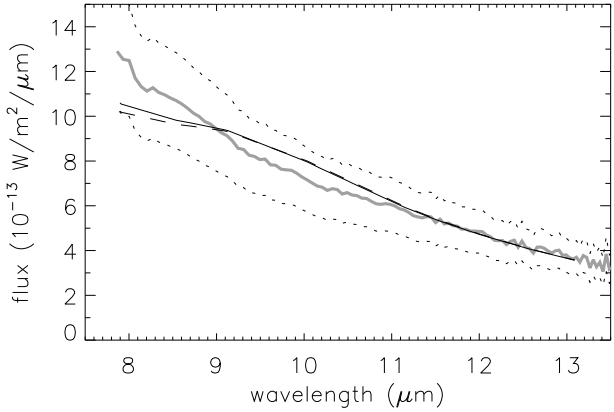}
\caption{VLTI/MIDI flux of CPD-57\degr\,2874 (thick solid grey curve) and the $\pm20\%$ adopted uncertainty (dots). The thin solid and dashed curves are the best-fit model fluxes for assumed distances of $d=1.7$~kpc and $d=2.5$~kpc, respectively. The wavelengths used for the fit are those from Fig.~\ref{fig:midi_vis_cpd}. \label{fig:midi_flux_cpd}}
\end{figure}

Figure~\ref {fig:fracs_img_cpd}  shows the intensity map of the model corresponding to $\chirmin^2$ (best-fit model) for our distance estimate of 1.7~kpc, which also corresponds to the lowest $\chirmin^2$. The visibilities and fluxes for the best-fit model are shown, together with the observations, in Figs.~\ref{fig:midi_vis_cpd} and \ref{fig:midi_flux_cpd}. These plots show that the model well reproduces most observations within their uncertainties for both adopted distances (1.7 and 2.5~kpc). In particular the slightly curved shape of the visibilities is well reproduced by FRACS. The models indicate that this curved shape is probably caused by the combined fact (1) that the intensity maps have different relative contributions from the central source and from the dusty CSE at different wavelengths, (2) that the optical properties of the adopted dust grains are wavelength-dependent even if there is no strong silicate feature seen in the spectrum, and (3) that the angular resolution significantly changes along the observed wavelengths. 

The model parameters at $\chirmin^2$ and their uncertainties are listed in Table~\ref{tab:fit_parameters}. The derived parameters are almost independent of the adopted distance, except of course for those scaling with the distance. The uncertainties of the parameters have been estimated from $\chir^2$ maps calculated with $21\times21$ points for each pair of free parameters (45 pairs) and centred on the $\chirmin^2$ position. All 45 $\chir^2$ maps are shown in Figs.~\ref{fig:chi2_maps_1} to \ref{fig:chi2_maps_3} for $d=1.7$~kpc. These maps show that the $\chir^2$ space presents a well defined $\chirmin^2$, without showing several local minima in the parameter domain explored. Additionally, they provide visual and direct information on the behaviour of the model parameters in the vicinity of $\chirmin^2$, revealing, for instance, potential correlations between certain parameters.


We have estimated the parameter uncertainties in a conservative way by searching for the maximum parameter extension in all $\chir^2$ maps corresponding to $\chirmin^2+\Delta\chi^2$, where $\Delta\chi^2=1$ (see contours in Figs.~\ref{fig:chi2_maps_1} to \ref{fig:chi2_maps_3}). This choice of $\Delta\chi^2$ sets a lower limit confidence region of $\simeq60\%$ to the parameter uncertainties. This limit results from two extreme assumptions about the data:

\begin{itemize}
  \item Data points per baseline are completely dependent (correlated): because the same set of stars is used to calibrate all visibilities of a given baseline, we can consider a limiting case where all these visibilities are correlated. This assumption implies that only 10 independent visibility observations are available (this corresponds to the number of baselines). The flux at each spectral channel can still be considered to be independent. This pessimistic assumption leads to the lower limit of $\simeq60\%$ to the formal confidence level for $\Delta\chi^2=1$, corresponding to only 20 independent observations (10 baselines and 10 fluxes).
  \item All data points are completely independent (uncorrelated): an upper limit of $\simeq100\%$ of formal confidence level is obtained if we assume that all data points are independent. Then the uncertainties derived from $\Delta\chi^2=1$ are very conservative (overestimated).
\end{itemize}

Hence the parameter uncertainties given in Table~\ref{tab:fit_parameters} correspond to a confidence level  of at least $60\%$, but most probably they are somewhat overestimated. 

In the next section we present a physically motivated discussion of the derived model parameters of CPD-57\degr\,2874.

%

%
\begin{table}
\begin{minipage}[t]{\columnwidth}
\caption{Best-fit model parameters and uncertainties derived for CPD-57\degr\,2874 from a $\chi^2$ minimisation. The uncertainties were estimated from the $\chir^2$ maps.}
\label{tab:fit_parameters}
\centering
\renewcommand{\footnoterule}{}  
\renewcommand{\arraystretch}{1.4} 
\begin{tabular}{c | c c | c c }     
\hline
Adopted distance &  \multicolumn{2}{c} {$d=1.7$~kpc} &  \multicolumn{2}{| c} {$d=2.5$~kpc} \\
Reduced $\chi^2_\mathrm{min}$  &  \multicolumn{2}{c} {$\chirmin^2=0.54$}    &   \multicolumn{2}{| c} {$\chirmin^2=0.56$}    \\
\hline
Model parameters & value & error & value & error \\
\hline
$\Islref$ ($10^{5}\,\W\,\m^{-2}\,\micron^{-1}\,\str^{-1}$)  &
$ 2.2$ & $_{-0.7}^{+0.7}$ &   $ 4.2$ & $_{-1.4}^{+1.8}$   \\
$\alpha$  &
$ 2.4$ & $_{-1.2}^{+1.3}$ &   $ 2.4 $ & $_{-1.4}^{+1.4}$   \\
$\Tin$ (K)\footnote{To be physically consistent with the adopted dust composition the upper limit for $\Tin$ is $1500$~K, even 
though the $\chir^2$ maps were allowed to explore higher temperature values.}   &
$1498 $ & $_{-427}^{+1042}$ &   $ 1500 $ & $_{-535}^{+1050} $   \\
$\gamma$  &
$ 1.02 $ & $_{-0.29}^{+0.71} $  &  $ 0.86 $ & $_{-0.24}^{+0.43} $   \\
$\Rin$ (AU)  &  
$ 12.7$ & $_{-2.9}^{+3.6}$ &  $ 14.4 $ & $_{-4.0}^{+5.1}$   \\
$i$ (\degr)  &
$ 61.3 $ & $_{-18.2}^{+10.8}$ &   $ 59.6 $ & $_{-21.2}^{+11.8} $   \\
$\PAd$ (\degr)  &
$ 140.3 $ & $_{-14.0}^{+12.3} $ &  $ 139.4 $ & $_{-15.1}^{+13.7} $   \\
%
$A_2\;$\footnote{Not well constrained.}  &
$ -0.98$ & - &  $ -0.98 $ & -  \\
$\rhoin$ (m$^{-3}$)$^{\;b}$  &
  $ 0.30$ & - &   $ 0.33 $ & -  \\
$m^{\;b}$   &
$ 332$ & - &  $ 377$  & - \\
\hline
Other derived parameters & value & error & value & error \\
\hline
$\fs$ ($10^{-13}\,\W\,\m^{-2}\,\micron^{-1}$)\,\footnote{Observed mid-IR flux from the central region at $\lambda_0=10\micron$ (Eq.~\ref{eq:fsource}).} &
$ 3.5$ & $_{-1.1}^{+1.1}$ &   $ 6.8$ & $_{-2.3}^{+2.8}$   \\
$\Rin/d$ (mas)   & 
 $7.5$ & $_{-1.7}^{+2.1} $  & $5.8$  & $_{-1.6}^{+2.0} $   \\
$\thetadisc$ (\degr)  &
 $7.5$ & -  & $7.0$  &  -  \\
  \hline
\end{tabular}
\end{minipage}
\end{table}



%

%
%
%
\begin{figure*}[ht]
\centering
\includegraphics[width=0.33\hsize,draft=false]{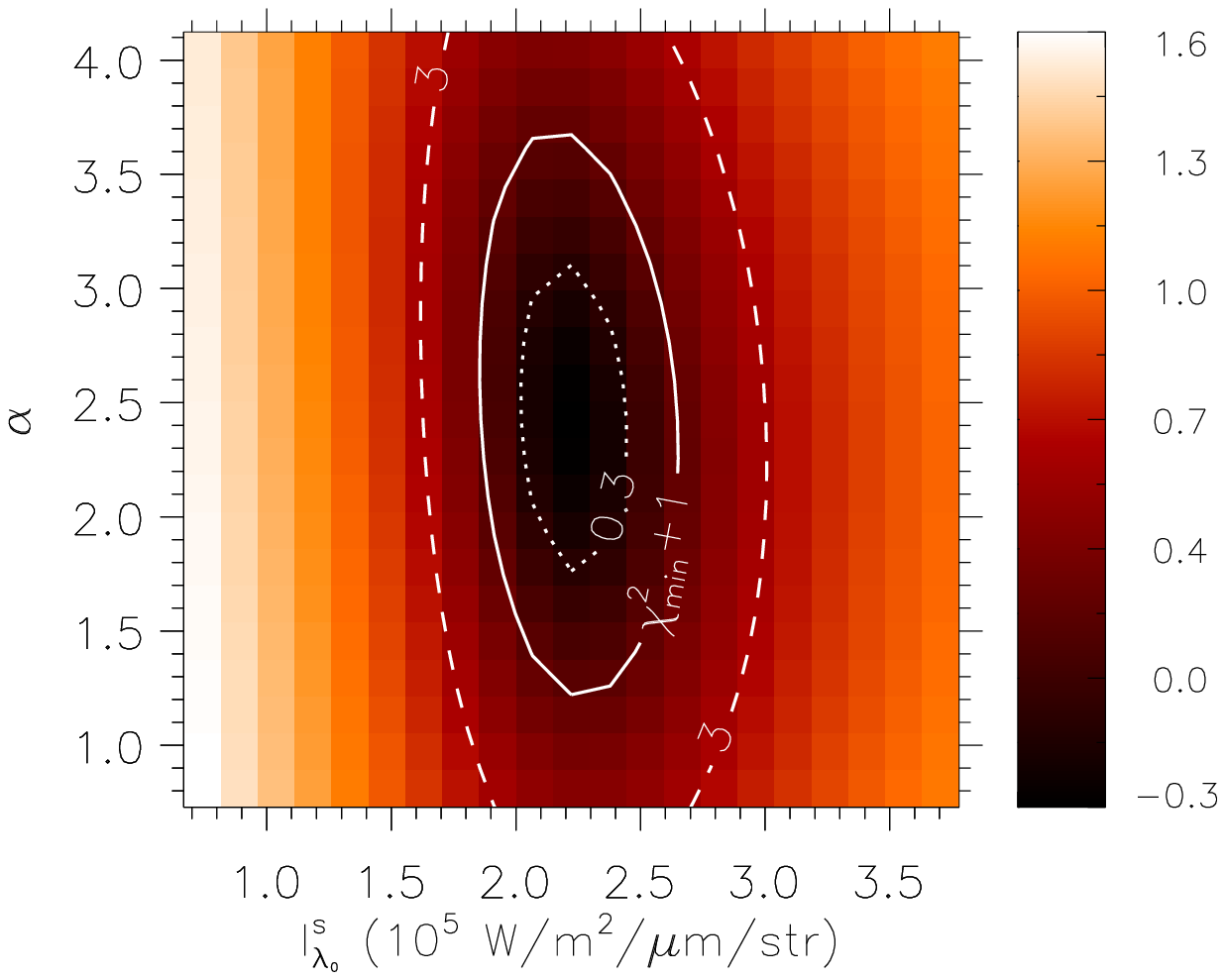}
\includegraphics[width=0.33\hsize,draft=false]{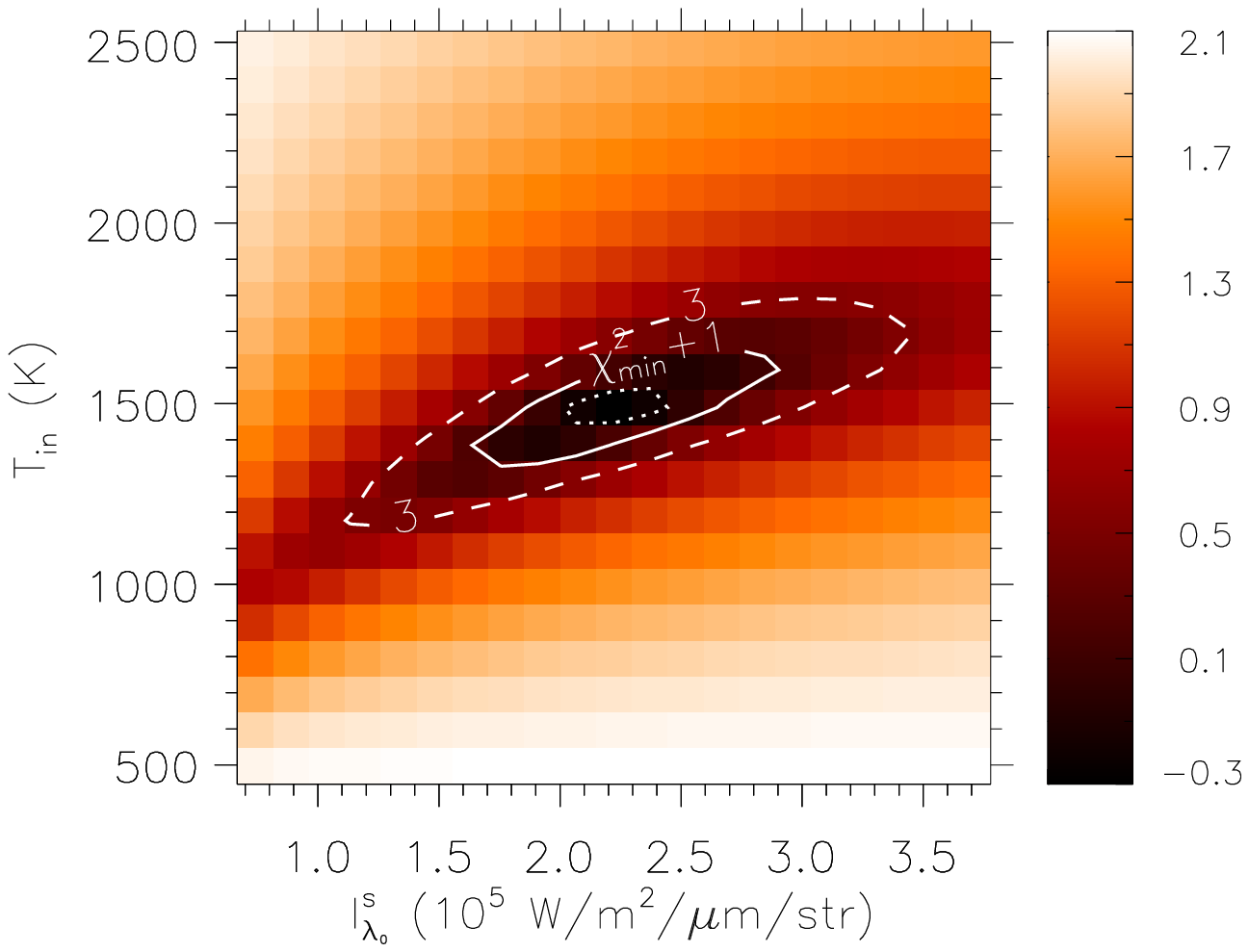}
\includegraphics[width=0.33\hsize,draft=false]{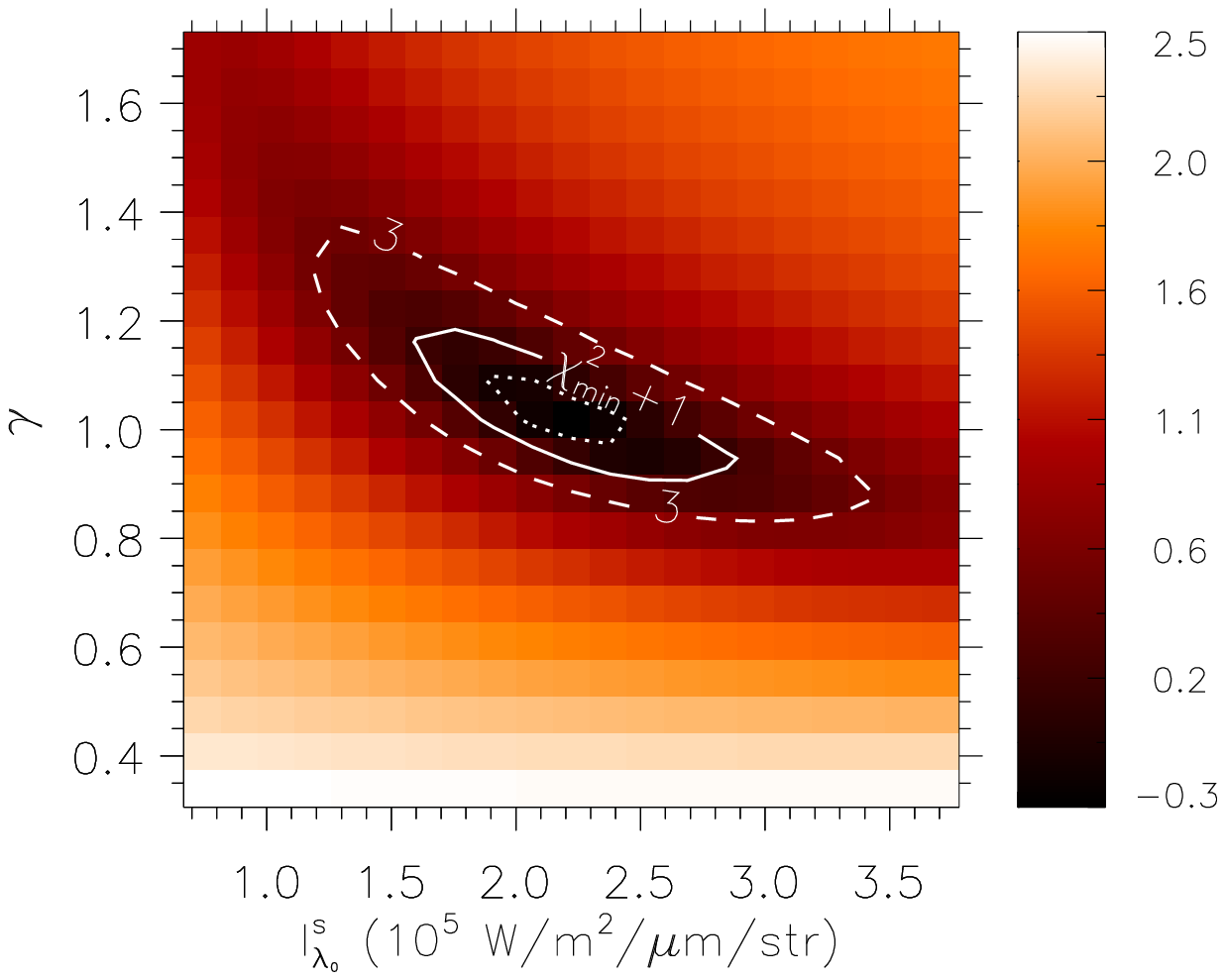}
\includegraphics[width=0.33\hsize,draft=false]{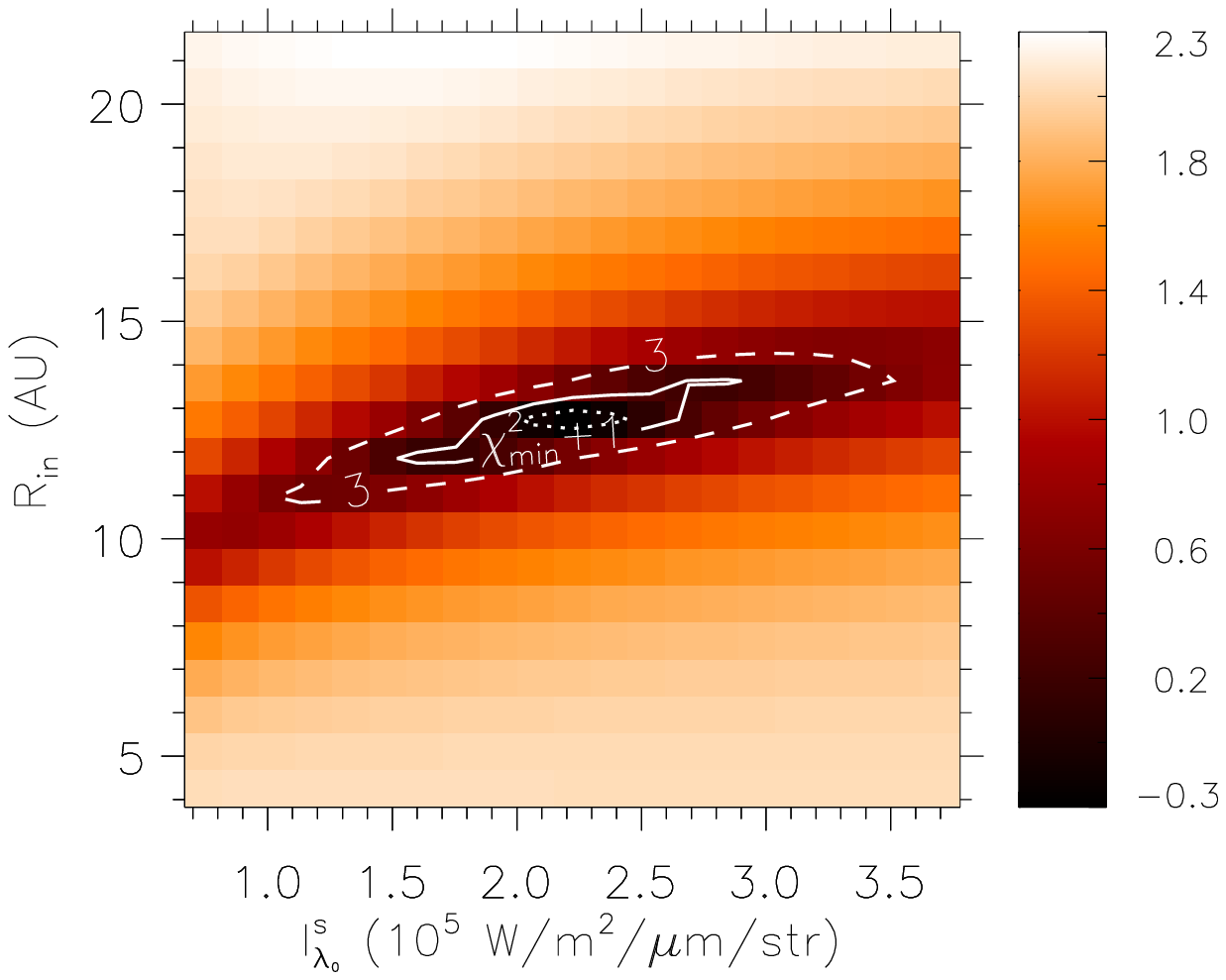}
\includegraphics[width=0.33\hsize,draft=false]{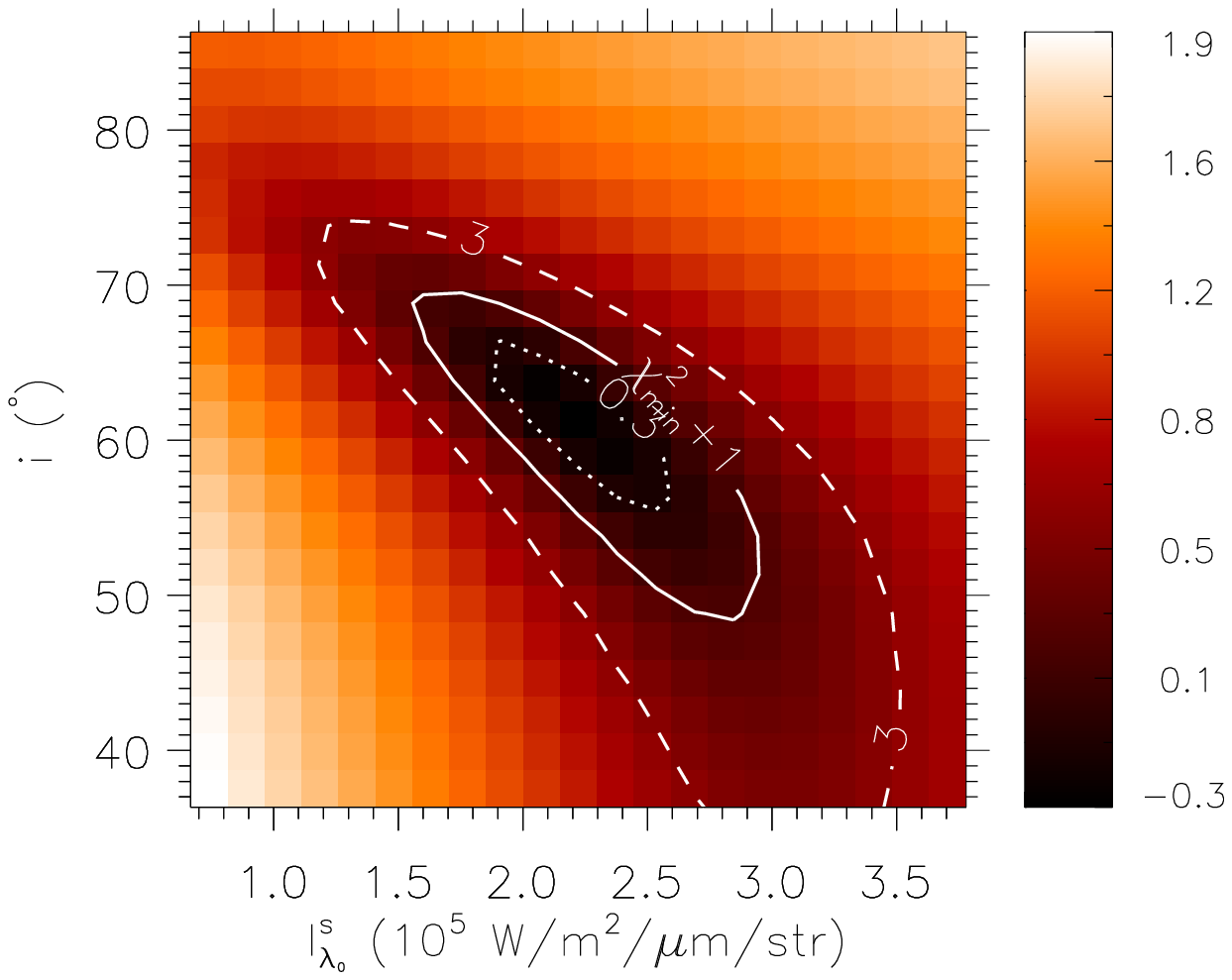}
\includegraphics[width=0.33\hsize,draft=false]{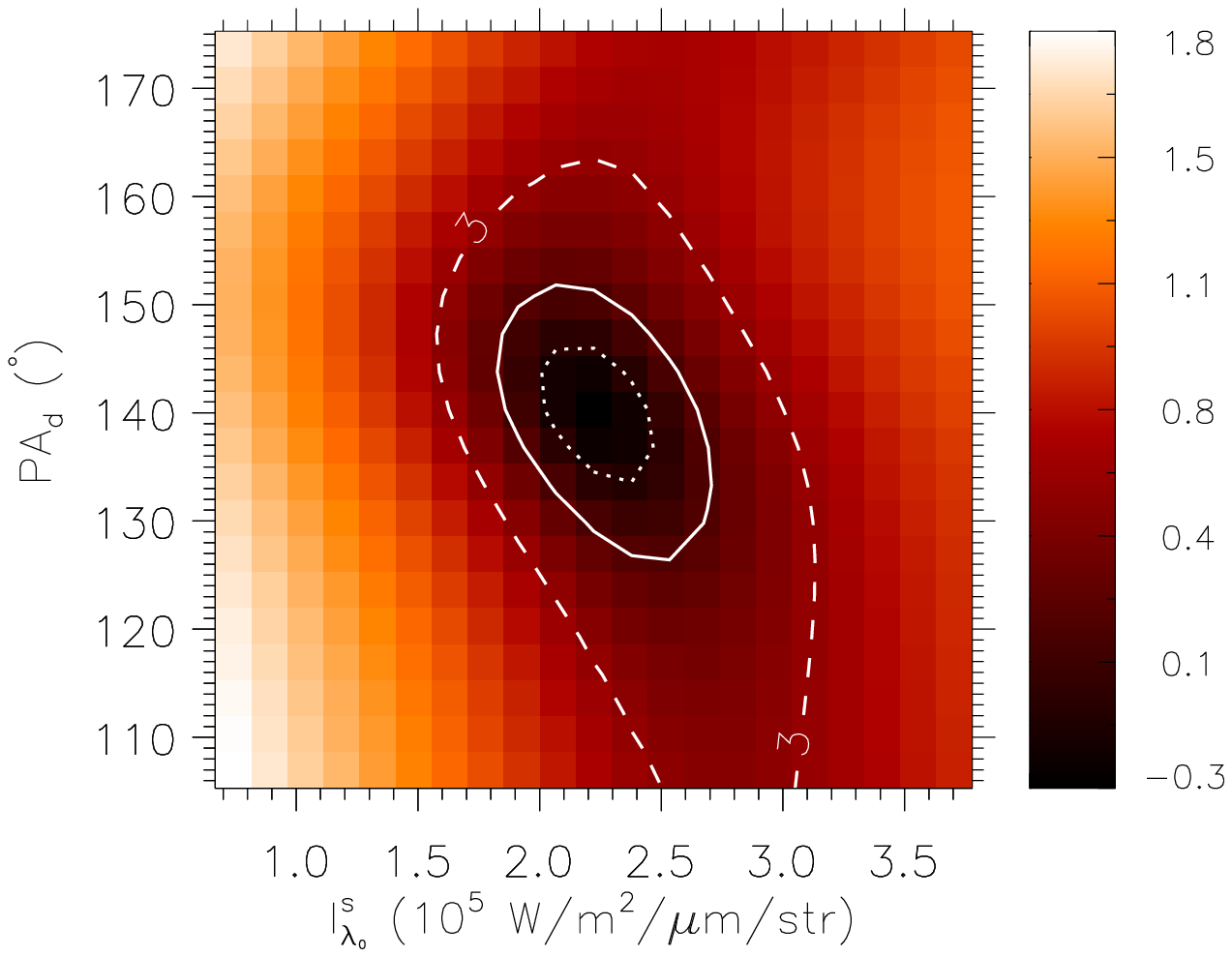}
\includegraphics[width=0.33\hsize,draft=false]{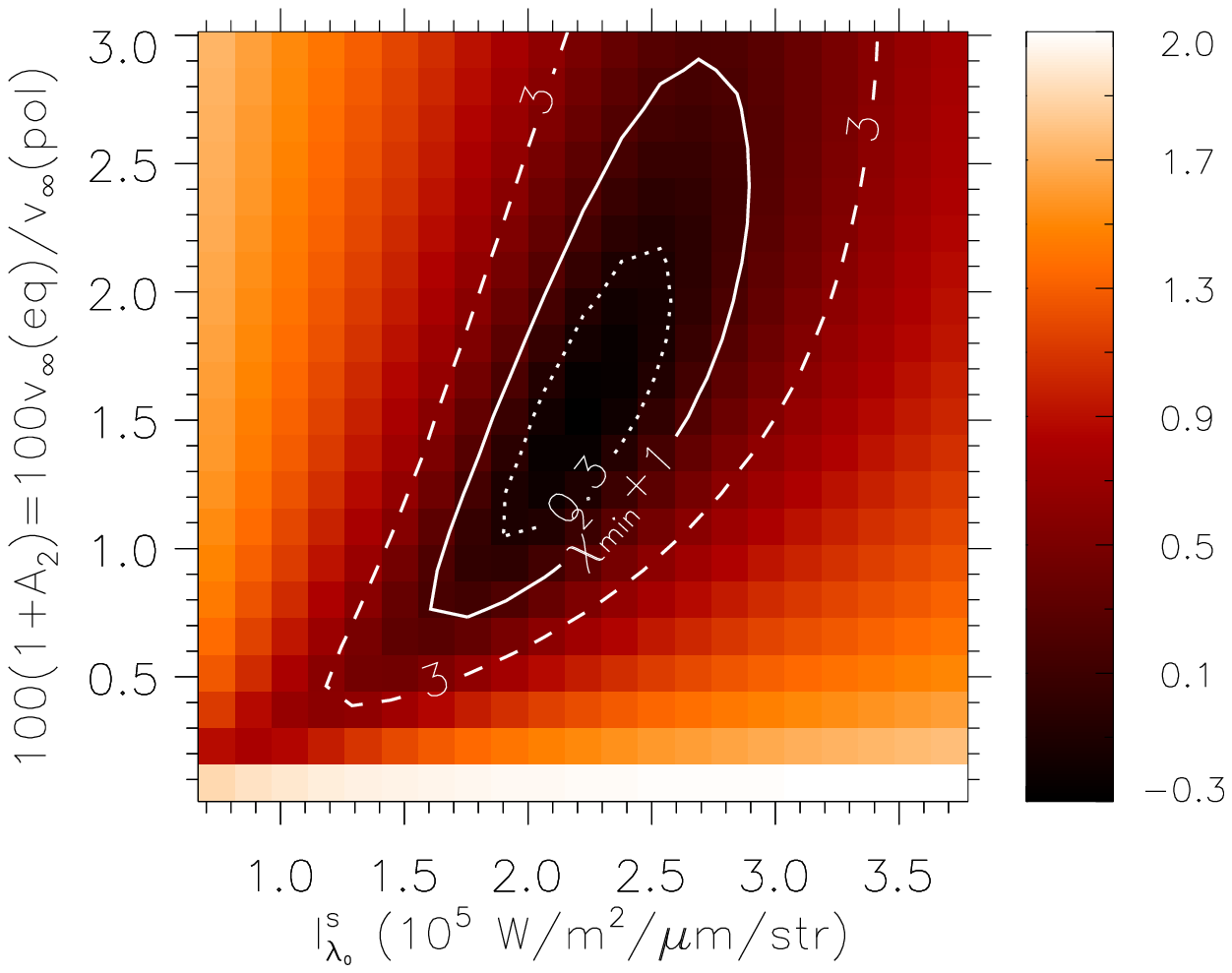}
\includegraphics[width=0.33\hsize,draft=false]{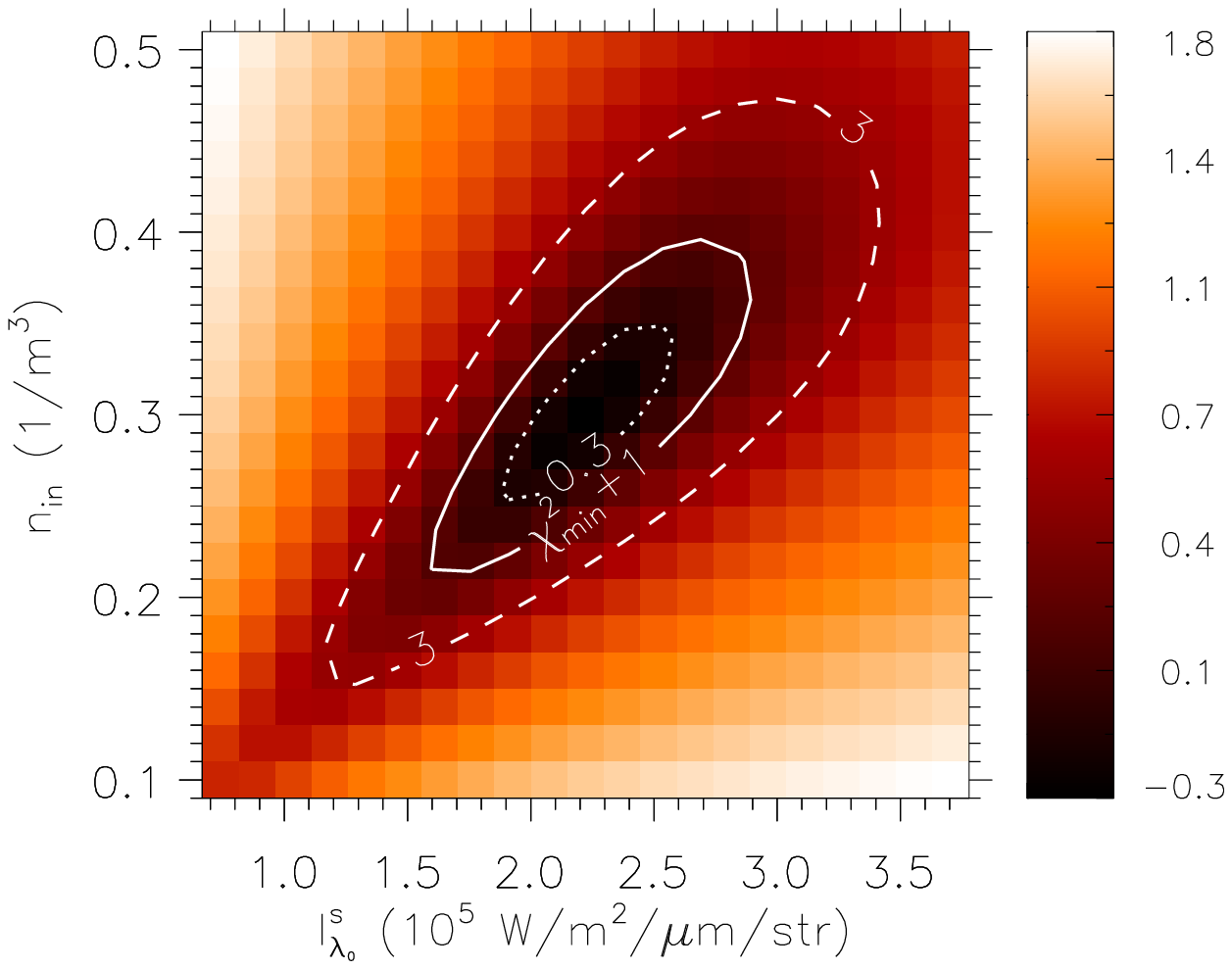}
\includegraphics[width=0.33\hsize,draft=false]{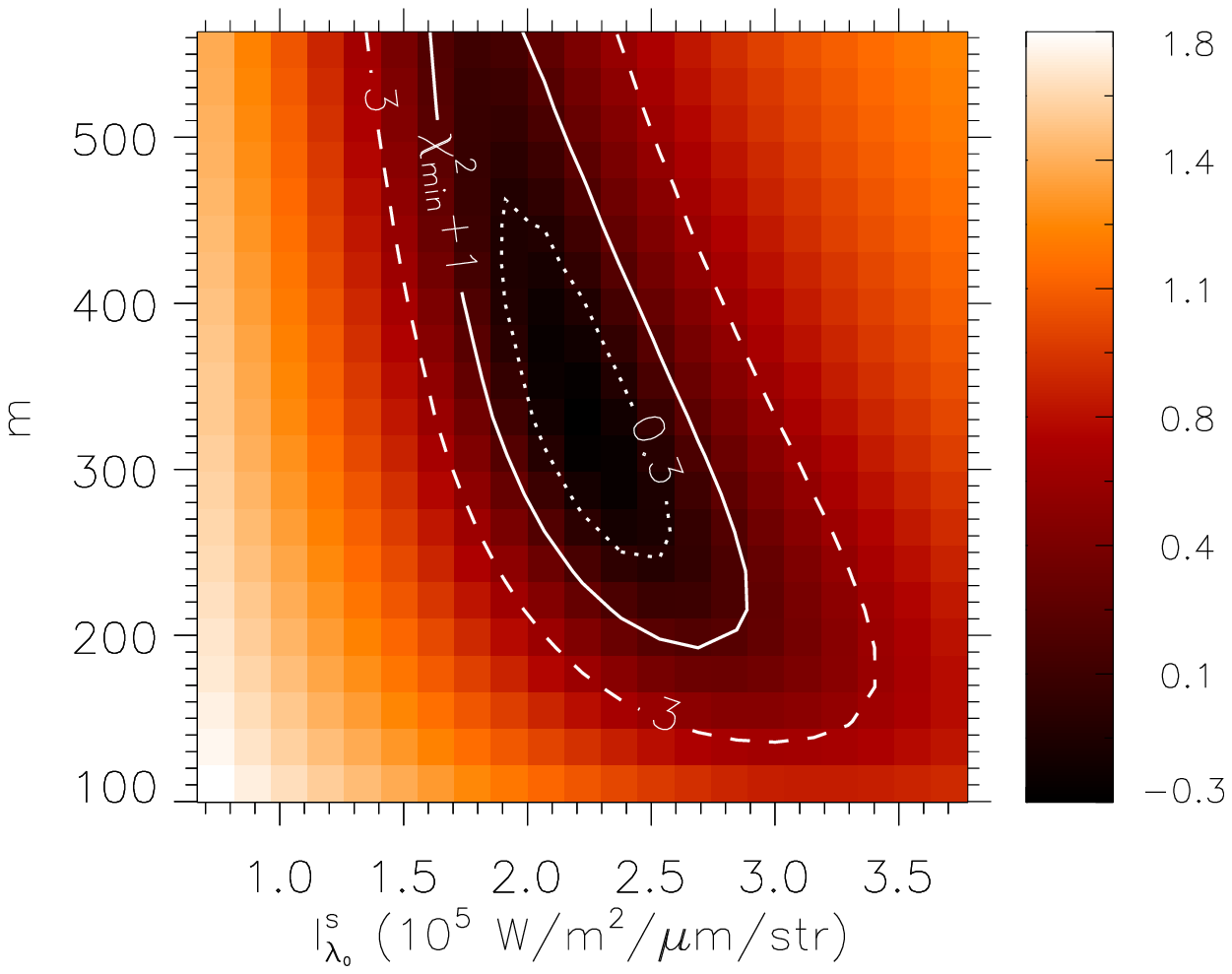}
\includegraphics[width=0.33\hsize,draft=false]{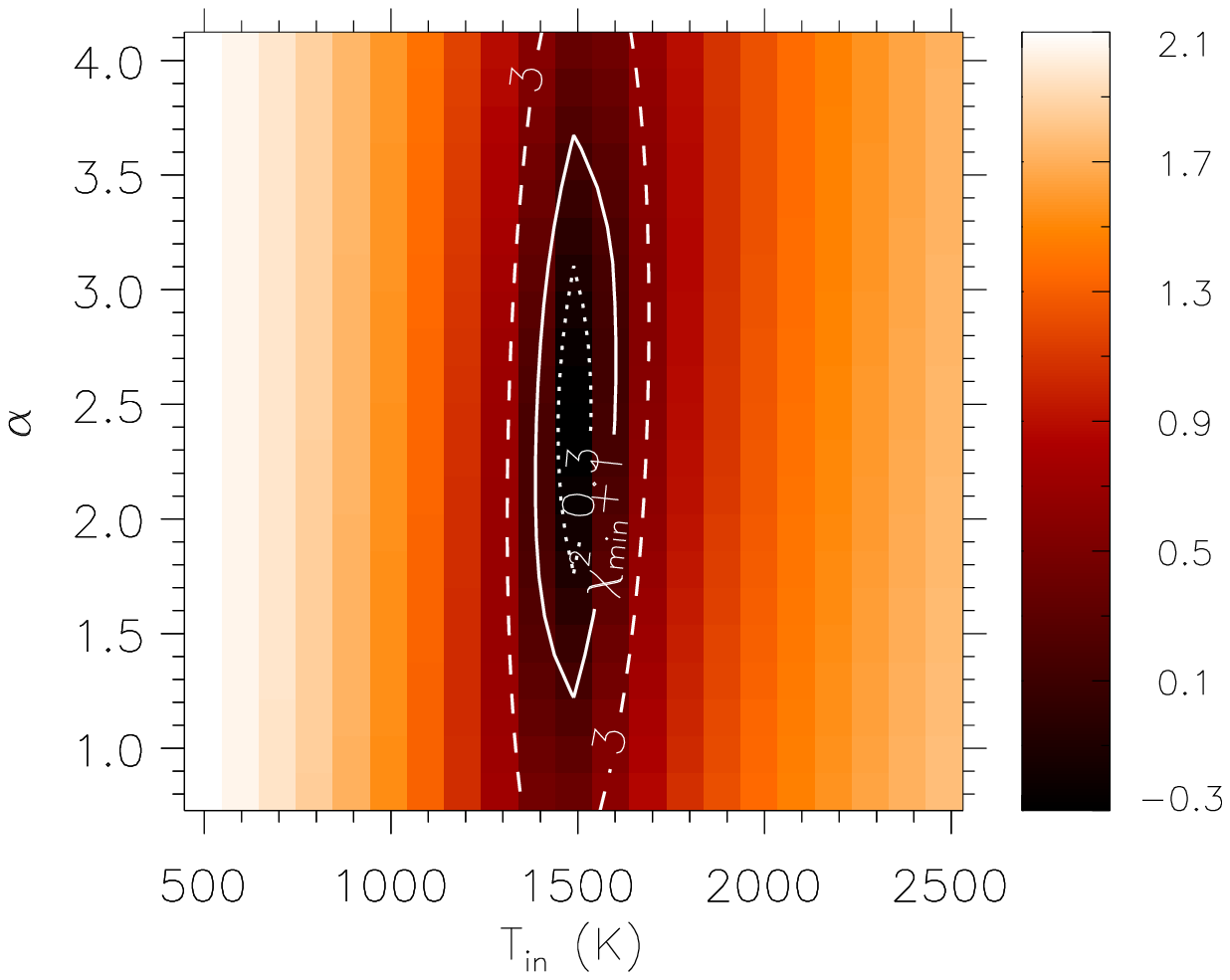}
\includegraphics[width=0.33\hsize,draft=false]{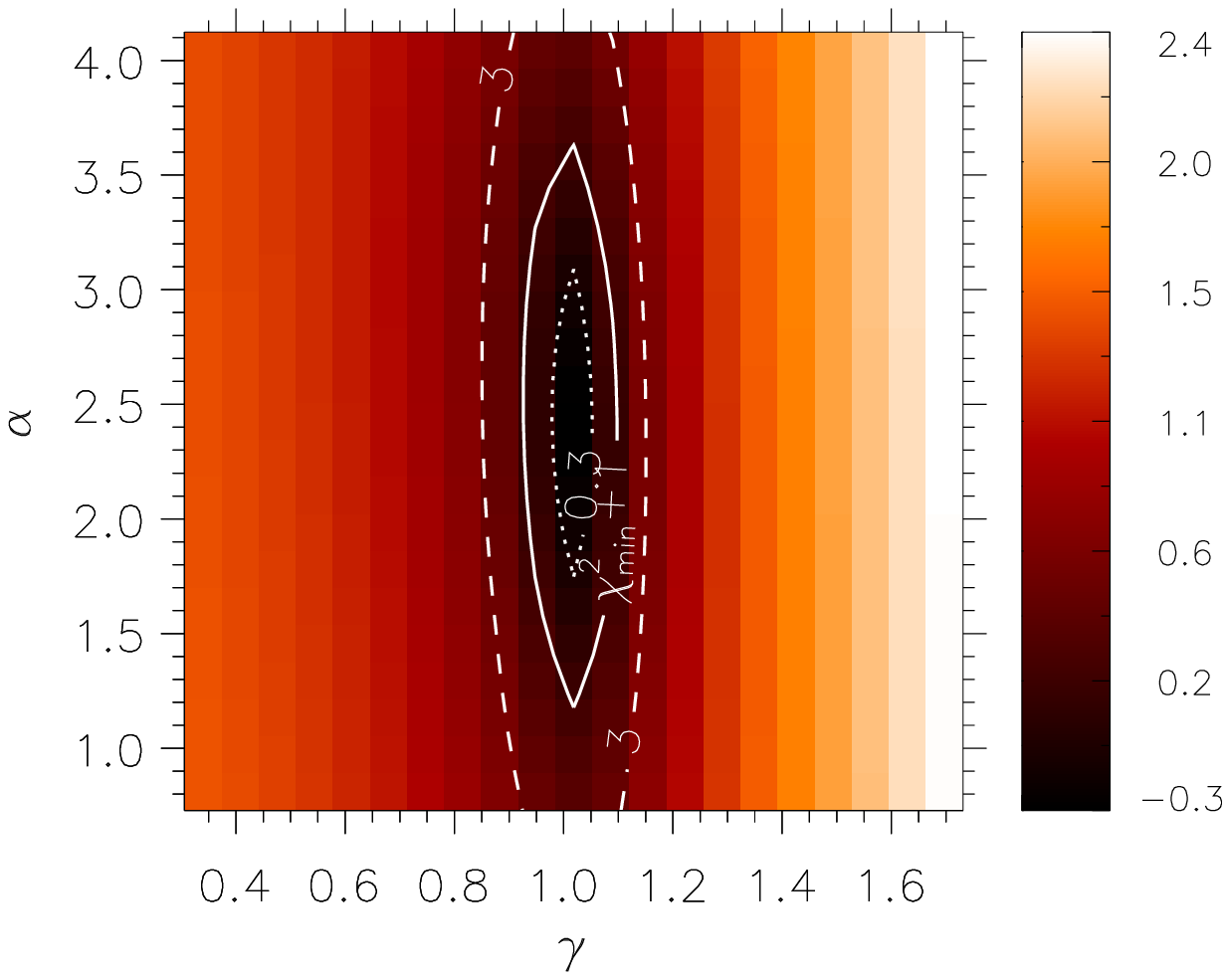}
\includegraphics[width=0.33\hsize,draft=false]{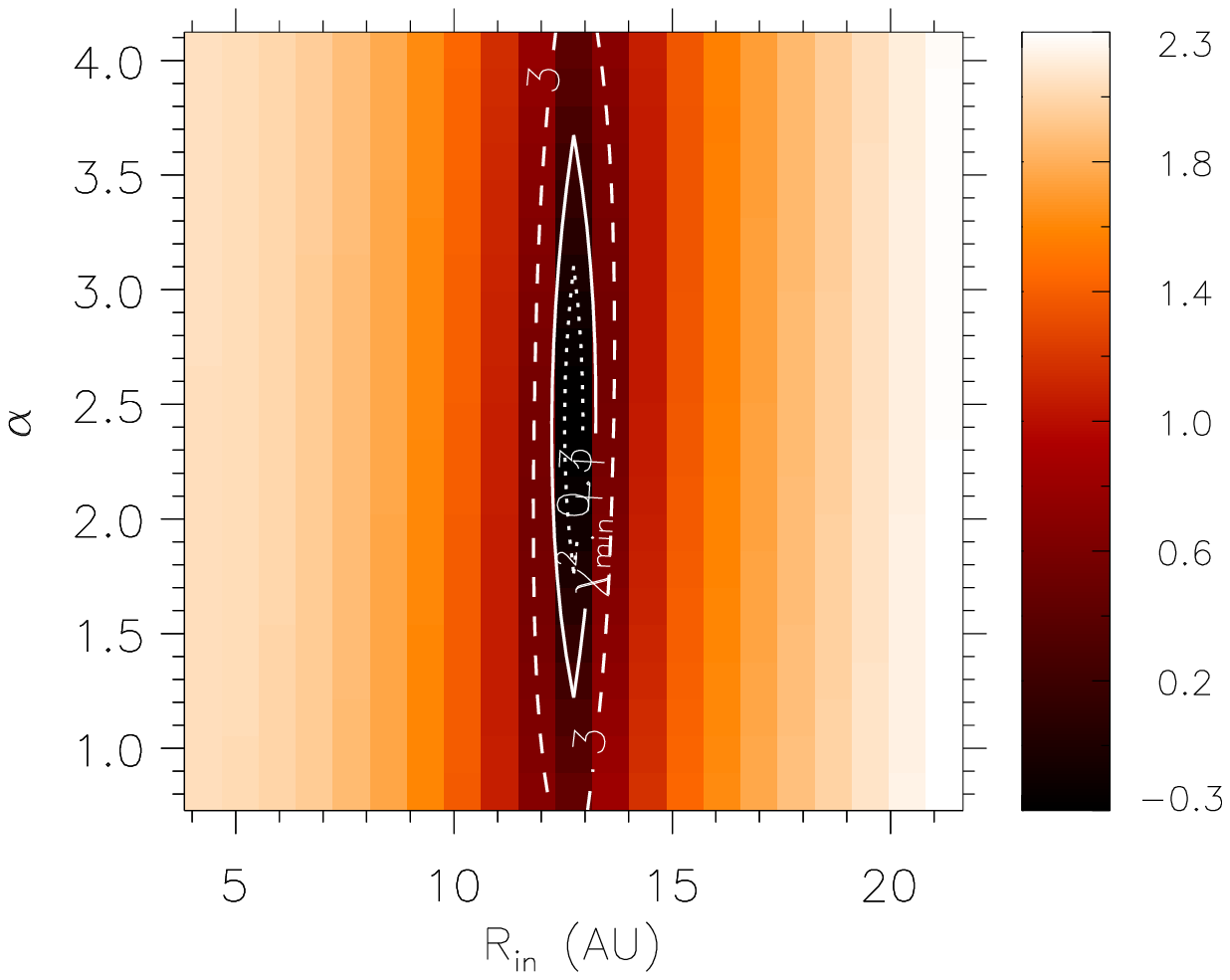}
\includegraphics[width=0.33\hsize,draft=false]{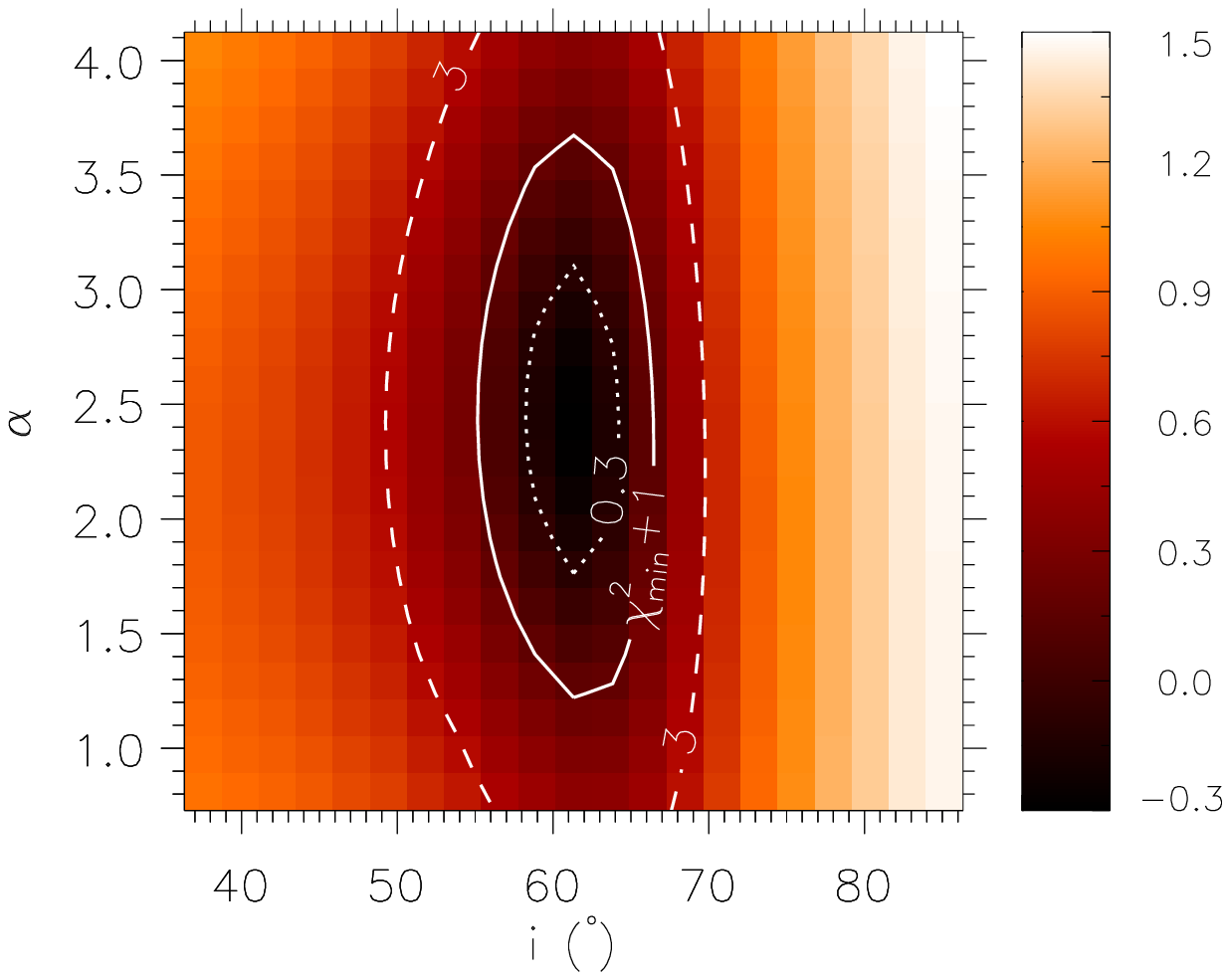}
\includegraphics[width=0.33\hsize,draft=false]{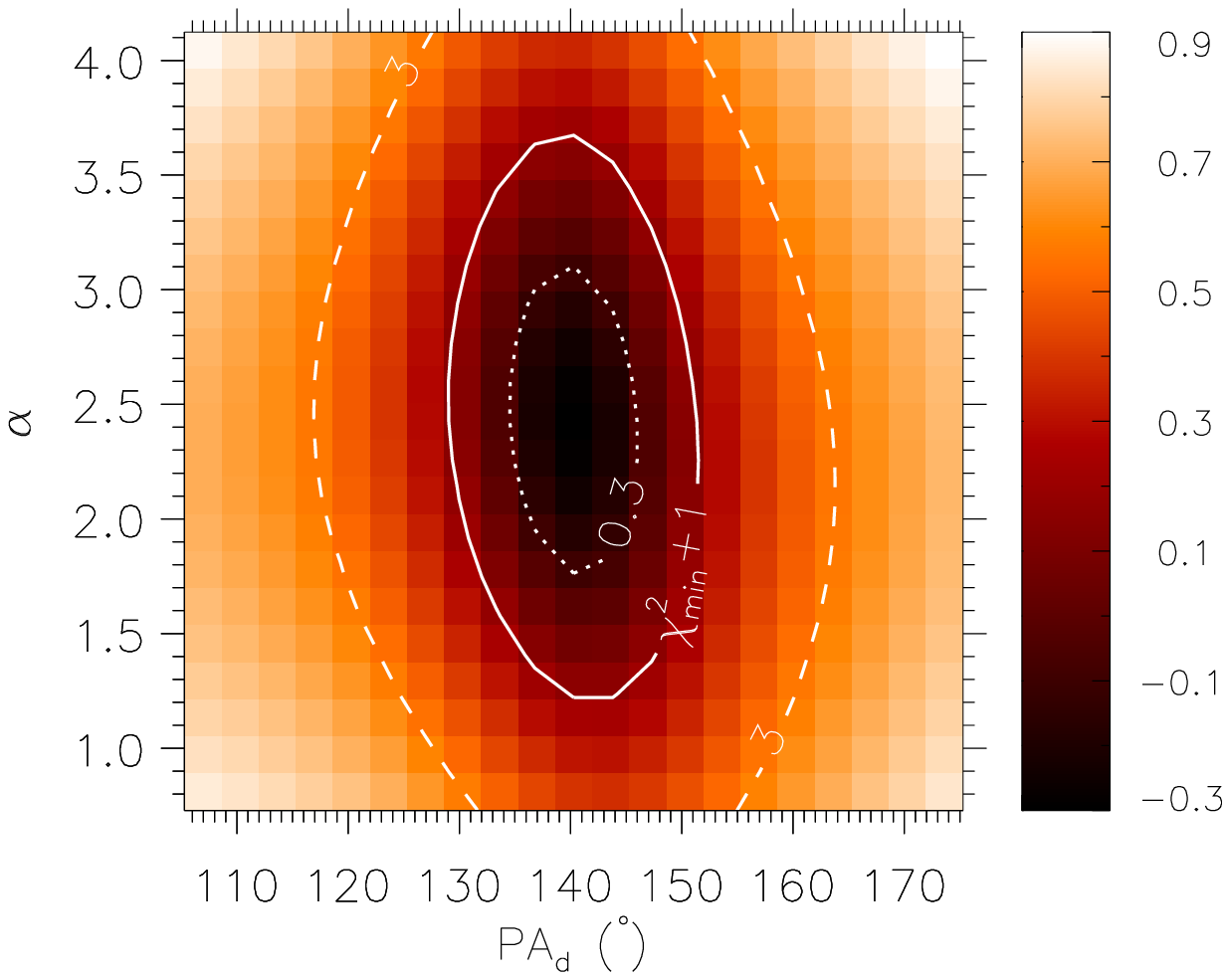}
\includegraphics[width=0.33\hsize,draft=false]{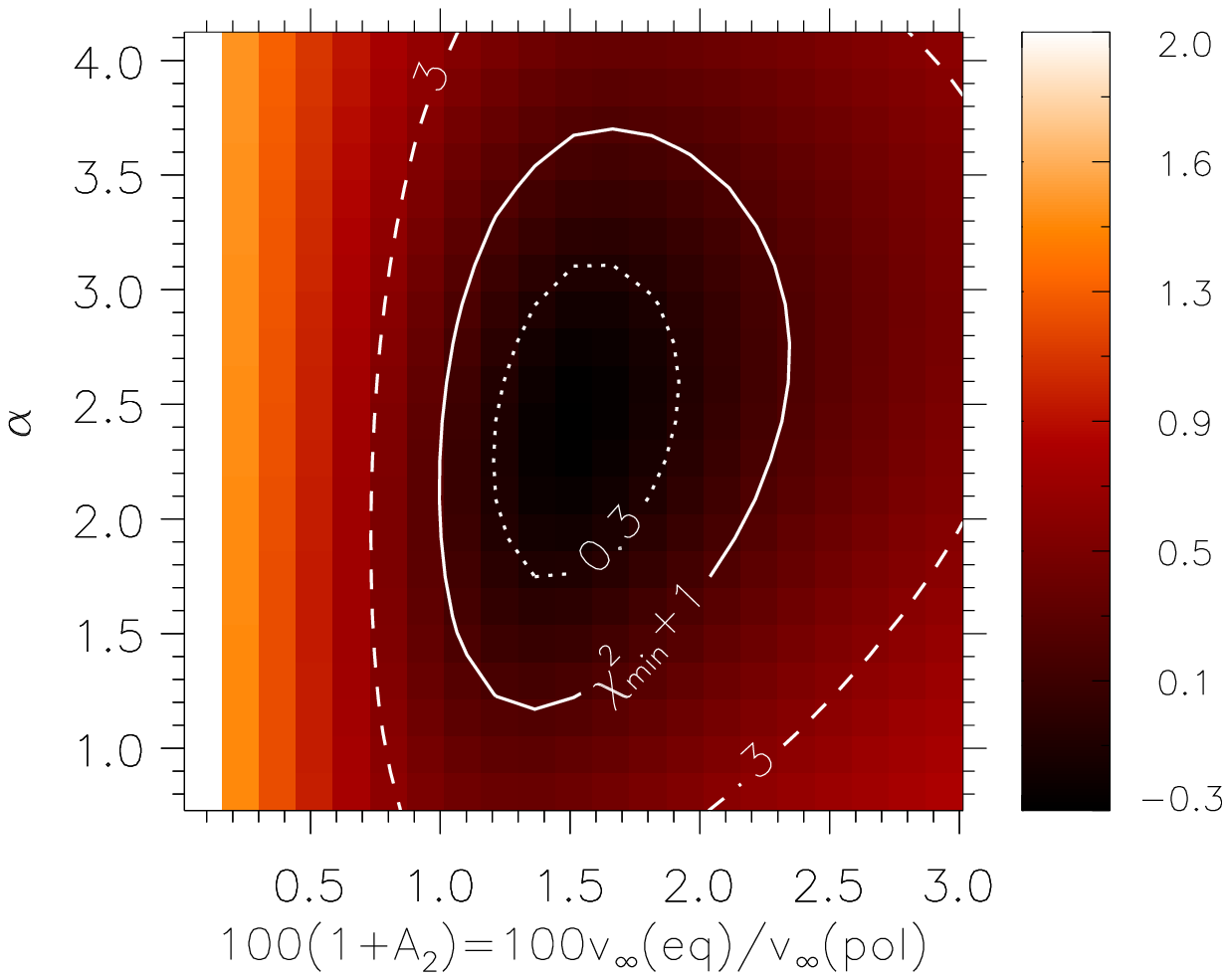}
\caption{Reduced $\chi^2$ maps from VLTI/MIDI observations of CPD-57\degr\,2874 for all 45 combinations of free parameters of our model (FRACS). The maps are centered on the $\chirmin^2(=0.54)$ position and correspond to the distance of 1.7~kpc. Contours are drawn for $\chirmin^2+\Delta\chi^2$, with $\Delta\chi^2=0.3\, ,1\, , 3$. The parameters are ordered as in Table~\ref{tab:fit_parameters}. The map scale is given in logarithmic units. Details on the model-fitting procedure are given in Sect.~\ref{data_analysis}. \label{fig:chi2_maps_1}}
\end{figure*}

\begin{figure*}[ht]
\centering
\includegraphics[width=0.33\hsize,draft=false]{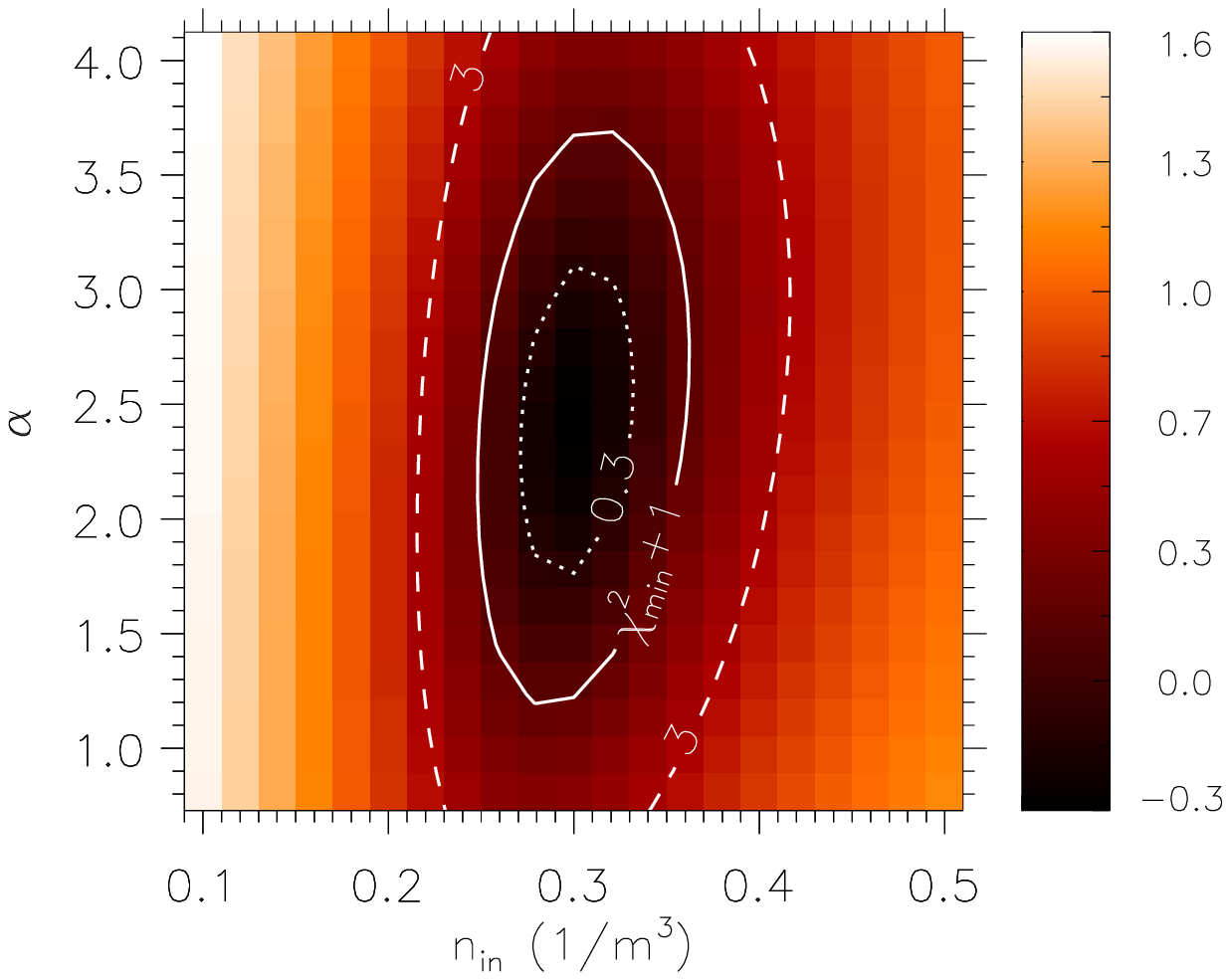}
\includegraphics[width=0.33\hsize,draft=false]{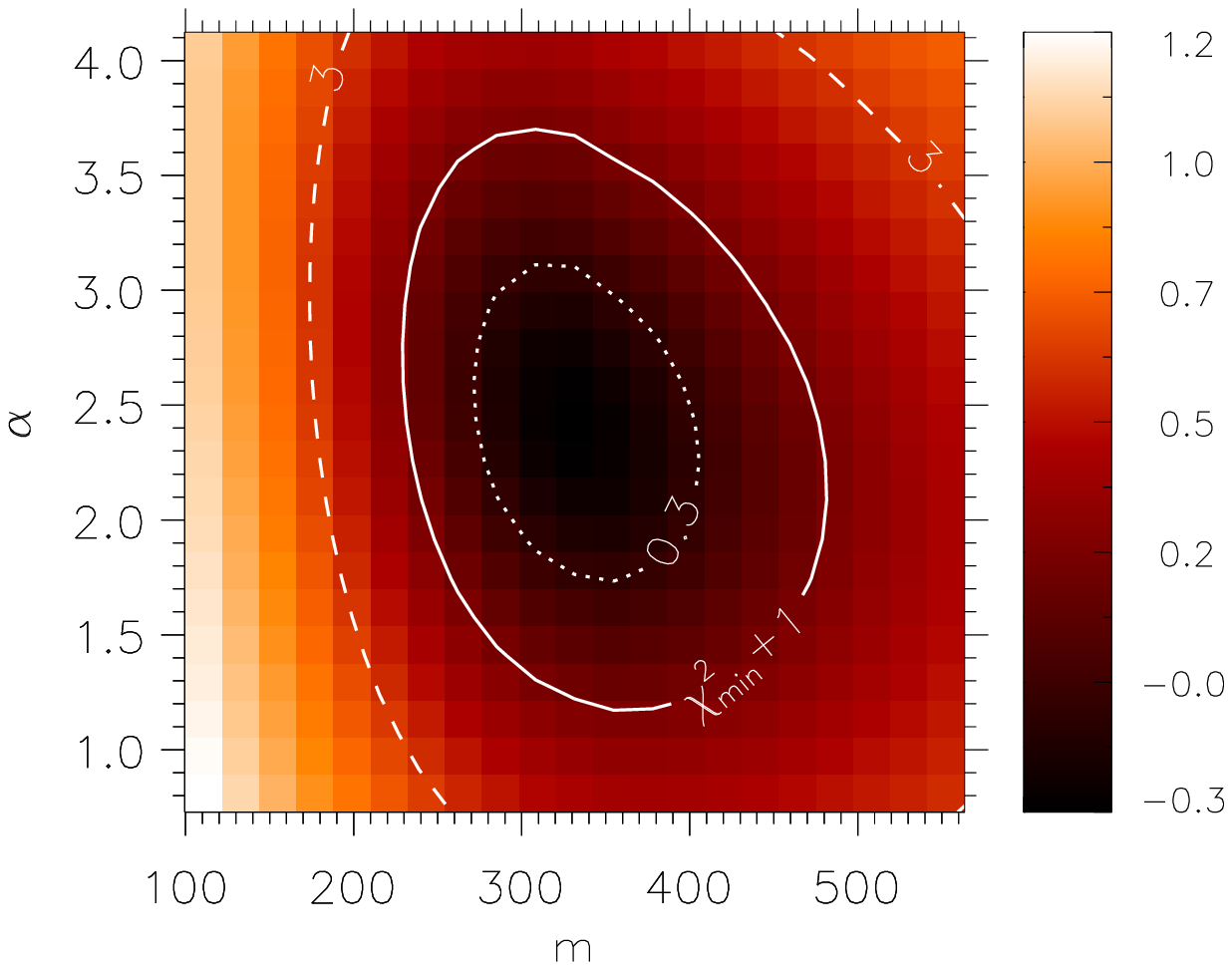}
\includegraphics[width=0.33\hsize,draft=false]{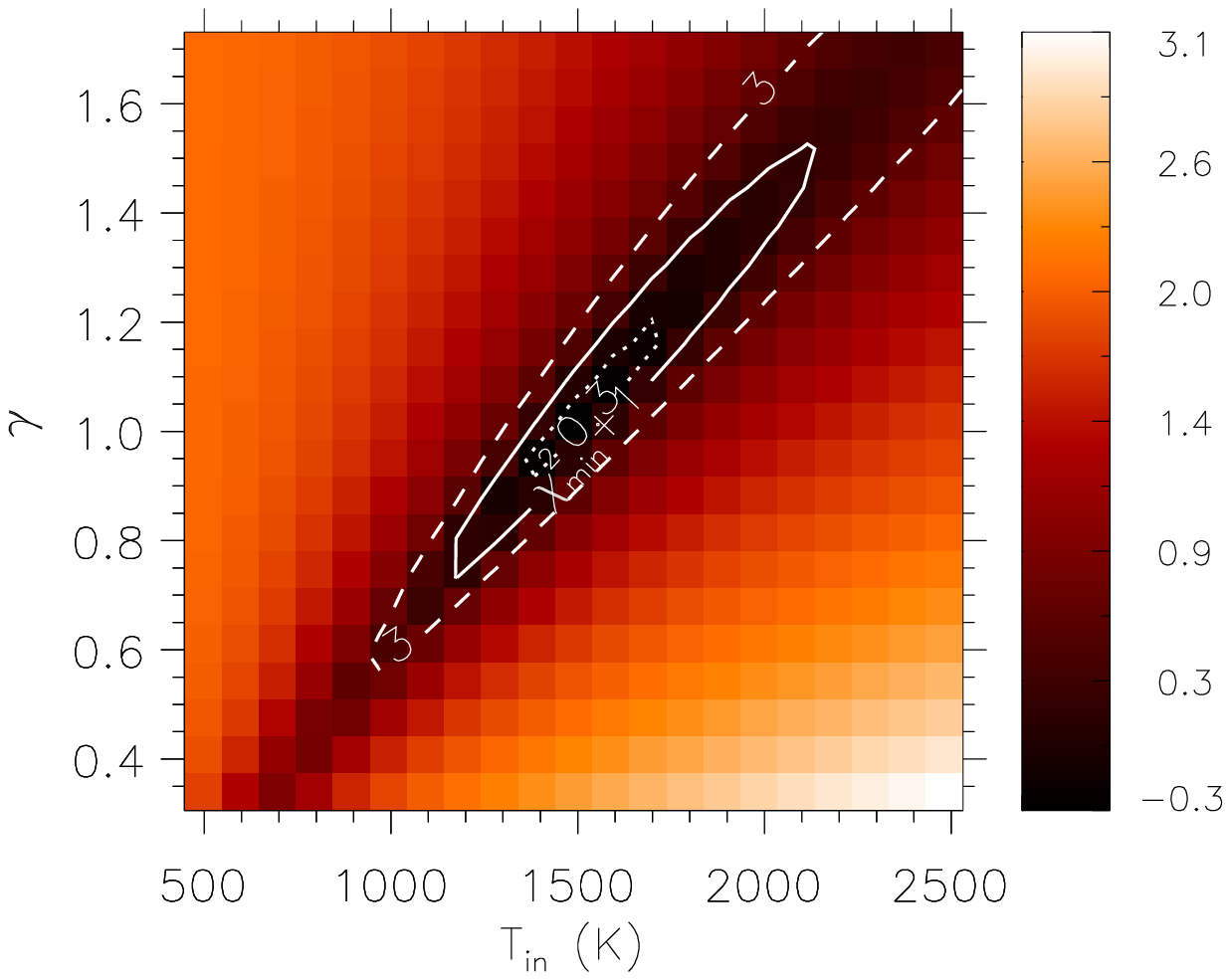}
\includegraphics[width=0.33\hsize,draft=false]{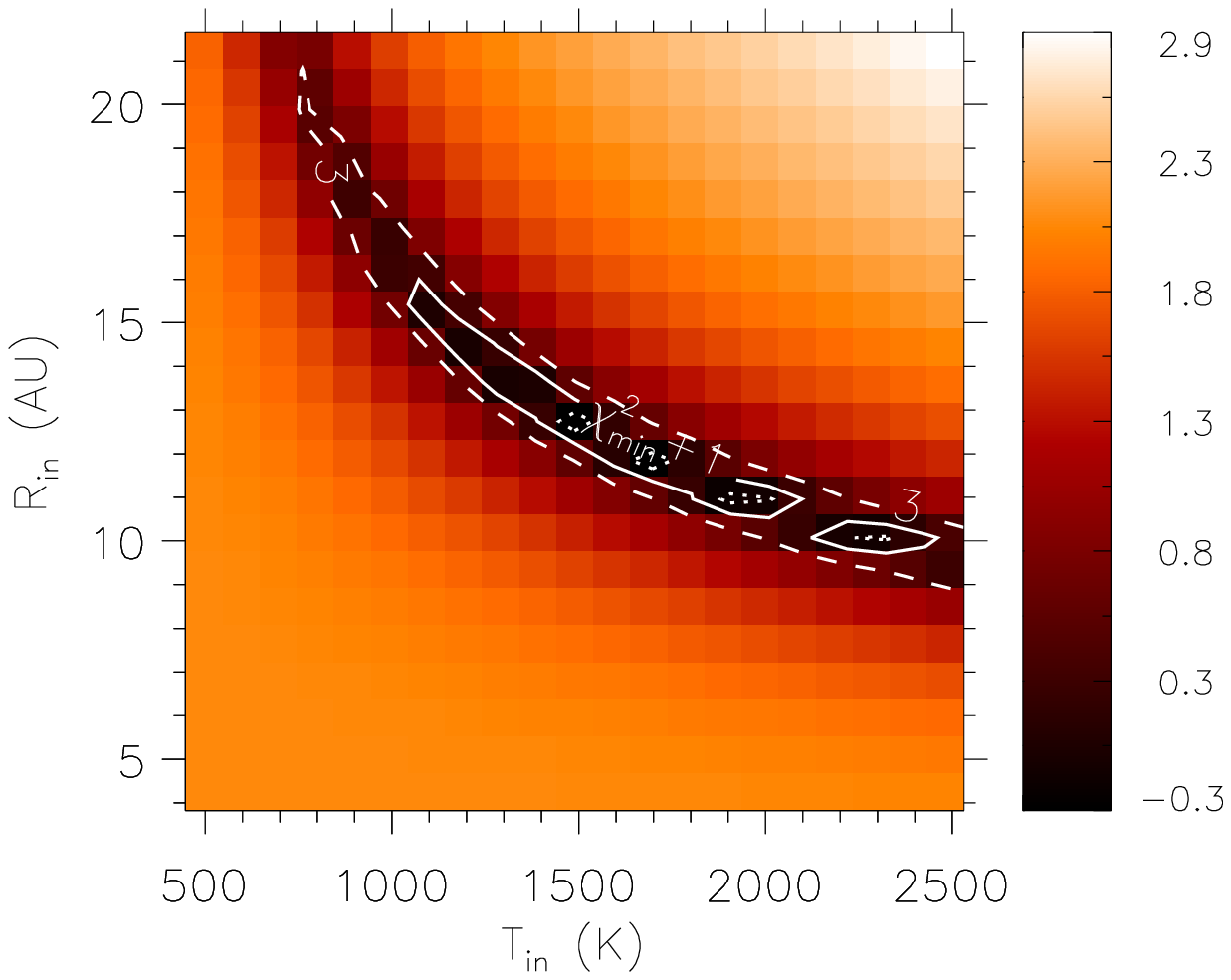}
\includegraphics[width=0.33\hsize,draft=false]{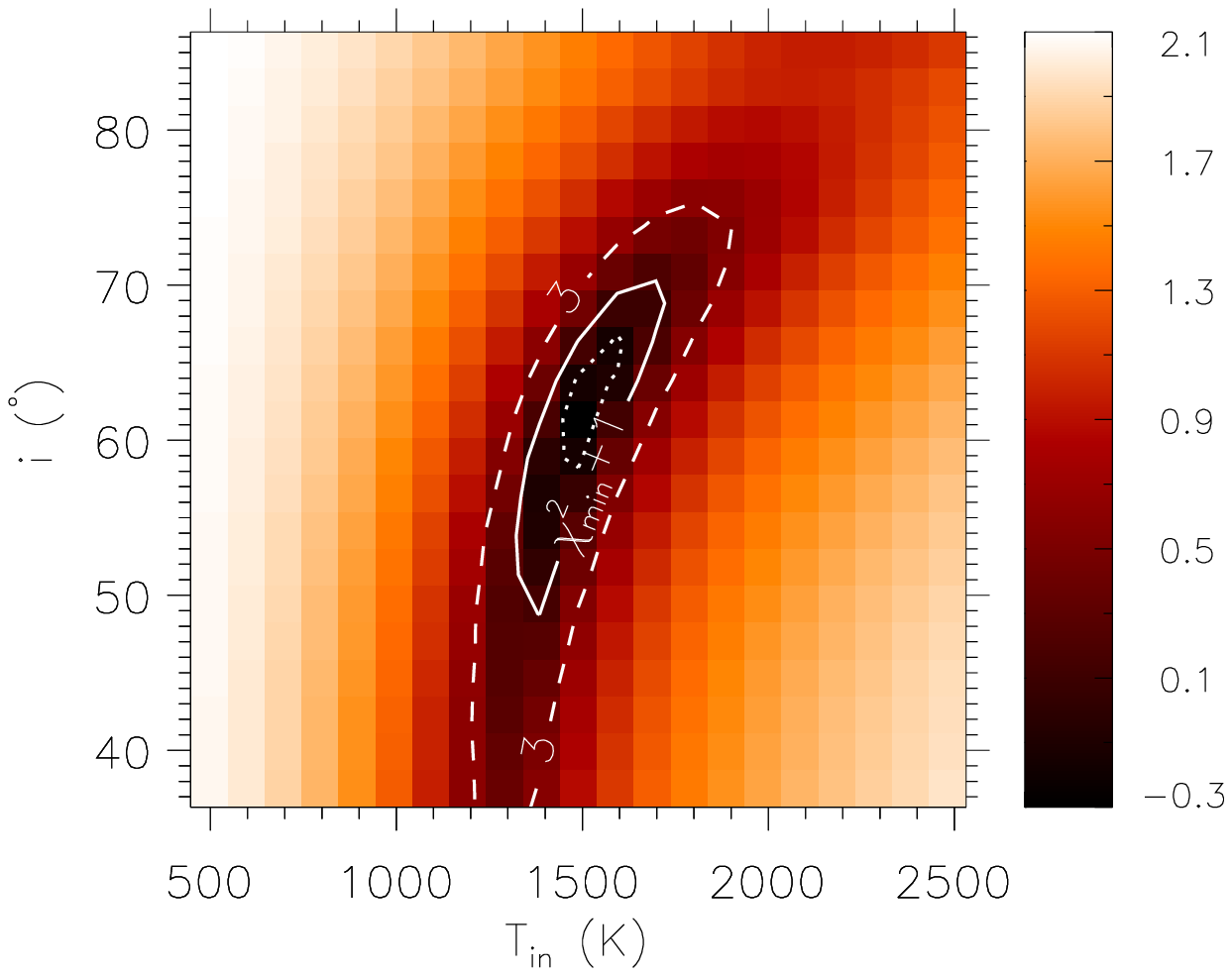}
\includegraphics[width=0.33\hsize,draft=false]{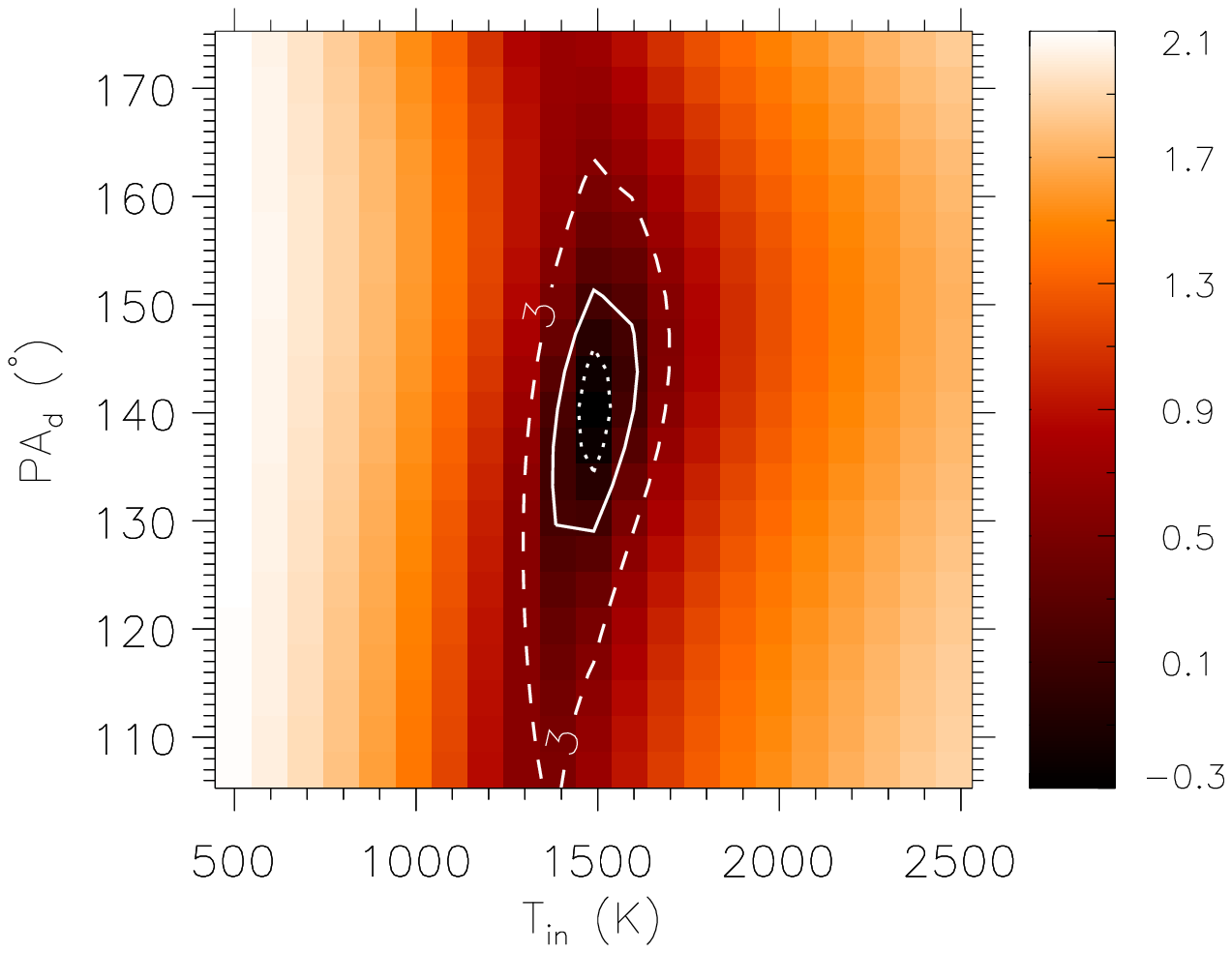}
\includegraphics[width=0.33\hsize,draft=false]{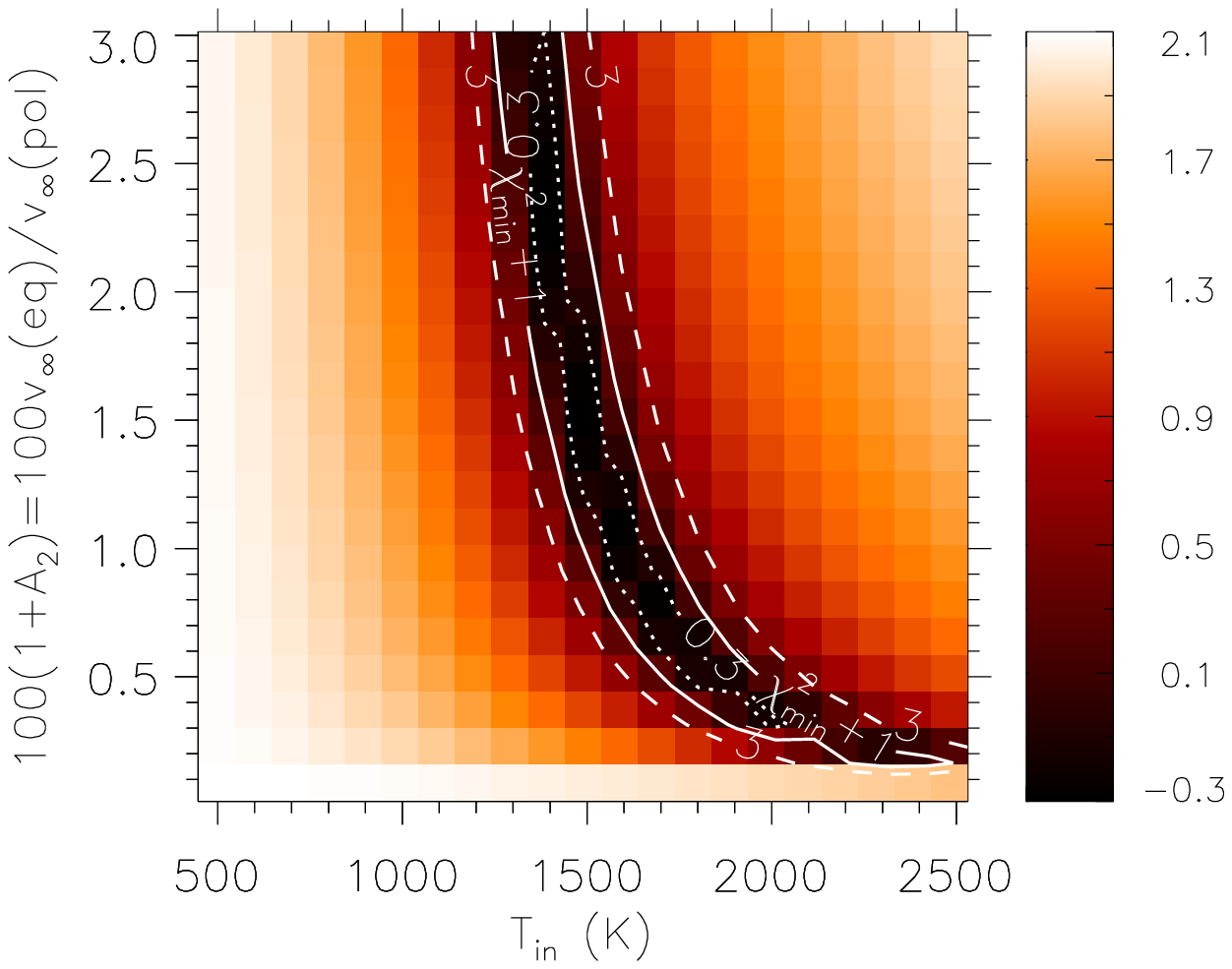}
\includegraphics[width=0.33\hsize,draft=false]{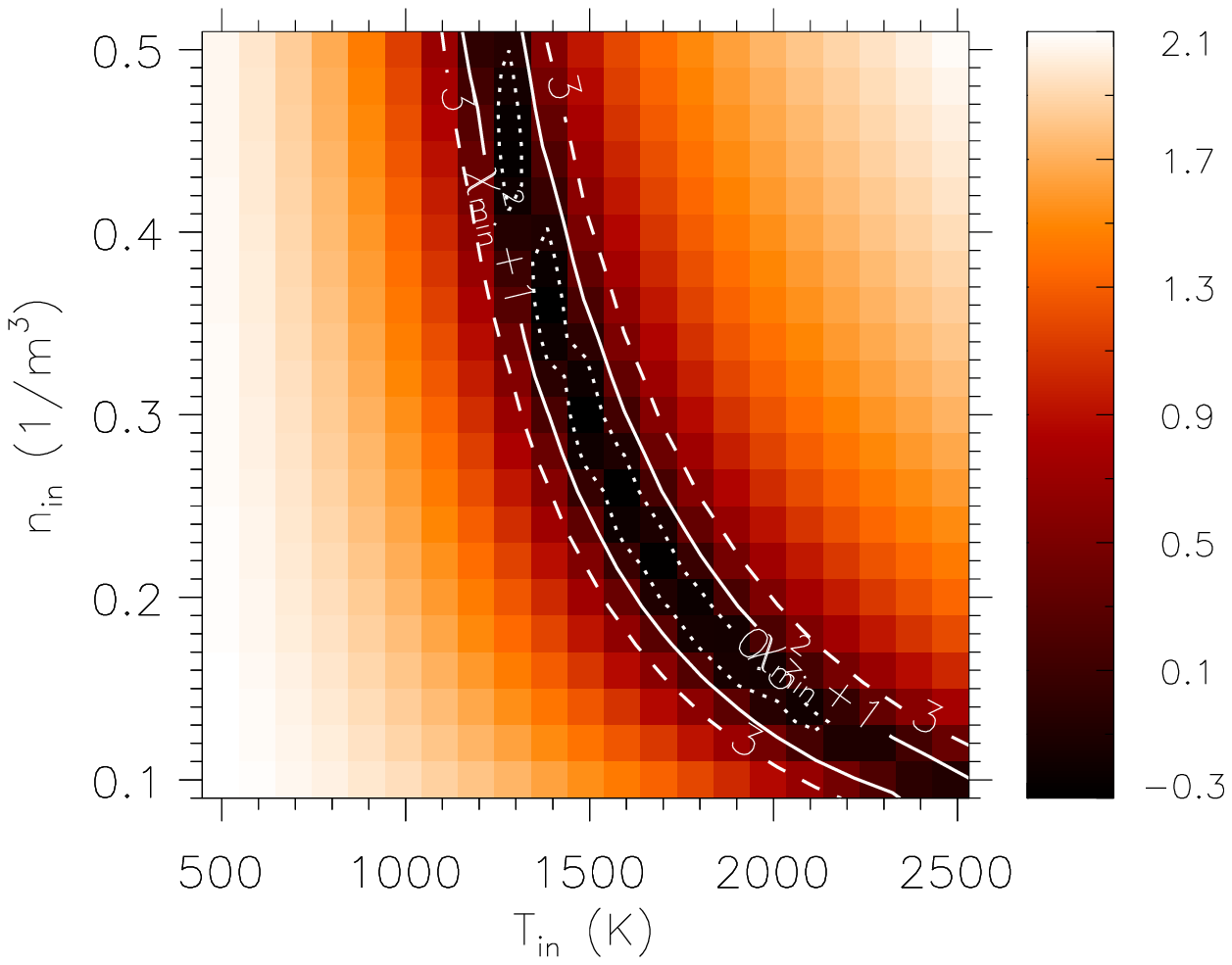}
\includegraphics[width=0.33\hsize,draft=false]{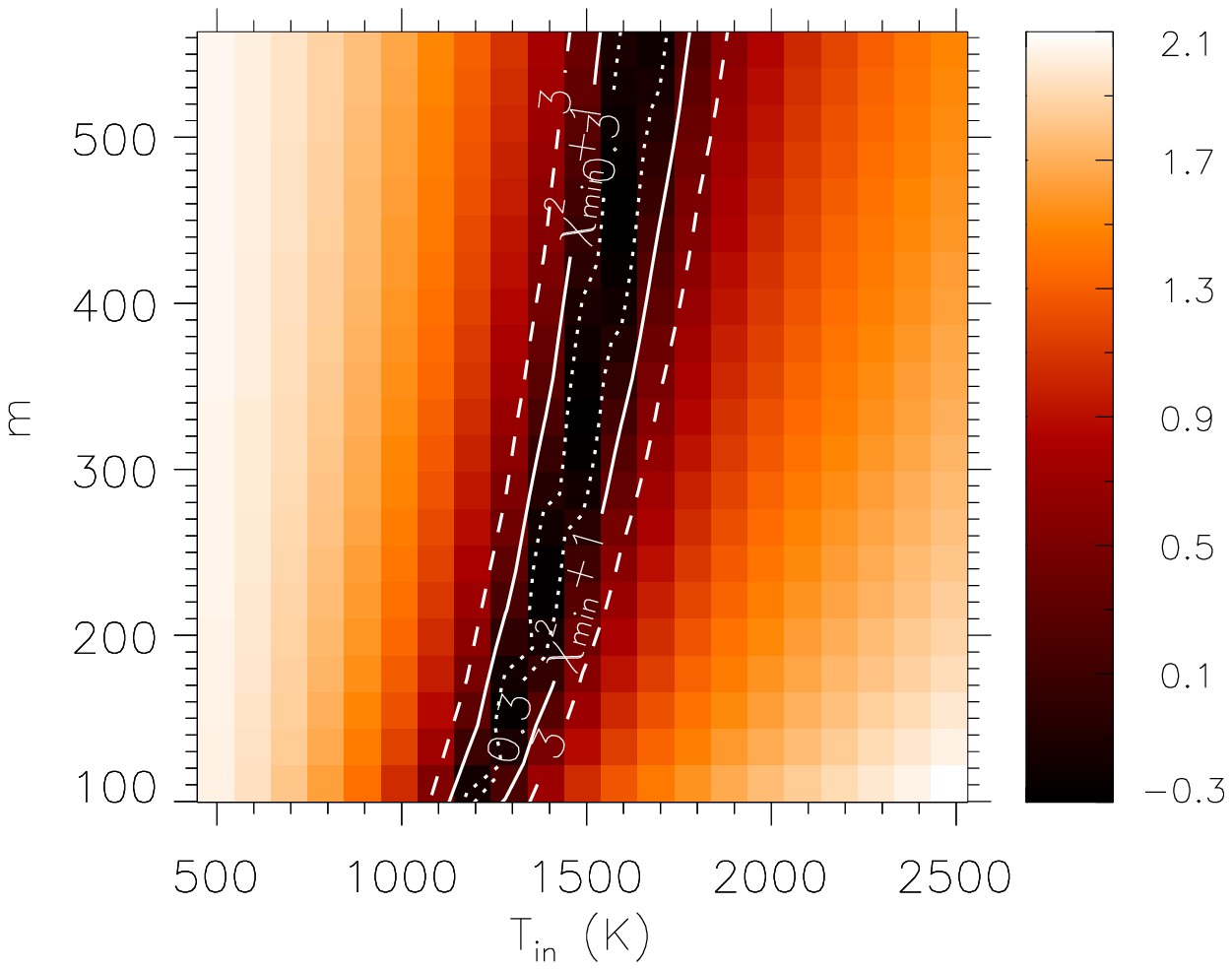}
\includegraphics[width=0.33\hsize,draft=false]{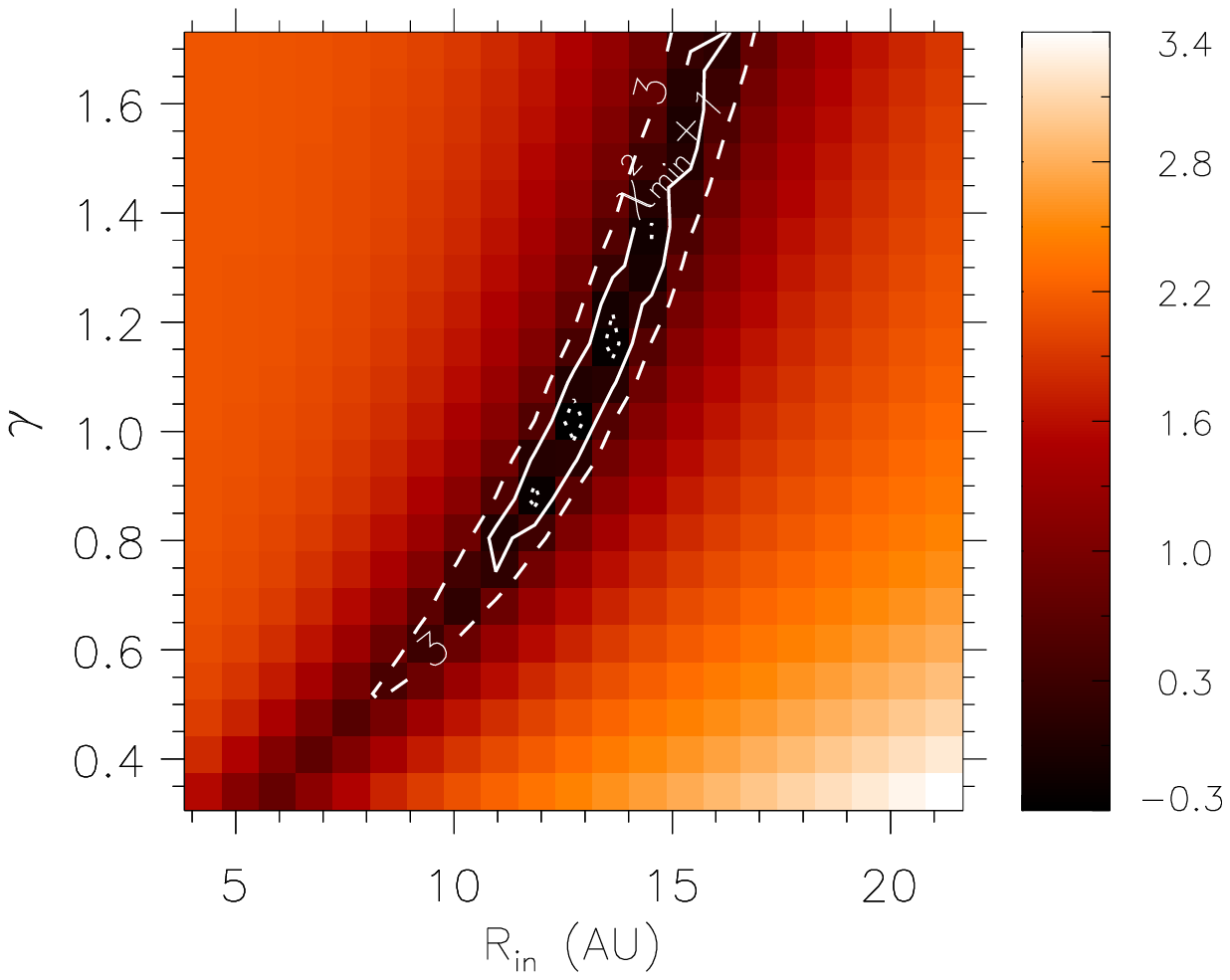}
\includegraphics[width=0.33\hsize,draft=false]{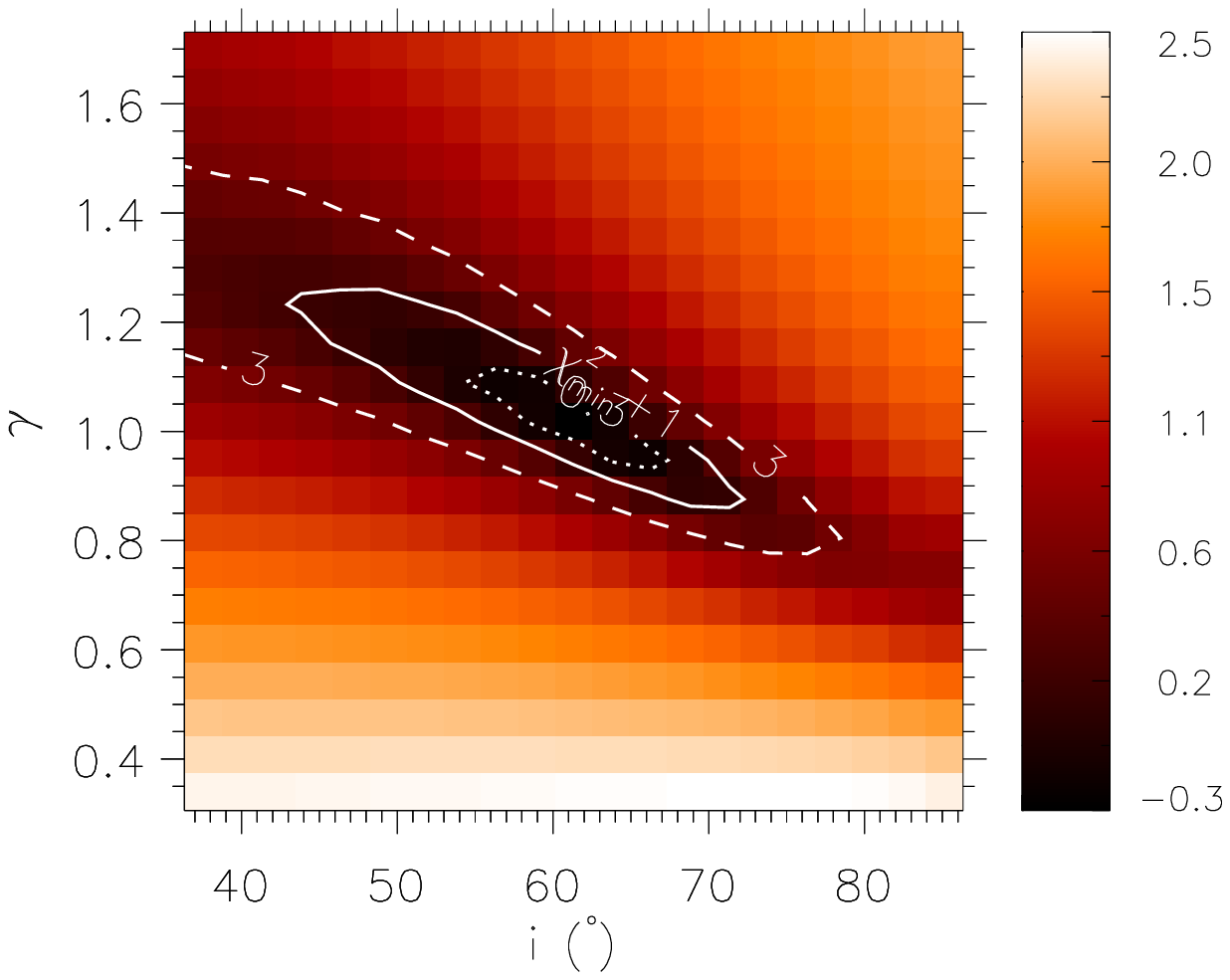}
\includegraphics[width=0.33\hsize,draft=false]{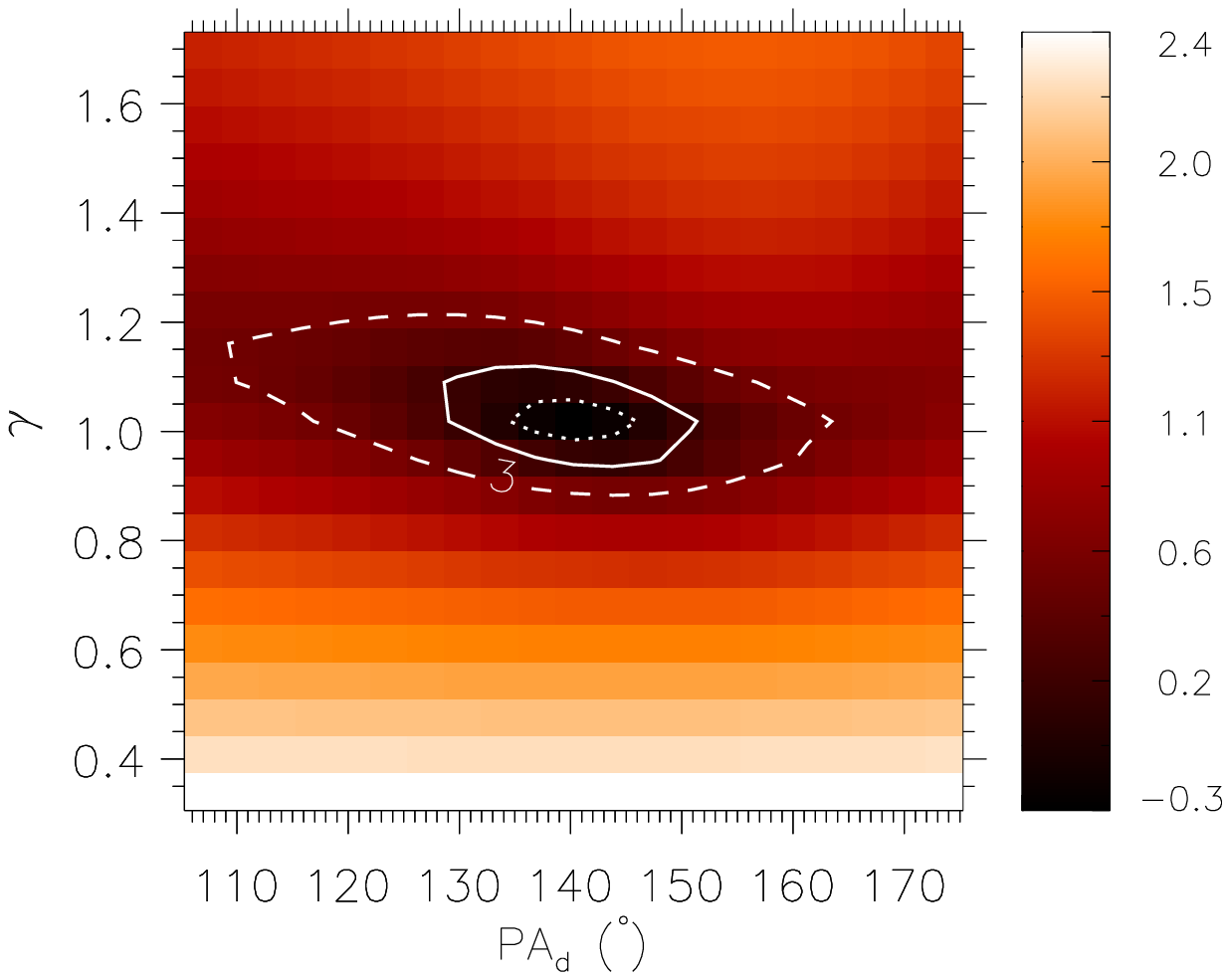}
\includegraphics[width=0.33\hsize,draft=false]{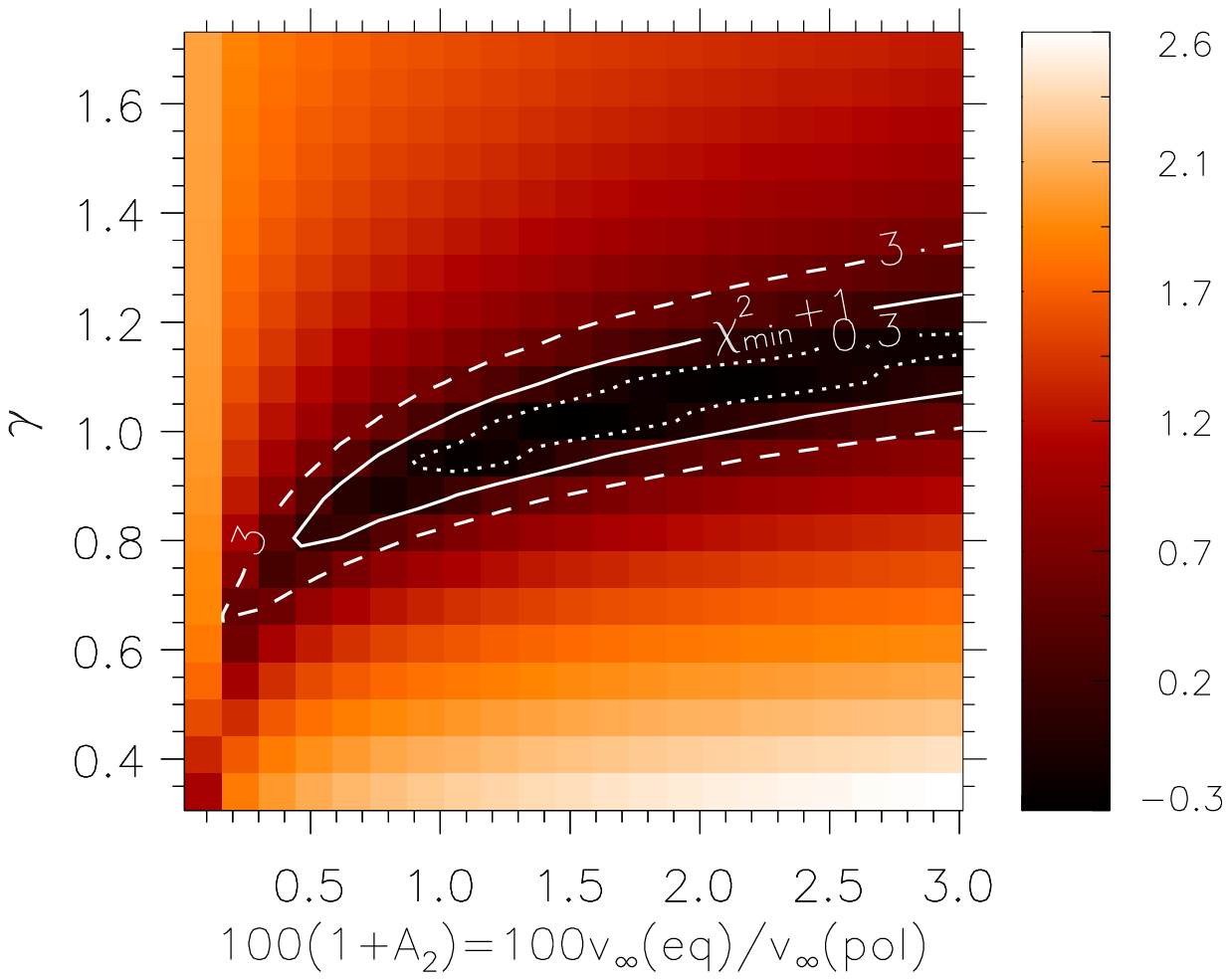}
\includegraphics[width=0.33\hsize,draft=false]{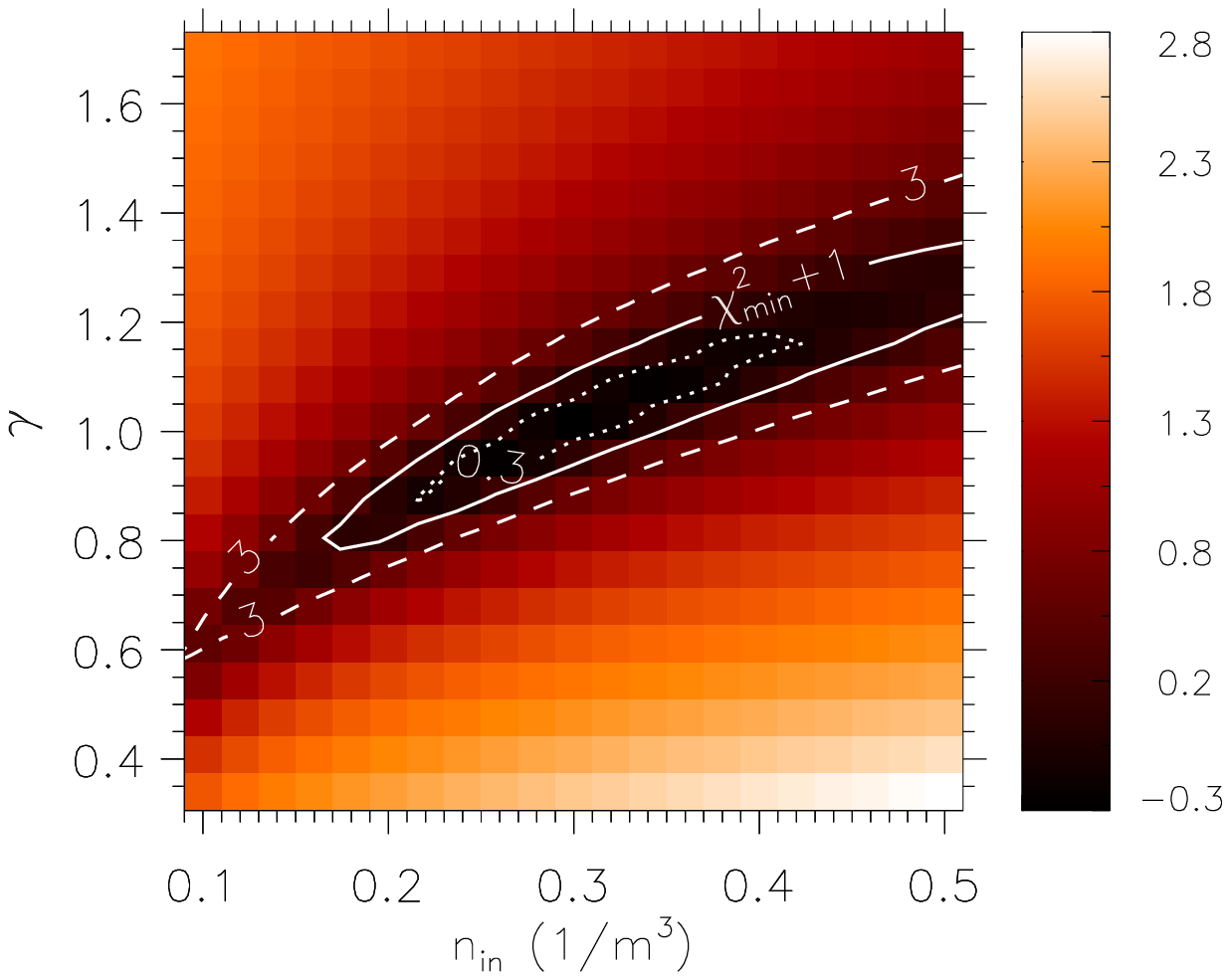}
\includegraphics[width=0.33\hsize,draft=false]{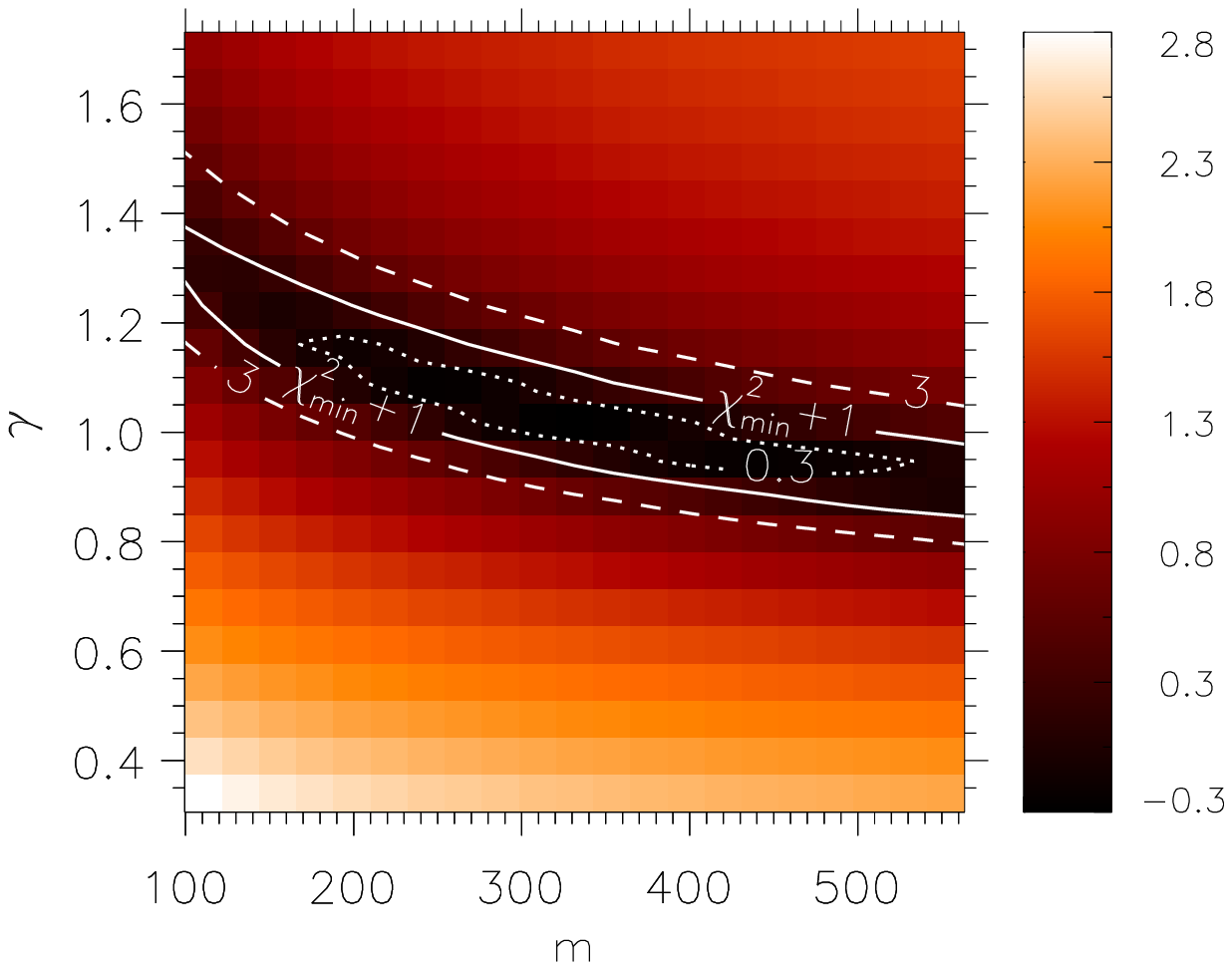}
\caption{Continuation of Fig.~\ref{fig:chi2_maps_1}. \label{fig:chi2_maps_2}}
\end{figure*}

\begin{figure*}[ht]
\centering
\includegraphics[width=0.33\hsize,draft=false]{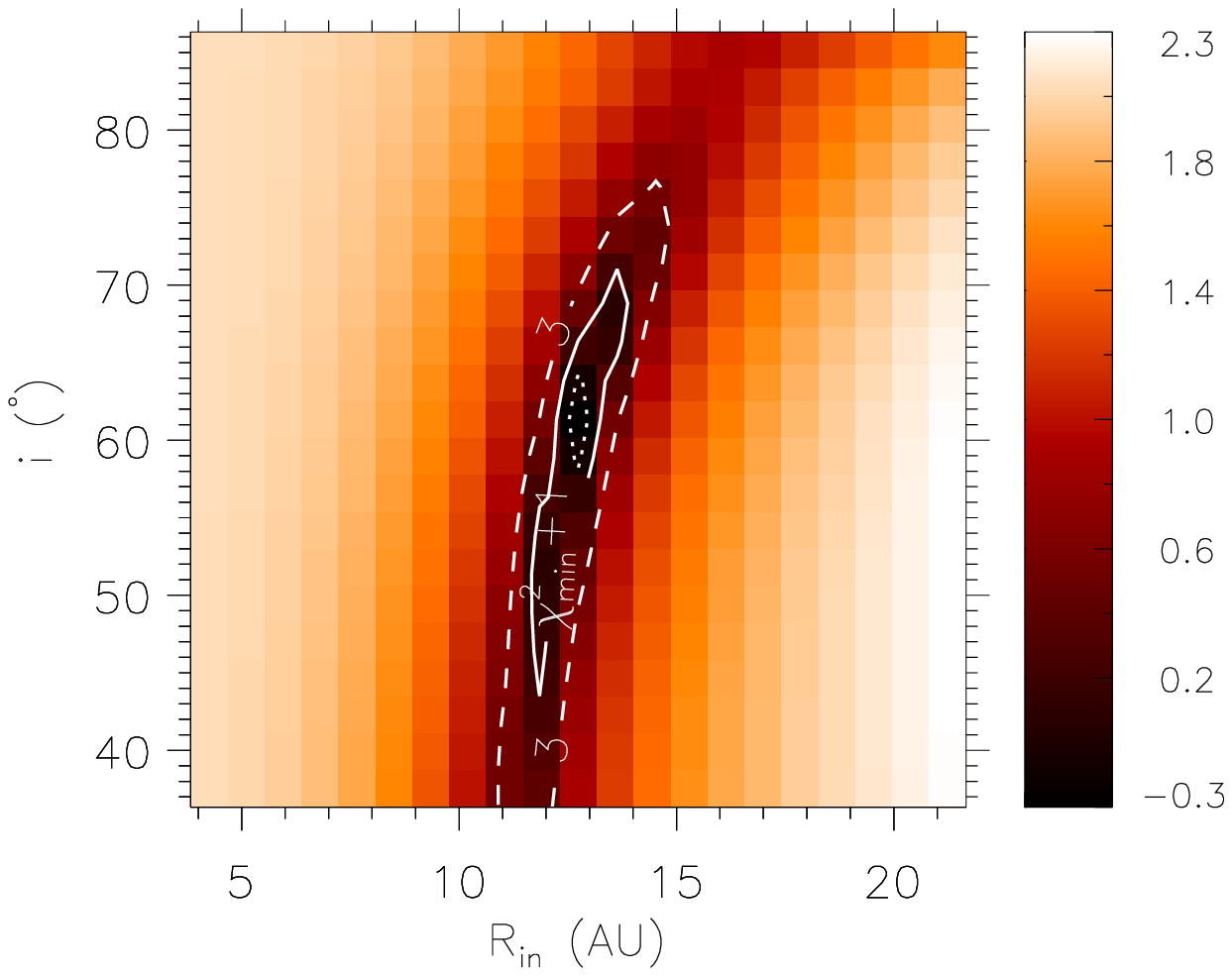}
\includegraphics[width=0.33\hsize,draft=false]{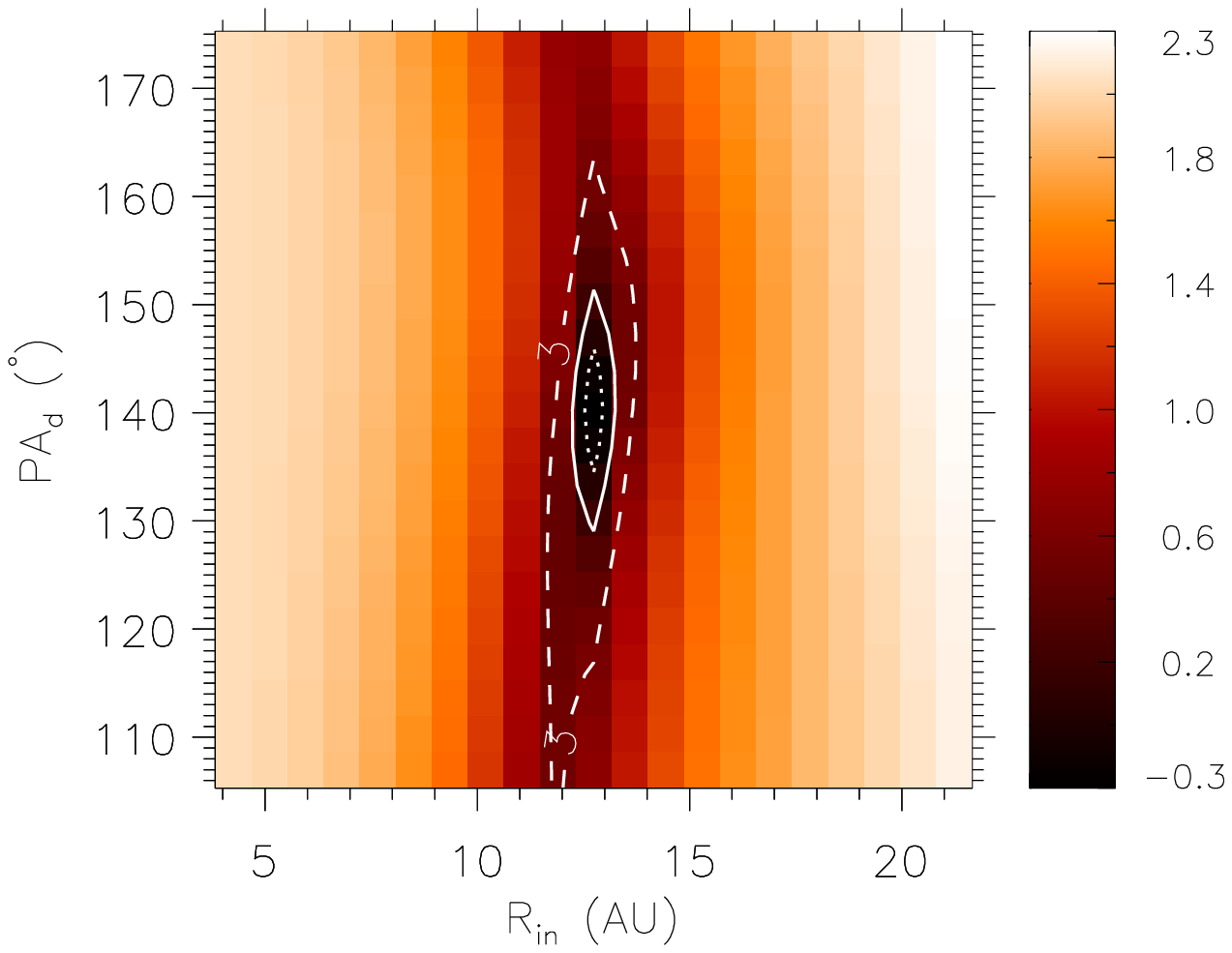}
\includegraphics[width=0.33\hsize,draft=false]{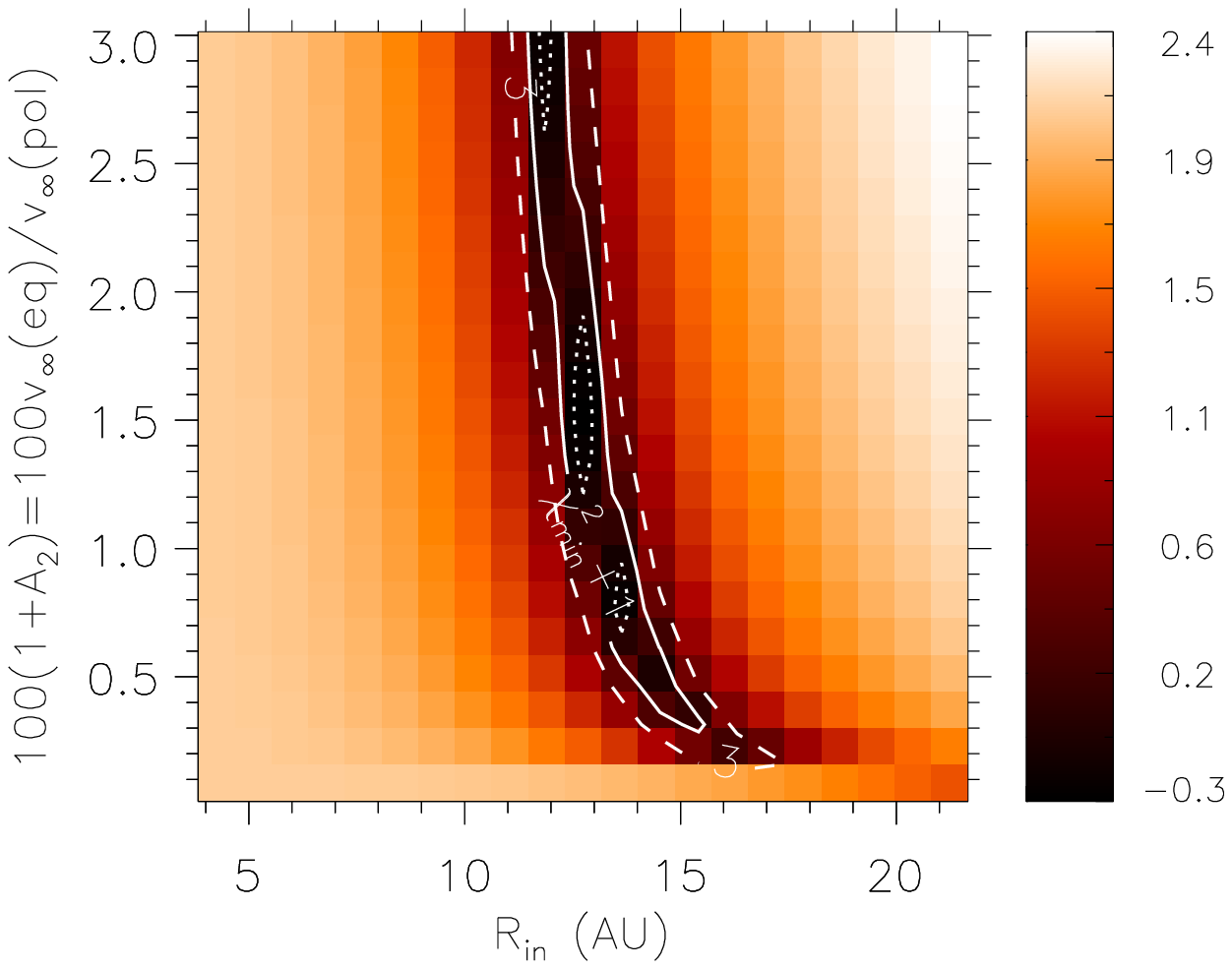}
\includegraphics[width=0.33\hsize,draft=false]{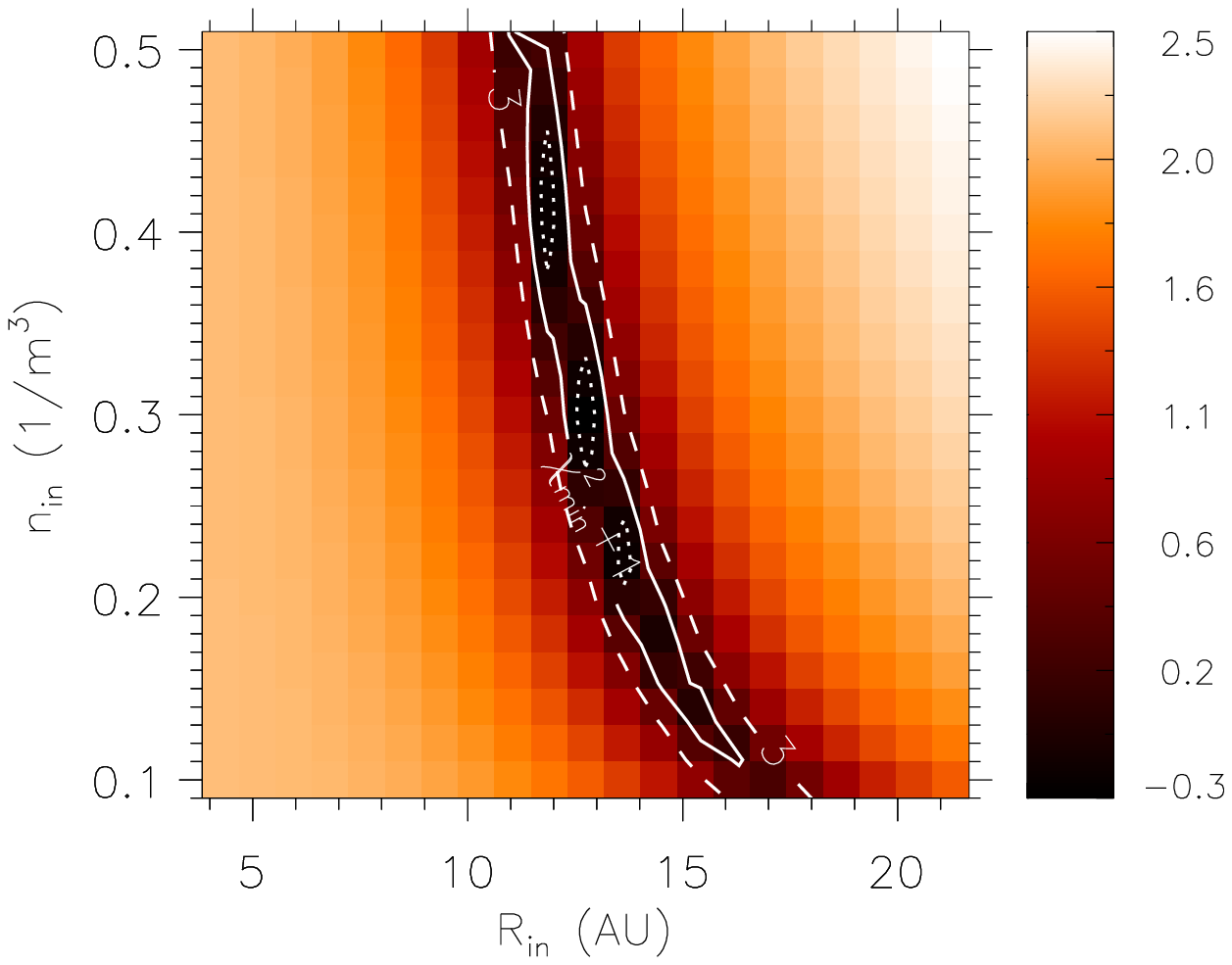}
\includegraphics[width=0.33\hsize,draft=false]{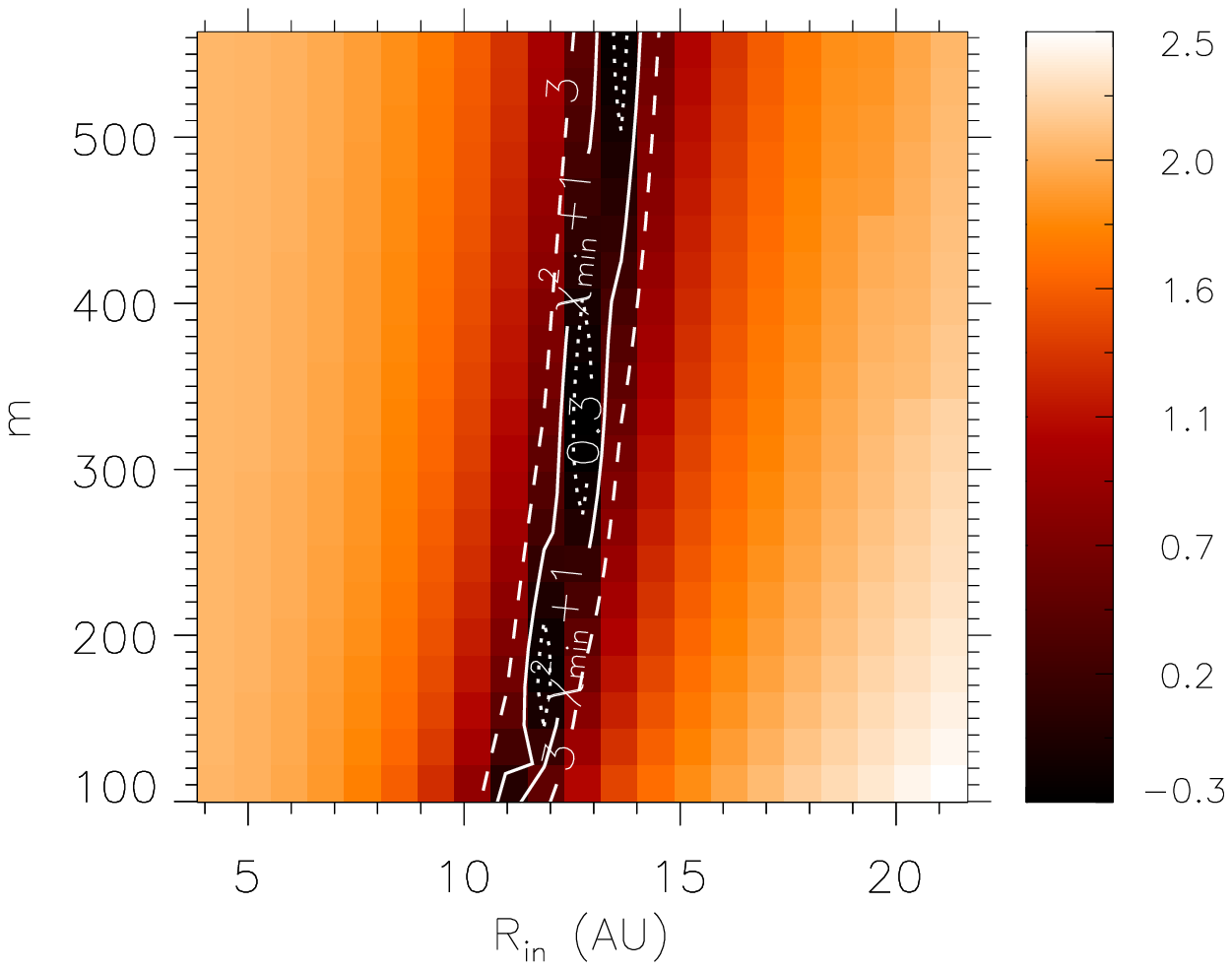}
\includegraphics[width=0.33\hsize,draft=false]{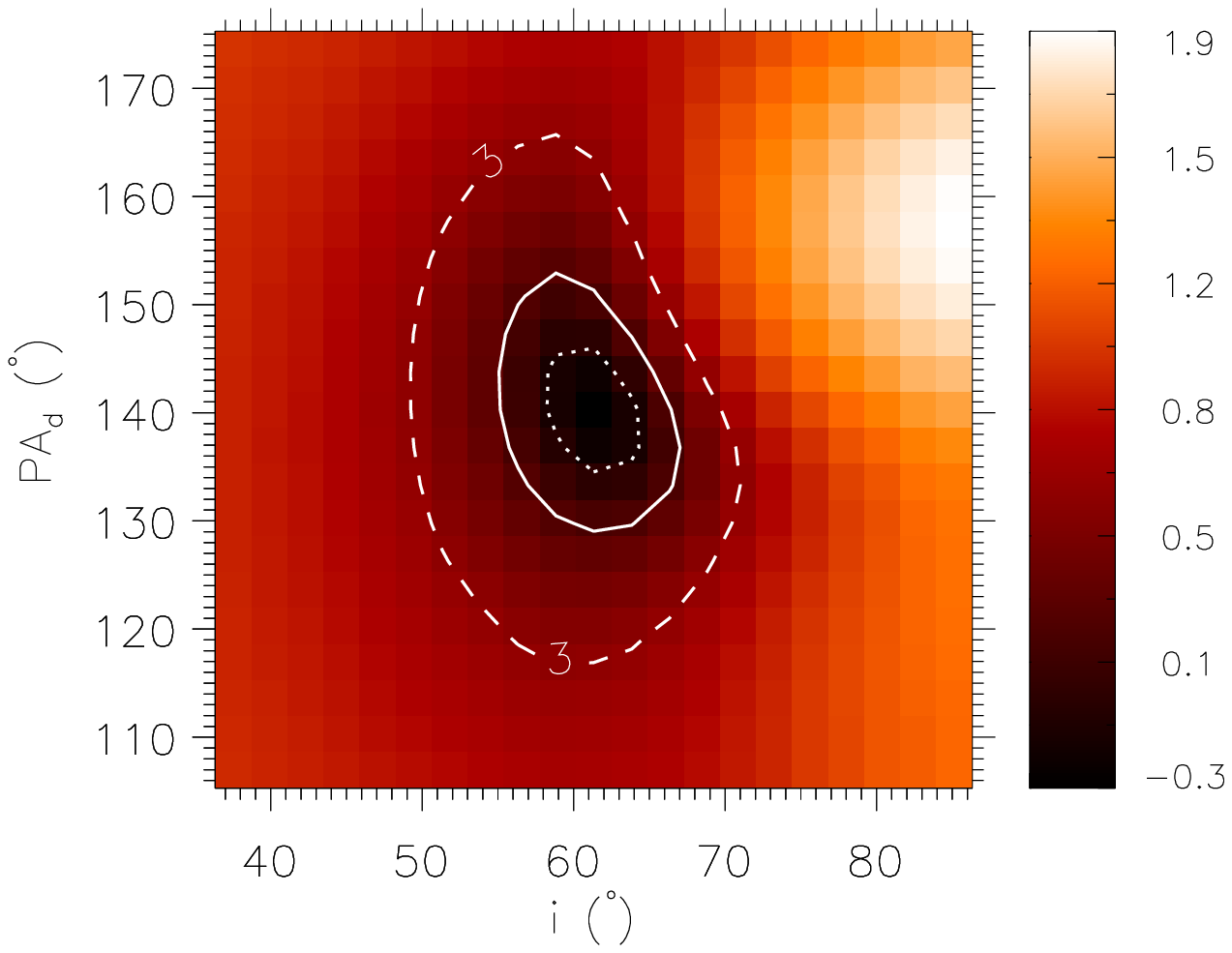}
\includegraphics[width=0.33\hsize,draft=false]{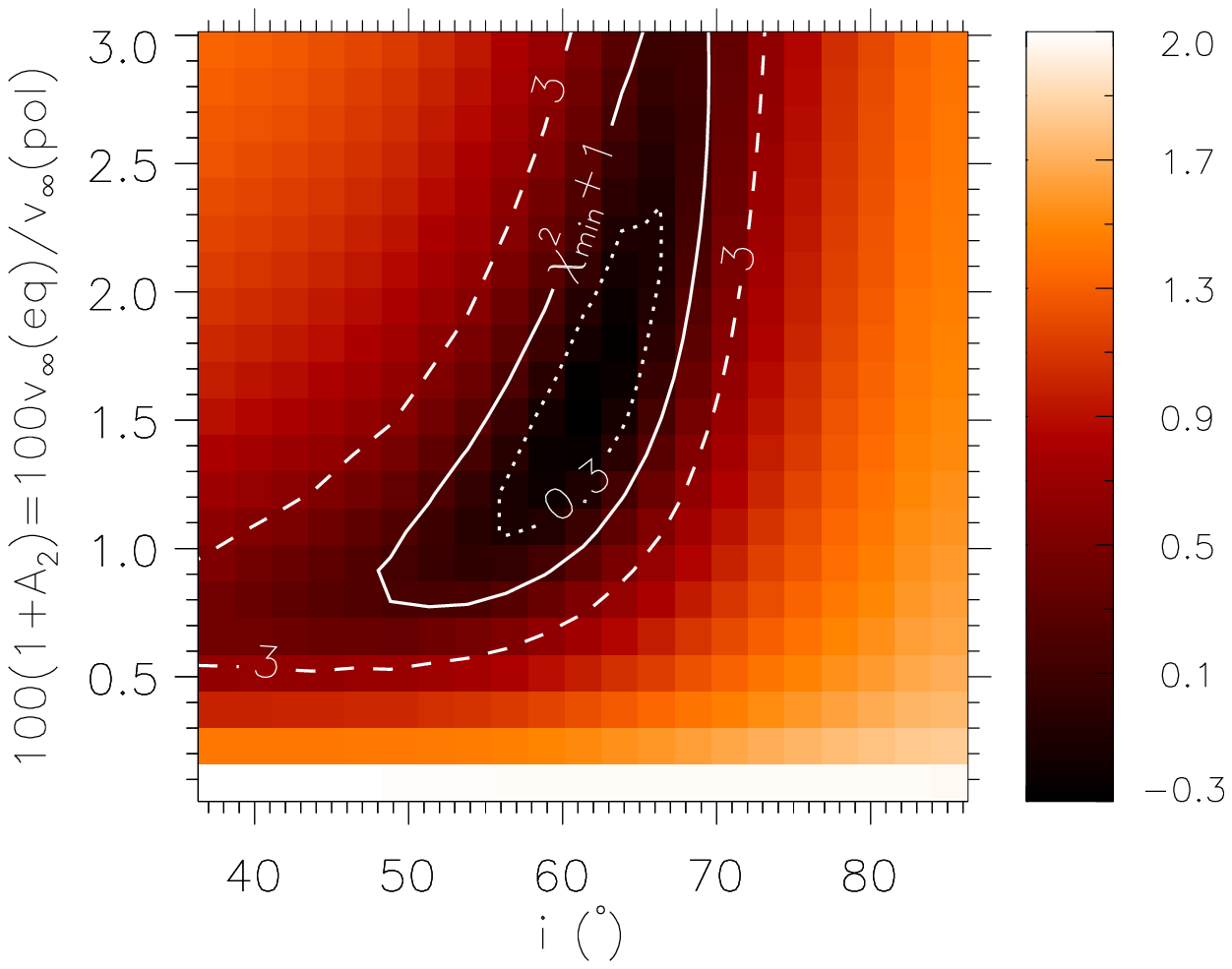}
\includegraphics[width=0.33\hsize,draft=false]{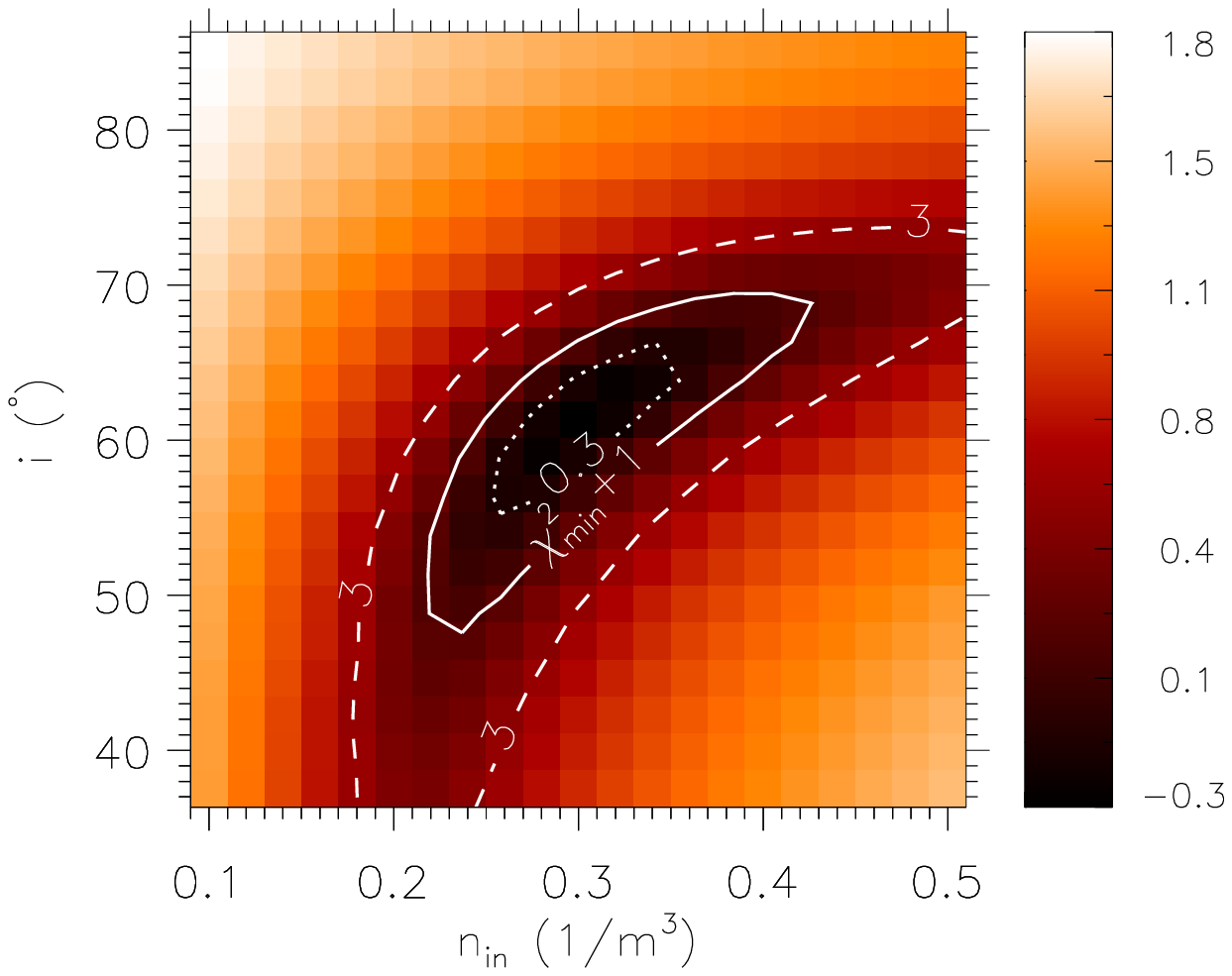}
\includegraphics[width=0.33\hsize,draft=false]{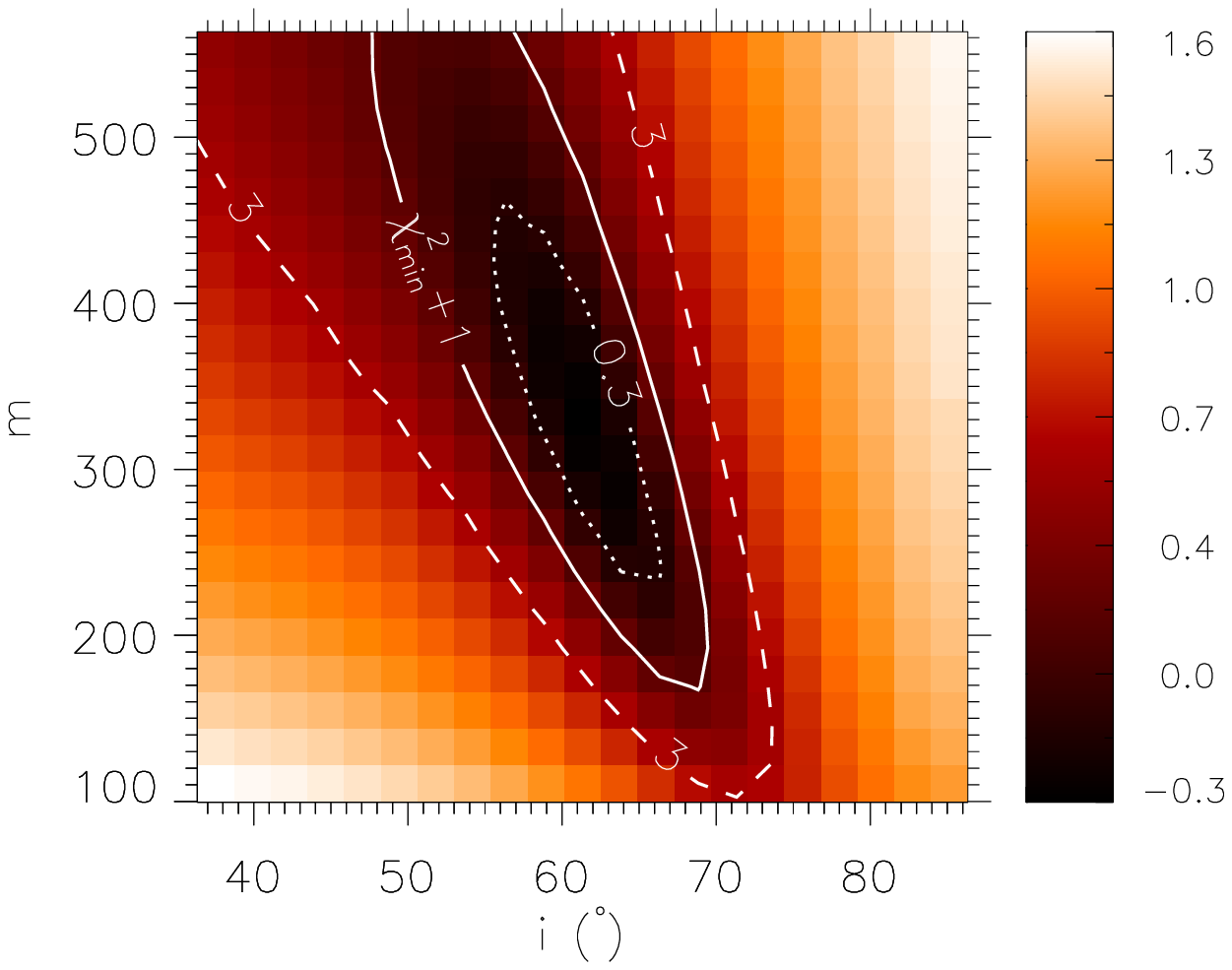}
\includegraphics[width=0.33\hsize,draft=false]{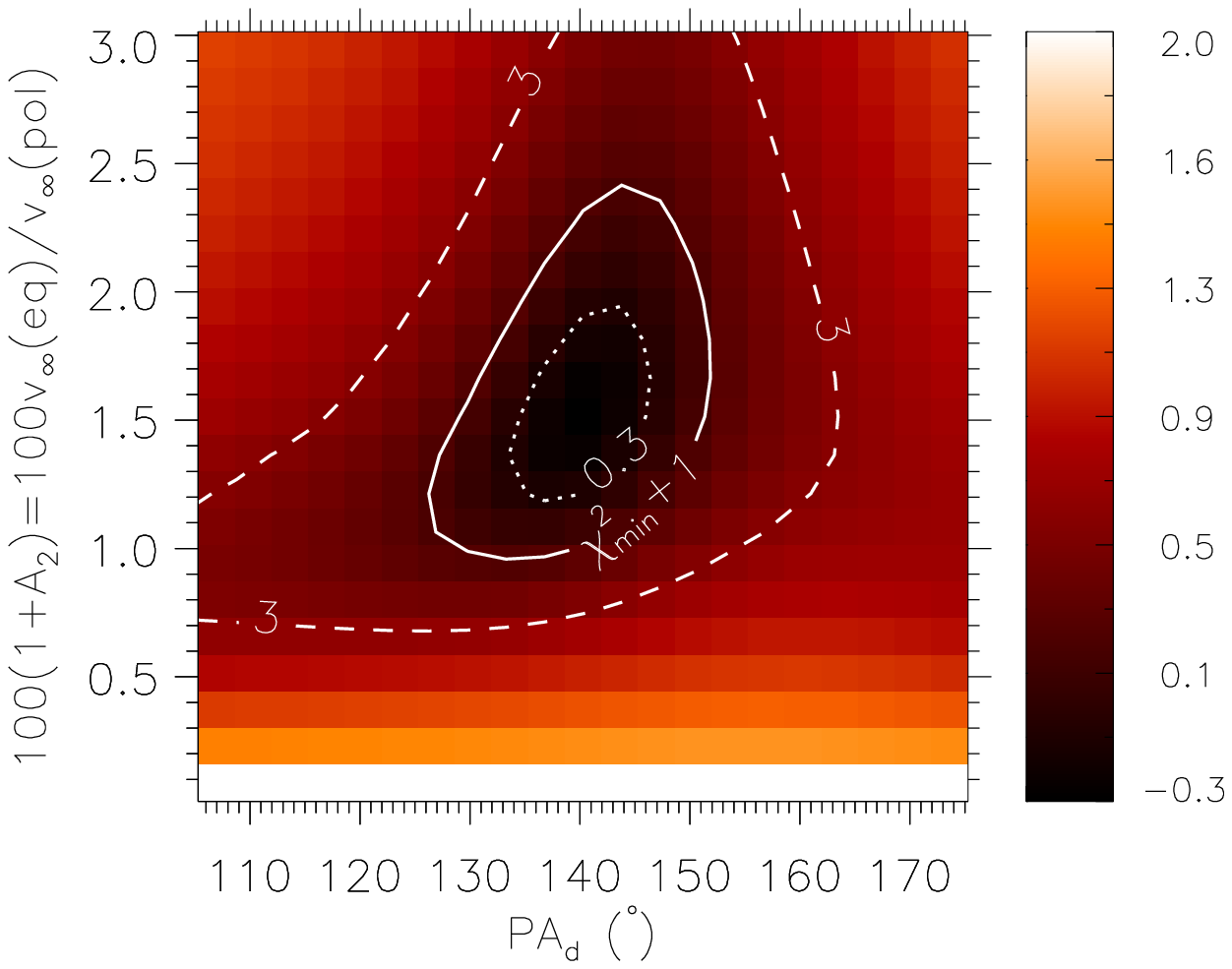}
\includegraphics[width=0.33\hsize,draft=false]{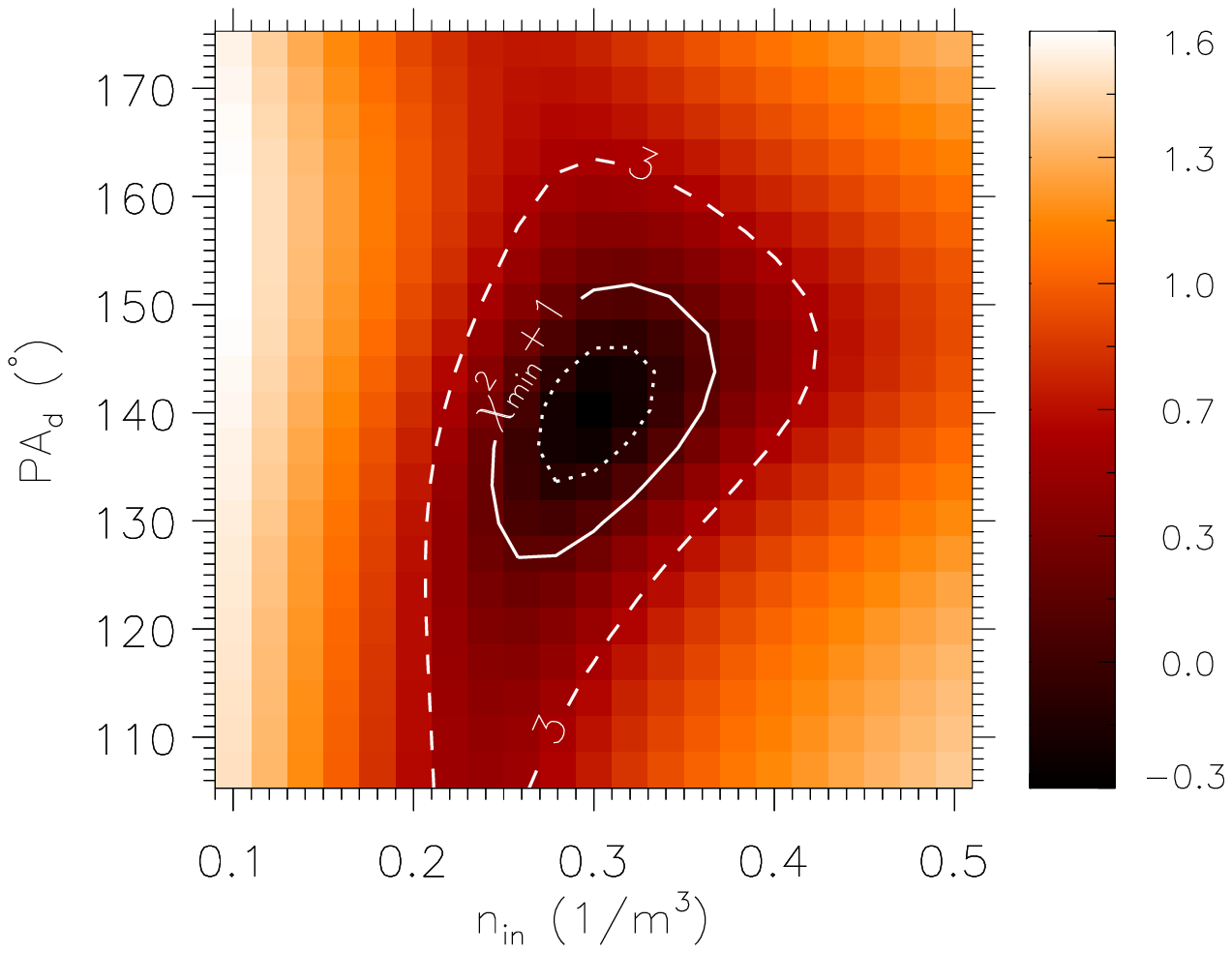}
\includegraphics[width=0.33\hsize,draft=false]{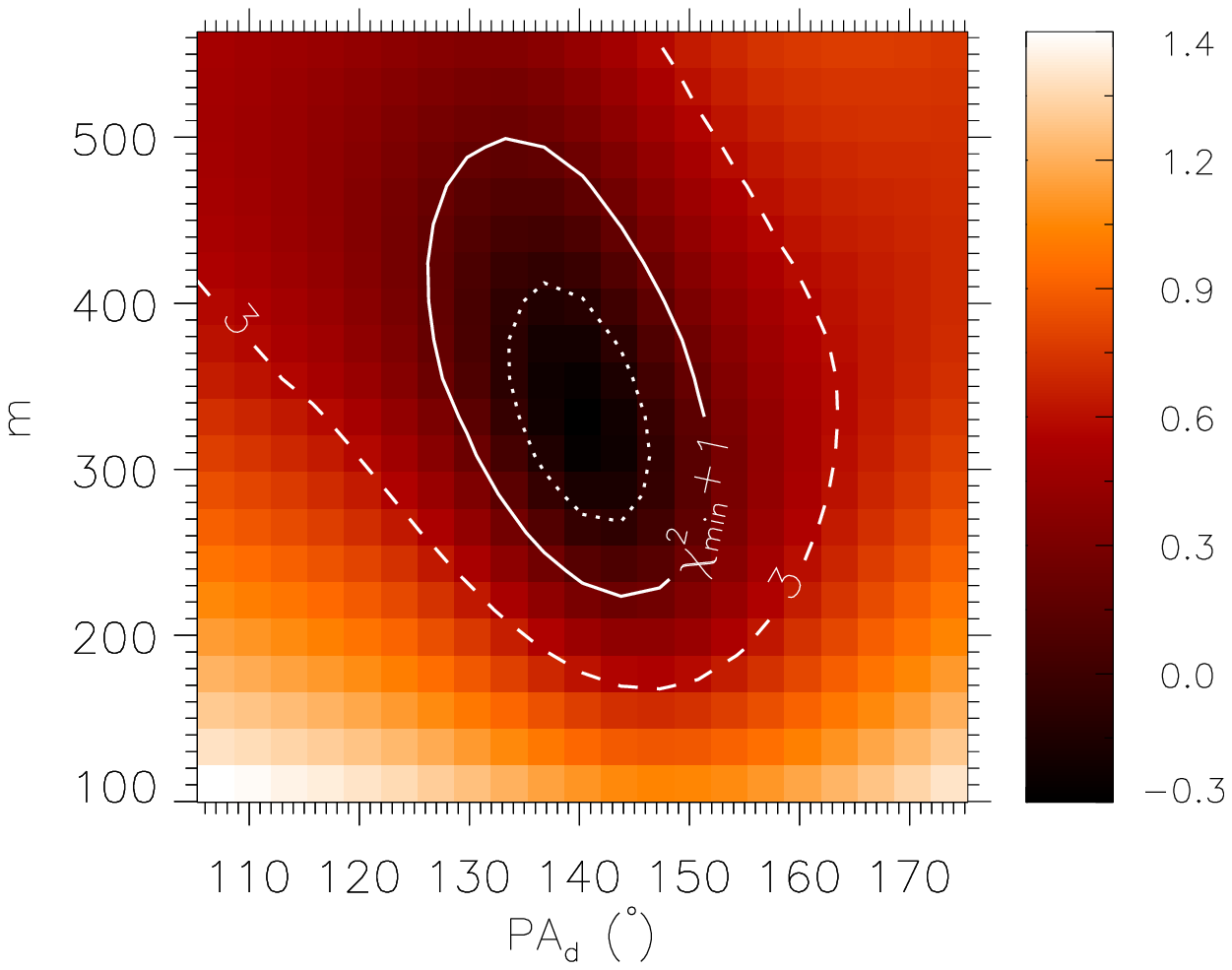}
\includegraphics[width=0.33\hsize,draft=false]{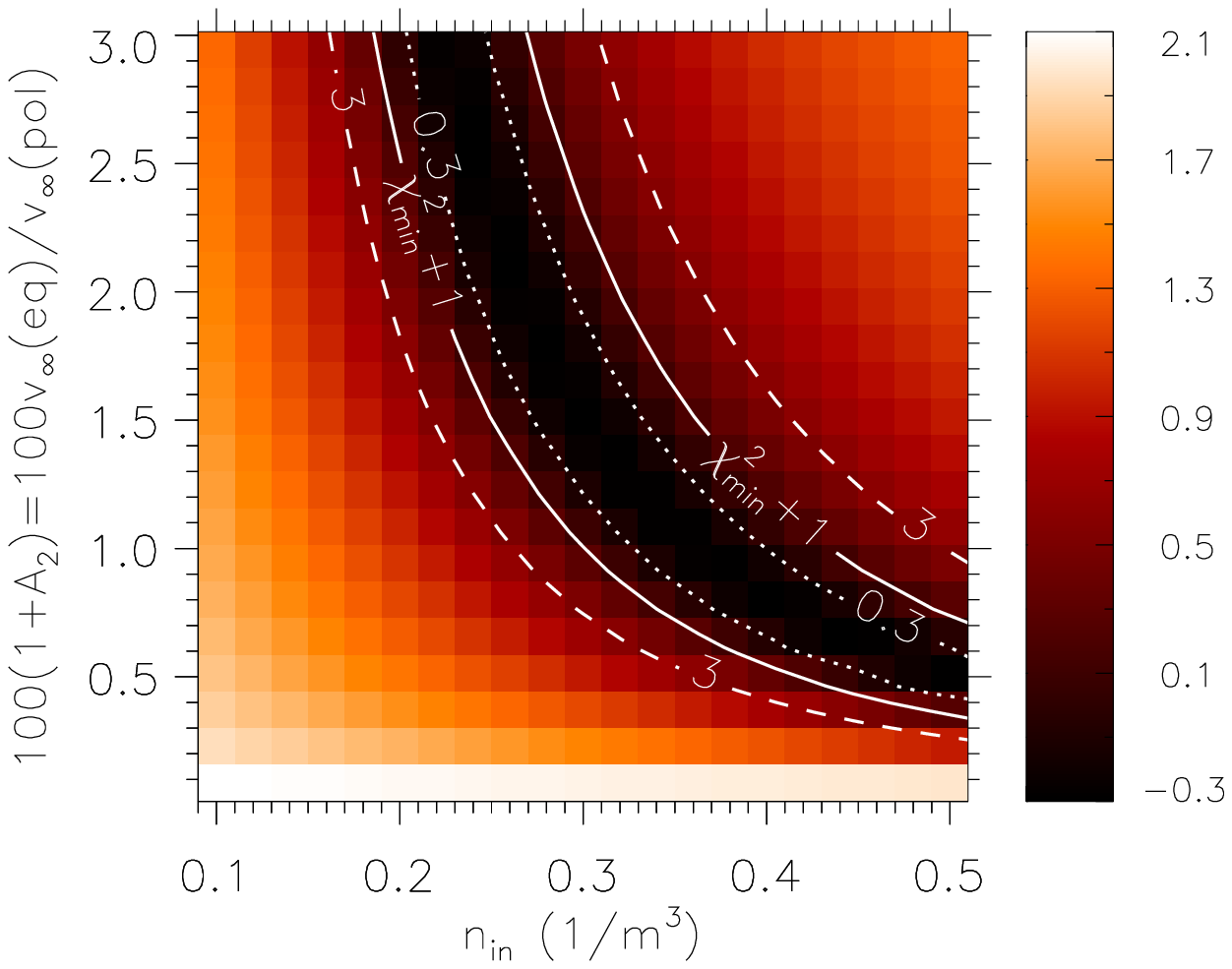}
\includegraphics[width=0.33\hsize,draft=false]{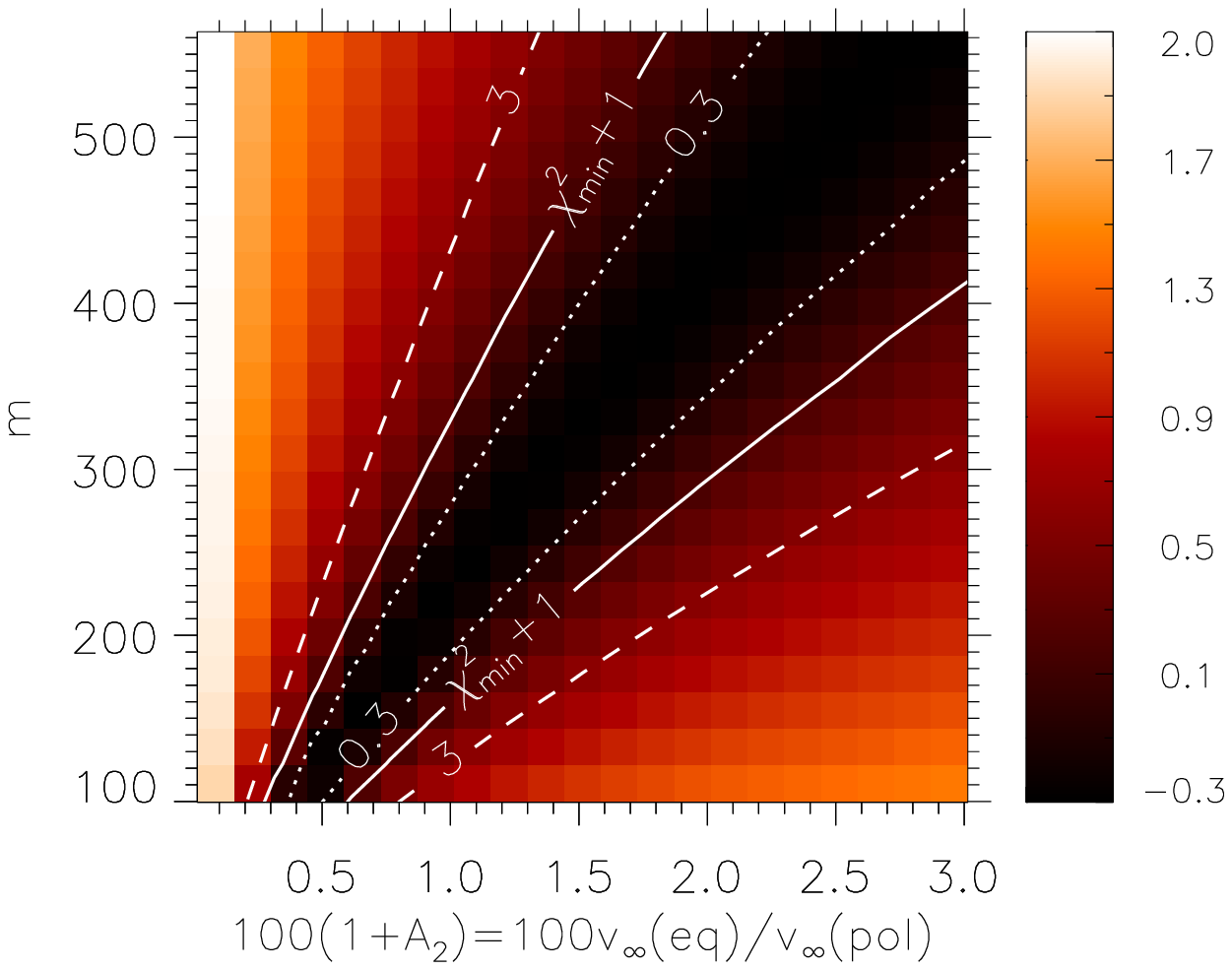}
\includegraphics[width=0.33\hsize,draft=false]{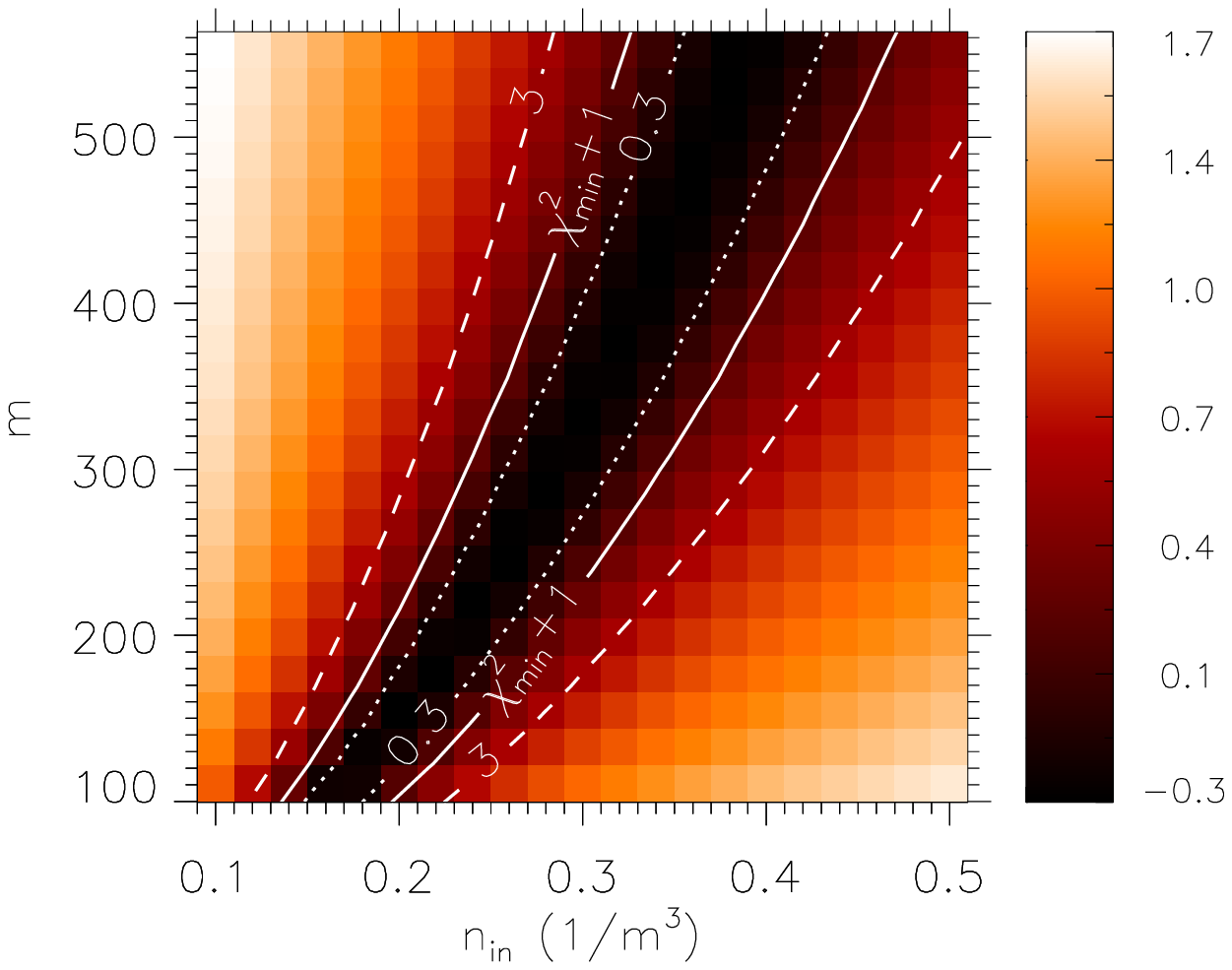}
\caption{Continuation of Fig.~\ref{fig:chi2_maps_2}. \label{fig:chi2_maps_3}}
\end{figure*}

\section{Discussion \label{discussion}}


\subsection{Geometrical parameters ($\PAd$, $\Rin$, and $i$)}

Let us first compare the derived geometrical parameters ($\PAd$, $\Rin$, and $i$) with those previously obtained by DS07 from elliptical Gaussian models fitted on a sub-set of the VLTI/MIDI data used here.

The geometrical parameter $\PAd$ can be directly compared with the major-axis position angle of the ellipse previously determined by DS07  ($\simeq143\degr-145$\degr). As expected, the two estimates of $\PAd$ are identical within their error bars.

Because the bulk of the thermal IR emission comes from the internal regions of the disc, one can expect the inner dust angular radius ($\Rin/d$) to be comparable to (or slightly smaller than) the major-axis half width at half maximum (HWHM) of an elliptical Gaussian. Indeed, the $\Rin/d$ derived here agrees with the major-axis HWHM ($=0.5$FWHM) given by DS07: $4.5 < \mathrm{HWHM~(mas)} < 8.0$.

Domiciano de Souza et al. (2007) estimated a CSE viewing angle $i\sim30\degr-60\degr$ from the minor- to major-axis ratio of the fitted elliptical Gaussian model. This estimate agrees fairly well with the more precise determination of this parameter given here.

The comparison of these parameters shows on one hand that the geometrical parameters obtained with FRACS agree with those obtained from a simpler approach using analytical models. On the other hand, this comparison clearly shows that, at the cost of somewhat higher but similar computing times, FRACS gives us access to physical parameters of CPD-57\degr\,2874 that cannot be extracted from simple geometrical analytical models.

\subsection{Continuum emission from the central source ($\Islref$ and $\alpha$)}

In contrast to what may be initially expected, we show below that the central source (star and continuum emission components such as free-free, free-bound) contributes to almost half of the total mid-IR radiation of CPD-57\degr\,2874. The total $10\micron$ flux of the best model for $d=1.7$~kpc is $\ftot=7.9\times10^{-13} \W\,\m^{-2}\,\micron^{-1}$ (see Fig.~\ref{fig:midi_flux_cpd}). The contribution of the dust CSE alone to the $10\micron$ flux computed with FRACS is $\fd=4.4\times10^{-13} \W\,\m^{-2}\,\micron^{-1}$. It follows therefore that the $10\micron$ flux from the CSE is $\simeq56\%\ftot$ and from the central source is $\simeq44\%\ftot$. Similar results are obtained for $d=2.5$~kpc.

We note that in the particular case of CPD-57\degr\,2874 where $i\simeq60\degr$ and where the dust is confined in a relatively narrow disc (opening angle $\sim7\degr$), these relative flux contributions can be directly obtained from $\fs$ (derived from $\Islref$; see Table~\ref{tab:fit_parameters}). This is valid if there is no absorption of the central regions by the CSE.

Although the uncertainties on the spectral index $\alpha$ are relatively high ($\simeq50\%$), the derived value ($=2.4$) suggests an important contribution of free-free continuum radiation from the central regions (Felli \& Panagia 1981).

\subsection{Temperature structure of the dusty CSE ($\Tin$ and $\gamma$)}

The dust temperature at the inner radius $\Tin$ is found to be $\simeq1500$~K (the imposed upper limit for the fit), which is consistent with the definition of $\Tin$ itself and with the chosen silicate dust composition. The large upper limit uncertainties in $\Tin$ and the fact that the best-fit $\Tin$ is 1500~K for $d=2.5$~kpc indicate that a $\Tin$ slightly higher than $1500$~K could still be compatible with the observations. A higher $\Tin$ value is consistent with, for instance, different dust compositions. However, we prefer to keep our choice of dust composition and to have a relatively large upper limit uncertainty in  $\Tin$, since the present observations do not provide strong constraints on the exact dust composition.

The derived value for the coefficient of the temperature profile $\gamma$ indicates an almost linear decrease of the disc temperature as a function of $r$. The steepness of this temperature profile lies between those expected for a non irradiated, adiabatically cooling disc ($T \propto r^{-4/3}$) and a reprocessing disc ($T \propto r^{-3/4}$) (c.f. Porter 2003). This implies that a non negligible part of the reprocessed radiation from the inner parts of the disc escapes without being re-absorbed, so that the disc cools down faster than highly optically thick discs, without of course reaching the limit of a purely adiabatic cooling.

\subsection{Parameters related to the density law ($\rhoin$, $A_2$, and $m$)}

Our results suggest that the observed mid-IR visibilities and fluxes cannot strongly constrain each individual parameter related to the density law: $\rhoin$, $A_2$, and $m$. The $\chi^2$ maps show a significant correlation between these three parameters, indicating that they are degenerated for the available VLTI/MIDI observations. Even if their uncertainties are significant, the values of these parameters at $\chi^2_\mathrm{min}$ suggest

\begin{itemize}
  \item a low inner density $\rhoin$ corresponding to relatively low CSE optical depth in the mid-IR ($\la0.2$ along the line of sight around $10\micron$).
  \item a polar-to-equatorial terminal velocity ratio $v_\infty(0\degr)/v_\infty(90\degr)=1/(1+A_2)\simeq50$, compatible with the values found in the literature (e.g. Zickgraf 2003).
  \item a high value for $m$, translating into a quite narrow opening angle ($\leq10\degr$) for the dust disc.
\end{itemize}

Because $\rhoin$, $A_2$, and $m$ are not well constrained, we fitted the observations by fixing these parameters to their values in Table~\ref{tab:fit_parameters} in order to investigate their influence on the remaining parameters. We have also fixed the $\Tin$ to $1500$~K. The fit was performed for $d=1.7$~kpc, starting from slightly different values from those in Table~\ref{tab:fit_parameters} .  The $\chi^2$ and values obtained for the free parameters ($\Islref$, $\alpha$, $\gamma$, $\Rin$, $i$, $\PAd$) are essentially the same as in Table~\ref{tab:fit_parameters} (differences are only a small fraction of the parameter uncertainties). We have also checked that the uncertainties on the other parameters are not affected by the fact that $\rhoin$, $A_2$, and $m$ are not well constrained.

%
\begin{table}[!t]
\begin{minipage}[t]{\linewidth}
\caption{Best-fit model parameters derived for CPD-57\degr\,2874 from a $\chi^2$ minimisation using a model with fewer free parameters then the initial one (c.f. Sect.~\ref{simple_model}). Uncertainties were estimated from the Levenberg-Marquardt (LM) algorithm and  can be considered as lower limits to the errors on the derived parameters (see discussion in the text).}
\label{tab:fit_parameters2}
\centering
\renewcommand{\footnoterule}{}  
\renewcommand{\arraystretch}{1.4} 
\begin{tabular}{c | c  c | c  c }     
\hline
Adopted distance &  \multicolumn{2}{c} {$d=1.7$~kpc} &  \multicolumn{2}{| c} {$d=2.5$~kpc} \\
Reduced $\chi^2_\mathrm{min}$  &  \multicolumn{2}{c} {$\chirmin^2=0.54$}    &   \multicolumn{2}{| c} {$\chirmin^2=0.56$}    \\
\hline
Model parameters & value  &  error   &   value  & error \\
                             &           &  LM      &             &  LM \\
\hline
$\Islref$ ($10^{5}\,\W\,\m^{-2}\,\micron^{-1}\,\str^{-1}$)  &
$ 2.1$  & $_{-0.1}^{+0.1}$ &   $ 4.2$  & $_{-0.1}^{+0.1}$  \\
$\alpha$  &
$ 2.4$  &  $_{-0.2}^{+0.2}$ &  $ 2.4 $ &  $_{-0.2}^{+0.2}$  \\
$\gamma$  &
$ 0.92 $ & $_{-0.07}^{+0.07} $  &  $ 0.85 $ & $_{-0.05}^{+0.05} $   \\
$\Rin$ (AU)  &  
$ 11.0$ & $_{-2.0}^{+2.0}$ &  $ 13.9 $ & $_{-2.4}^{+2.4}$  \\
$i$ (\degr)  &
$ 60.5 $ & $_{-1.5}^{+1.5}$ &   $ 59.3 $ & $_{-1.5}^{+1.5}$  \\
$\PAd$ (\degr)  &
$ 139.8$ & $_{-1.0}^{+1.0} $ &  $ 139.3 $ & $_{-1.0}^{+1.0}$ \\
%
$\rhoin$ (m$^{-3}$)  &   
$ 0.09$ &  $_{-0.06}^{+0.06} $ & $ 0.11 $ &   $_{-0.07}^{+0.07} $\\
$\thetadisc$ (\degr)  &  
$7.5$ & $_{-4.4}^{+4.4} $ & $5.9$  &  $_{-4.4}^{+4.4} $\\
%
\hline
\end{tabular}
\end{minipage}
\end{table}



\subsection{Data analysis from a model with fewer free parameters  \label{simple_model}}

Thanks to the data analysis performed here we found that some parameters of the sgB[e] model adopted for CPD-57\degr\,2874 cannot be well constrained from the available VLTI/MIDI data. Of course this is not necessarily a general conclusion because it depends on the nature of the studied target and on the spectro-interferometric data available. 

The results from our analysis of CPD-57\degr\,2874 indicate that it is justified to consider a simplified version of the model described in Sect~\ref{fracs}. As shown in the previous section, the parameters related to the density law are the less constrained by the data. Let us thus consider an alternative density law where the number density of dust grains is given by
\begin{equation}
  \label{eq:density2}
  n(r,\theta)=\left\{
  \begin{array}{l@{\, ; \;}l}
\rhoin\,\left(\frac{\Rin}{r}\right)^2 & 90\degr-0.5\thetadisc \leq \theta \leq 90\degr+0.5\thetadisc \\
0 & \theta < 90\degr-0.5\thetadisc \,\mathrm{and}\, \theta >  90\degr+0.5\thetadisc \\
\end{array} \right.
  \end{equation}

The parameters $A_1$, $A_2$, and $m$ are not present in this simpler density prescription. Only $\rhoin$ and $\thetadisc$ are necessary to define the density structure. Based on our previous results we have also fixed $\Tin$ to $1500$~K. The number of free parameters is thus reduced from 10 to 8, namely, $\Islref$, $\alpha$, $\gamma$, $\Rin$, $i$, $\PAd$, $\rhoin$, and $\thetadisc$.

Following the previous procedure we have performed a $\chi^2$ minimisation on the visibilities and fluxes using a LM algorithm. The results are shown in Table~\ref{tab:fit_parameters2}. Note that the best-fit parameters completely agree (well within the uncertainties) with the previous values obtained with the initial model (Table~\ref{tab:fit_parameters}). 

Only $\rhoin$ shows an important difference compared to the previously tested model. Contrarily to the previous density law (Eq.~\ref{eq:density}), the new one (Eq.~\ref{eq:density2}) assumes that the density is constant along a given $r$ inside the dust disc. In order to obtain similar fluxes and optical depths as before, $\rhoin$ has to be somewhat smaller than the previous value.

Table~\ref{tab:fit_parameters2} also provides the parameter uncertainties estimated with the LM algorithm, which are smaller than those estimated from the $\chir^2$ maps (given in Table~\ref{tab:fit_parameters}). These smaller errors appear because all data points are assumed to be independent and no covariance matrix is used in the error estimations. The estimated errors from LM in Table~\ref{tab:fit_parameters2} can then be considered as lower limits, while the errors from the $\chir^2$ maps can be considered as upper limits to the parameter uncertainties (see discussion in Sect.~\ref{data_analysis}).

\section{Conclusions \label{conclusions}}

The dusty CSE of the Galactic sgB[e] CPD-57\degr\,2874 was spatially resolved thanks to mid-IR spectro-interferometric observations performed with the VLTI/MIDI instrument. Several physical parameters and corresponding uncertainties of this star were derived from a $\chi^2$ minimisation and from the analysis of $\chi^2$ maps. The physical quantities derived include the inner dust radius, relative flux contribution of the central source and of the dusty CSE, dust temperature profile, and disc inclination (refer to Table~\ref{tab:fit_parameters} and Sect.~\ref{discussion} for details).

To our knowledge, this is the first direct determination of physical parameters of the dusty CSE of a B[e] supergiant based on interferometric data and using a model-fitting approach from a $\chi^2$ minimisation. This was possible thanks to FRACS, which adopts a parametrised description of the central object and of the dusty CSE combined to a simplified radiative transfer (no scattering). Its main advantage is computation speed ($<10$~s per monochromatic image with $300\times300$ pixels). Because it is fast, FRACS allows us (1) to explore a large parameter space domain for a true global $\chi^2$ minimisation and (2) to more realistically estimate the uncertainties on the best-fit parameters. We would like to recall that contrarily to a model such as FRACS,  simple geometrical models do not allow a simple and direct access to physical parameters and uncertainties of the dusty CSE.

{Future complementary observations could be included to measure new CSE parameters and/or to reduce the uncertainties and the correlations on some parameters that were not strongly constrained by the VLTI/MIDI observations alone. Consistently with the domain of validity of FRACS, these complementary data should be obtained at wavelengths above the near-IR, where the dust starts to contribute to a significant amount of the stellar flux (UV to visible observations cannot be consistently modelled by FRACS).  For example, the closure phase information from the MATISSE beam combiner (second generation instrument on VLTI; Lopez et al. 2009, Wolf et al. 2009) could inform on the disc inclination and opening angle, which are parameters that can influence the symmetry of the mid-IR intensity distribution. Also, near-future interferometric observations in the millimetre with ALMA will be more sensitive than MIDI and MATISSE to the thermal emission from the colder regions of the dusty CSE. Thus, these observations will probably better constrain the temperature profile of the dust and also provide direct information on the structure and actual size of the CSE, allowing also for mass estimates. Moreover, this information on the colder dust is very important for the study of the evolutionary history of sgB[e] stars (e.g. mass and angular momentum losses, chemical enrichment and interaction with the close interstellar medium).}

\begin{acknowledgements}
This research used the SIMBAD and VIZIER databases at the CDS, Strasbourg (France), and NASA's ADS bibliographic services. M.B.F. acknowledges financial support from the Programme National de Physique Stellaire (PNPS-France) and the Centre National de la Recherche Scientifique (CNRS-France). M.B.F. also acknowledges Conselho Nacional de Desenvolvimento Cient\'ifico e Tecnol\'ogico (CNPq-Brazil) for the post-doctoral grant. We thank the CNRS for financial support through the collaboration program PICS France-Brazil. We also thank the referee for his useful and constructive comments, which helped us to improve the quality of this work.
\end{acknowledgements}

\listofobjects

\end{document}